\documentclass{ptephy}
\usepackage{url} 
\pdfoutput=1

\newcommand{\be}{\begin{equation}}
\newcommand{\ee}{\end{equation}}
\newcommand{\beq}{\begin{eqnarray}}
\newcommand{\eeq}{\end{eqnarray}}
\newcommand{\bra}[1]{\ensuremath{\langle #1 |}}
\newcommand{\ket}[1]{\ensuremath{| #1 \rangle}}

\def\1{\mathchoice{\rm 1\mskip-4.2mu l}{\rm 1\mskip-4.2mu l}{\rm
        1\mskip-4.6mu l}{\rm 1\mskip-5.2mu l}}
 \def\1{\mathchoice{\rm 1\mskip-4.2mu l}{\rm 1\mskip-4.2mu l}{\rm
        1\mskip-4.6mu l}{\rm 1\mskip-5.2mu l}}
\def\ket#1{\vert#1\rangle}
\def\bra#1{\langle #1\vert}
\def\braket#1#2{\langle #1\vert#2\rangle}
\def\ketbra#1#2{\vert#1\rangle\langle#2\vert}

\def\3ph{\frac{3\pi}{2}}

\def\and#1{\hspace{#1mm}\textrm{and}\hspace{#1mm}}
\def\ar#1{\longmapsto\hspace{-7mm}^{\rm{#1}}\hspace{3mm}}

\begin{document}

\title{Fundamental phenomena of quantum mechanics explored with neutron interferometers}

\author{
\name{J\"{u}rgen Klepp}{1,\dag}, \name{Stephan Sponar}{2,\dag}, and \name{Yuji Hasegawa}{2,\ast,}
\thanks{These authors contributed equally to this work.}
}

\address{
\affil{1}{University of Vienna, Faculty of Physics, Boltzmanngasse 5, A-1090 Wien, Austria}
\affil{2}{Atominstitut, Vienna University of Technology, A-1020 Wien, Austria}
\email{Hasegawa@ati.ac.at}
}

\begin{abstract}%
Ongoing fascination with quantum mechanics keeps driving the development of the wide field of quantum-optics, including its neutron-optics branch.  
Application of neutron-optical methods and, especially, neutron interferometry and polarimetry has a long-standing tradition for experimental investigations of fundamental quantum phenomena.  
We give an overview of related experimental efforts made in recent years. 
\end{abstract}

\subjectindex{xxxx, xxx}

\maketitle

\section{Introduction}

Since the early days of quantum mechanics, the peculiarities of this theory have not only fascinated and upset physicists, but have also become an issue of popular science. Within the realm of scientific everyday-life, the formalism of quantum mechanics is often merely applied as a well-oiled tool: As long as it makes correct predictions, many of its users do not need to reflect too much on what is going on behind the scenes. However, since living with the implications of quantum mechanics can constantly pose an intellectual challenge, its puzzling experimental consequences keep popping up continuously in physics literature. That is to say, because of the bewilderment quantum mechanics produces, numerous experiments have been carried out to put its predictions to the test. Therefore, it is probably one of the best-verified theories of physics. So far, it has not failed, but -- on the contrary -- experiments have unambiguously demonstrated the existence of a great many of weird phenomena. 
Among such quantum-optics experiments are those using electrons \cite{tonomuraRMP1987,sonnentagPRL2007}, photons \cite{panRMP2012}, ions \cite{leibfriedRMP2003,winelandRMP2013}, atoms \cite{cornellRMP2002,ketterleRMP2002,croninRMP2009}, large molecules \cite{hornbergerRMP2012}, optomechanical devices \cite{kippenbergScience2008,aspelmeyerPT2012}, superconducting circuits \cite{devoretLesHouches2004} and cavities \cite{raimondRMP2001}.  

A field that was early involved in investigations of quantum mechanics and also inspired many of the later undertakings mentioned above is neutron optics, in particular, neutron interferometry. 
Since its invention in 1974 \cite{rauchPLA1974}, numerous pioneering experiments have been carried out doing perfect-crystal neutron interferometry, taking advantage of the macroscopic beam separation of several centimeters to observe (and exploit) the wave-like aspect of neutrons \cite{rauchPetrascheck1978,bonseRauchBook1979,kleinRPP1983,
badurekPhysicaBC1988,rauchWernerBook2000,
kleinEPN2009,hasegawaNJP2011}. A method that -- due to its superior resilience against environmental disturbances -- complements split-beam experiments, is spin-interferometry \cite{mezeiLNIP1980}, a principle that bore neutron polarimetry. 
In the present paper, the latter term is understood as comprehending also the spin-manipulation techniques employed in 
spin-echo spectroscopy \cite{mezeiZP1972} and zero-field spin-echo spectroscopy \cite{golubPLA1987}. With neutron polarimeters, the interference between spin eigenstates or its entangled degrees of freedom is observed, mostly without spatial beam separation.

The purpose of the present article is to give an overview of the last 15 years' progress and development in experimental quantum physics, using neutron optics, with emphasis on neutron interferometry and neutron polarimetry. 
Here, for space-reasons, relevant ongoing investigations using similar or related methods \cite{fedorovPB1997,schoenPRC2003,blackPRL2003,huffmannPRC2004,pushinAPL2007,
pushinPRL2008,huberPRL2009,piegsaPRL2009,lemmelPRB2007,
lemmelPRA2010,lemmelACA2013,iannuzziPRL2006,
iannuzziPRA2011,frankJOPCS2012,ichikawaPRL2014}
had to be spared. 

The paper is organized as follows: 
In Sec.\,\ref{sec:devices}, neutron-optical devices and techniques with relevance to the rest of the paper are discussed. 
Section\,\ref{sec:History} offers a summary of earlier neutron-interferometry experiments, the majority of them beautiful textbook-like demonstrations of quantum-mechanical phenomena. 
Section\,\ref{sec:contextuality} is dedicated to investigations of quantum contextuality (of which quantum non-locality is a special case) and multi-partite entanglement of single-neutrons.
Ongoing studies of geometric-phase properties are explained in Sec.\,\ref{sec:TopologicalAndBerryPhases}, while Sec.\,\ref{sec:others} is devoted to special topics, such as a polarimeter study of an error-disturbance relation to complement Heisenberg's famous uncertainty relation.
Conclusion and outlook are offered in Sec.\,\ref{sec:concl}.

\section{Neutron-optical devices for investigation of quantum-mechanical phenomena}\label{sec:devices}

\subsection{Perfect-crystal neutron interferometer}\label{sec:PerfectCrystalNIFM}

\begin{figure}
\begin{center}
\includegraphics[width=100mm]{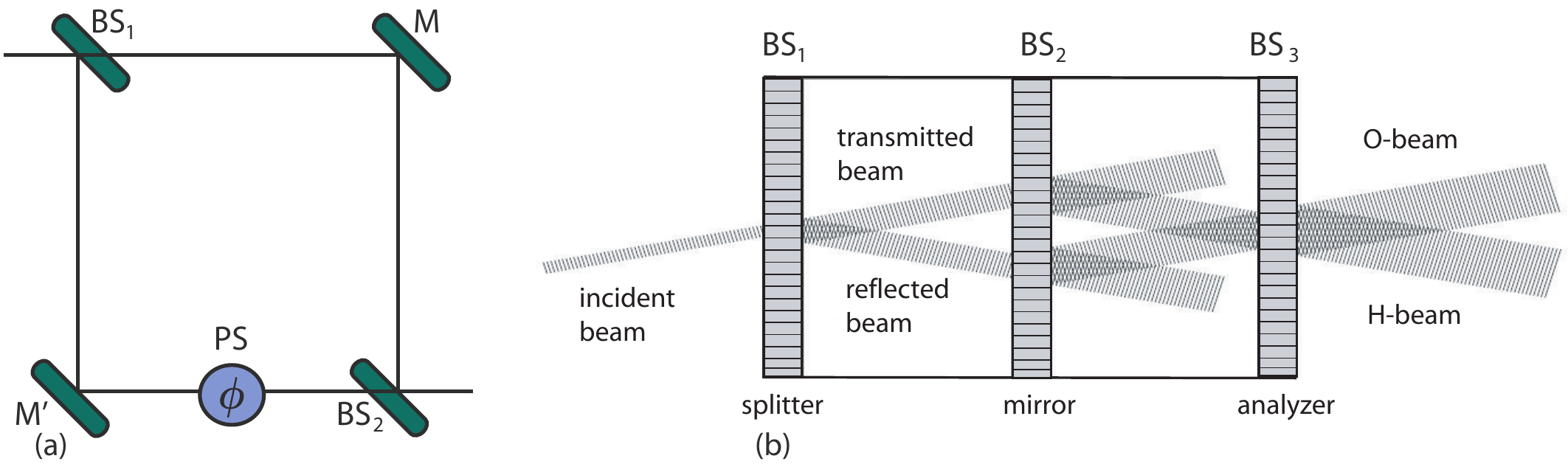}
\caption{(a) Optical Mach-Zehnder IFM. (b) Perfect-crystal neutron IFM of
triple-Laue (LLL) type.}
\label{fig:IFMTheory}
\end{center}
\end{figure}

In 1965, when semiconductor technology had advanced sufficiently to produce large monolithic silicon perfect-crystal ingots, Bonse and Hart
conceived a single-crystal interferometer (IFM) for X-rays \cite{bonseAPL1965}. This type of
IFM was then applied to neutrons resulting in the first
interference fringes observed in 1974 by Rauch, Treimer and Bonse
\cite{rauchPLA1974} at the 250kW TRIGA-reactor in Vienna.
In the experiment, a beam of neutrons -- massive particles -- is split by amplitude division, and superposed coherently after passing through different regions of
space. During this space-like separation of typically a few
centimeters the neutron wavefunction can be modified in phase and
amplitude in various ways. It can be manipulated via nuclear, magnetic, electric or gravitational potentials. Many different types of optical devices can be inserted. In the IFMs, neutrons exhibit \emph{self-interference}, since at most one single neutron propagates through the IFM at a given time.
The IFM is geometrically analogous to a common Mach-Zehnder IFM in light optics, as illustrated in 
Fig.\,\ref{fig:IFMTheory}.

\begin{figure}
\begin{center}
\includegraphics[width=150mm]{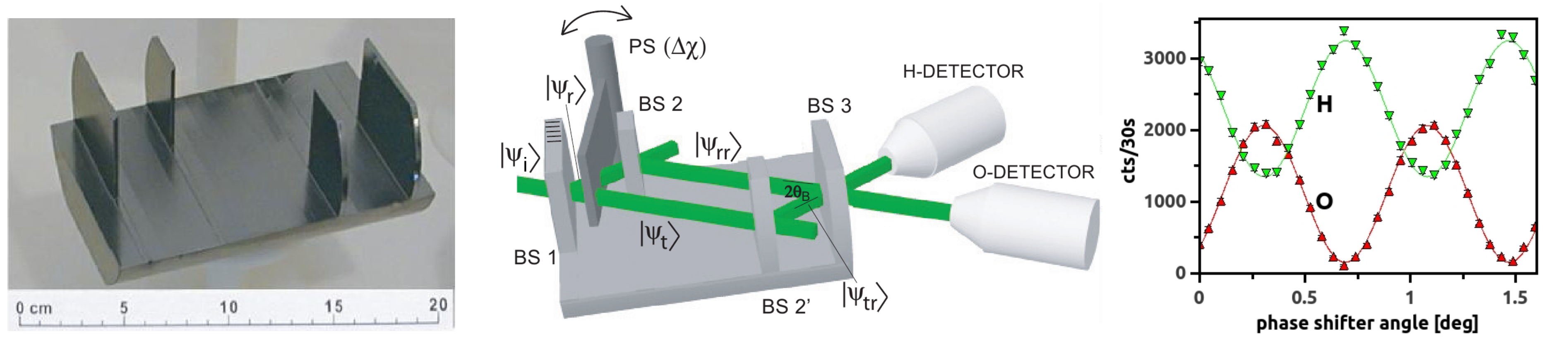}\caption {Left: Skew-symmetric perfect-crystal neutron
IFM. Center: Skew-symmetric perfect-crystal IFM with the incident beam $\ket{\Psi_i}$ being split at beamsplitter 1
(BS 1) and recombined at BS 3. The phase shifter PS induces a
relative phase difference ($\triangle\chi$) between the two
sub-beams to obtain interference fringes. Right: Interference fringes due to rotation of the phase shifter, observed in the 0- and
H-beams.}\label{fig:IFMPrinzipOsci}
\end{center}
\end{figure}

A neutron IFM consists of a single silicon
perfect-crystal, cut in
such a way that the incoming neutron beam is split by Bragg diffraction at the first
plate. For thermal neutrons -- with de Broglie wavelength of about 2\,\AA\, -- and silicon-crystals absorption is negligible. The sub-beams are split again by the
second plate. 
In analogy to the Mach-Zehnder IFM for light, the second
plate is often referred to as a mirror.
Elaborate geometries have been used: For instance, the skew-symmetric IFM in
Fig.\,\ref{fig:IFMPrinzipOsci}\,(left and center) has a split second plate to provide
more space for samples or neutron-optical devices to be inserted in one of the paths. 

At some point in the IFM, before the beams are recombined at the last plate (analyzer plate), a phase shifter is
inserted such that it is traversed by both beams. Its rotation changes the optical path-difference between the sub-beams and yields intensity oscillations behind the IFM [see
Fig.\,\ref{fig:IFMPrinzipOsci}\,(center and right)]. Usually the beams after the IFM are referred to as 0- and H-beam. The quantum states of 0- and H-beam are both superpositions of the states in the IFM paths as will be shown later in this Section. 
In the standard configuration, a monolithic triple-plate IFM [triple-Laue (LLL) geometry: surfaces of the plates
are parallel to each other and perpendicular to the reflecting net planes] is used. 

An intrinsic feature of diffraction by perfect crystals is the extremely small
width of the reflection curves of only a few arcseconds. This
presents a particular challenge for the alignment of beamsplitter,
mirror and analyzer crystal. The diffracting planes in all involved crystal slabs must be parallel to a small fraction of an arcsecond and the distance between them must be the same to an accuracy of a few microns. Thus, an elegant way to assure correct alignment is to cut the whole IFM out of a single, monocrystalline silicon ingot.

Behind the first plate (beamsplitter -- BS 1) the wave function is
found in a superposition of the transmitted ($t$) and the reflected ($r$)
sub-beams, as shown in Fig.\,\ref{fig:IFMPrinzipOsci}\,(center). In terms of state vectors, that superposition state after BS\,1 is given by
$
r\ket{\Psi_i}+t\ket{\Psi_i}=\ket{\Psi_r}+\ket{\Psi_t}$. The factors $t$ and $r$ are probability amplitudes with $\vert
t\vert^2+  \vert r\vert^2 = 1$. As mentioned above, an additional perfect-crystal slab produces an adjustable phase shift and one obtains
$e^{i\chi_{\rm I}}\ket{\Psi_r}+e^{i\chi_{\rm II}}\ket{\Psi_t}$, 
where $\chi_{\mbox{\scriptsize{I,II}}}=-Nb_c\lambda D_{\mbox{\scriptsize{I,II}}}$, with the thickness of
the phase shifter plate $D_{\mbox{\scriptsize{I,II}}}$ in paths I and II, the neutron wavelength $\lambda$, the
coherent scattering length $b_c$ and the atom number-density
$N$ of the phase shifter plate. By rotating the plate,
$\chi_{\mbox{\scriptsize{I,II}}}$ can be varied, introducing a phase
difference $\triangle\chi=\chi_{\rm II}-\chi_{\rm I}$. After passing the mirrors [termed BS 2 and BS 2\rq{} in Fig.\,\ref{fig:IFMPrinzipOsci}\,(center)], the state leaving the
IFM at the third plate in directions parallel to the incident beam -- the so-called 0-beam -- is denoted as
$\ket{\Psi_0}=trr\ket{\Psi_i}+e^{i\triangle\chi}rrt\ket{\Psi_i}$.
This yields intensity oscillations described by
\begin{equation}\label{eq:IFMContrastO}
I_0=\braket{\Psi_0}{\Psi_0}
=2\langle\Psi_i|\Psi_i\rangle |r|^4|t|^2\left(1+\cos\Delta\chi\right).
\end{equation}

Similarly, the intensity expected in the H-beam is written as
\begin{equation}\label{eq:IFMContrastH}
I_\textrm{H}=\braket{\Psi_{\rm{H}}}{\Psi_{\rm{H}}}
=\langle\Psi_i|\Psi_i\rangle\left(|t|^4|r|^2+|r|^6-2|r|^4|t|^2\cos\Delta\chi\right),
\end{equation}
with
$
\ket{\Psi_{\rm{H}}}=rrr\ket{\Psi_i}
+e^{i\triangle\chi}trt\ket{\Psi_i}
$.
The fringe visibility
(or contrast) of the oscillations is calculated as \mbox{$(I_{\rm max}-I_{\rm min})/(I_{\rm max}+I_{\rm min})$} and can theoretically become 1 in the 0-beam, while it depends on $|r|$ and $|t|$ for the H-beam. In practice, it is always less than the above theoretical prediction because of unwanted scattering, vibrations, temperature instabilities and the like. 
The wave functions corresponding to the state vectors described here
are calculated using dynamical theory of diffraction, which is
discussed in detail in \cite{rauchPetrascheck1978,searsBook1989,rauchWernerBook2000}.
$^3$He- and BF$_3$-gas detectors with high efficiency ($>$99\,\%) are used for detection of thermal neutrons. In these detectors, the nuclides of the high-pressure filling gas are converted into charged particles according to, for instance, the following reaction for $^3$He: n$+^3$He$\rightarrow ^3$H$+$p$+$0.764\,MeV.

\begin{figure}
\begin{center}
\includegraphics[width=90mm]{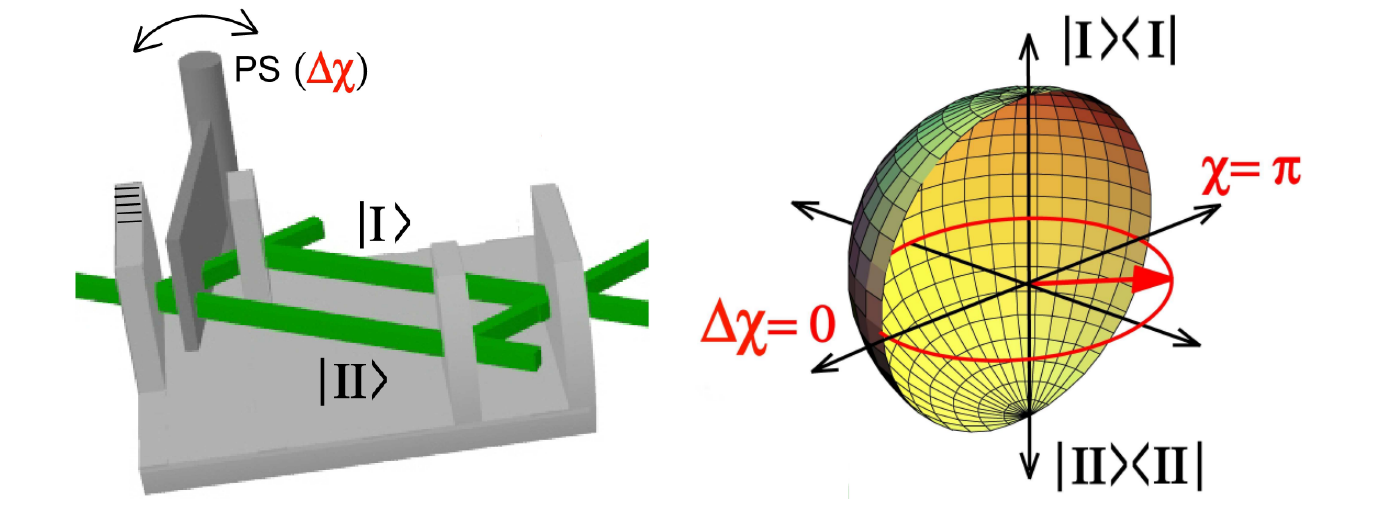}\caption {\label{fig:IFM2Level22}Bloch
 sphere representation of the neutron-IFM system (cf. Sec.\,\ref{sec:CoupledIFMLoops}). An equal superposition of 
$\ket{\textrm{I}} $ and $ \ket{\textrm{II}} $ is found on the equator of the Bloch sphere.}\label{fig:PathBlochSphere}
\end{center}
\end{figure}

It is important to note that the state of a neutron in an IFM can be treated formally as a two-level
system, where the two-dimensional Hilbert space is spanned by the
orthogonal states for paths $\ket{\textrm{I}}$ and $\ket{\textrm{II}}$, just like spin state-space is spanned by the spin-1/2 eigenstates $\ket{\!\Uparrow}$ and $\ket{\!\Downarrow}$. The
north- and south-pole of a Bloch-sphere are identified with
states $\ket{\textrm{I}}$ and
$\ket{\textrm{II}}$, respectively, each corresponding to a well-defined path. Thus, $\ket{\textrm{I}}$ and
$\ket{\textrm{II}}$ are eigenstates of the associated observables
$\ketbra{\textrm{I}}{\textrm{I}}$ and
$\ketbra{\textrm{II}}{\textrm{II}}$. An equally weighted
superposition of path eigenstates is therefore found on the equator
of the Bloch-sphere \cite{yurkePRA1986,hasegawaPRA1996}. The phase shifter induces a relative phase shift $\triangle\chi$
between the path states, denoted as
$\ket{\textrm{I}}+\ket{\textrm{II}} \ar{PS}
\ket{\textrm{I}}+e^{i\Delta\chi}\ket{\textrm{II}}$. $\Delta\chi$ determines the azimuthal angle on the Bloch-sphere, as illustrated in
Fig.\,\ref{fig:IFM2Level22}. 

\subsection{Manipulation of the neutron spin: spin-interferometry}\label{sec:SpinManipulation}

The neutron couples to magnetic fields via its permanent magnetic dipole-moment $\vec\mu$. The interaction is described by the Hamiltonian $H_{\rm mag}=-\vec\mu\cdot\vec B=-\mu\; \vec\sigma\cdot\vec B$, where $\mu=-1.91\,\mu_{\rm Nuc}$, with $\mu_{\rm Nuc}=5.051\times 10^{-27}$\,J/T (the nuclear magneton). $\vec\sigma=(\sigma_x,\sigma_y,\sigma_z)^T$ is the Pauli vector-operator consisting of the Pauli spin-matrices $\sigma_x$, $\sigma_y$ and $\sigma_z$. Stationary and/or time-dependent magnetic fields can be utilized for arbitrary spin rotations.

\subsubsection{Neutrons in a static magnetic field: Larmor precession}\label{Sec:TimeInDep}

\begin{figure}
\begin{center}
\includegraphics[width=80mm]{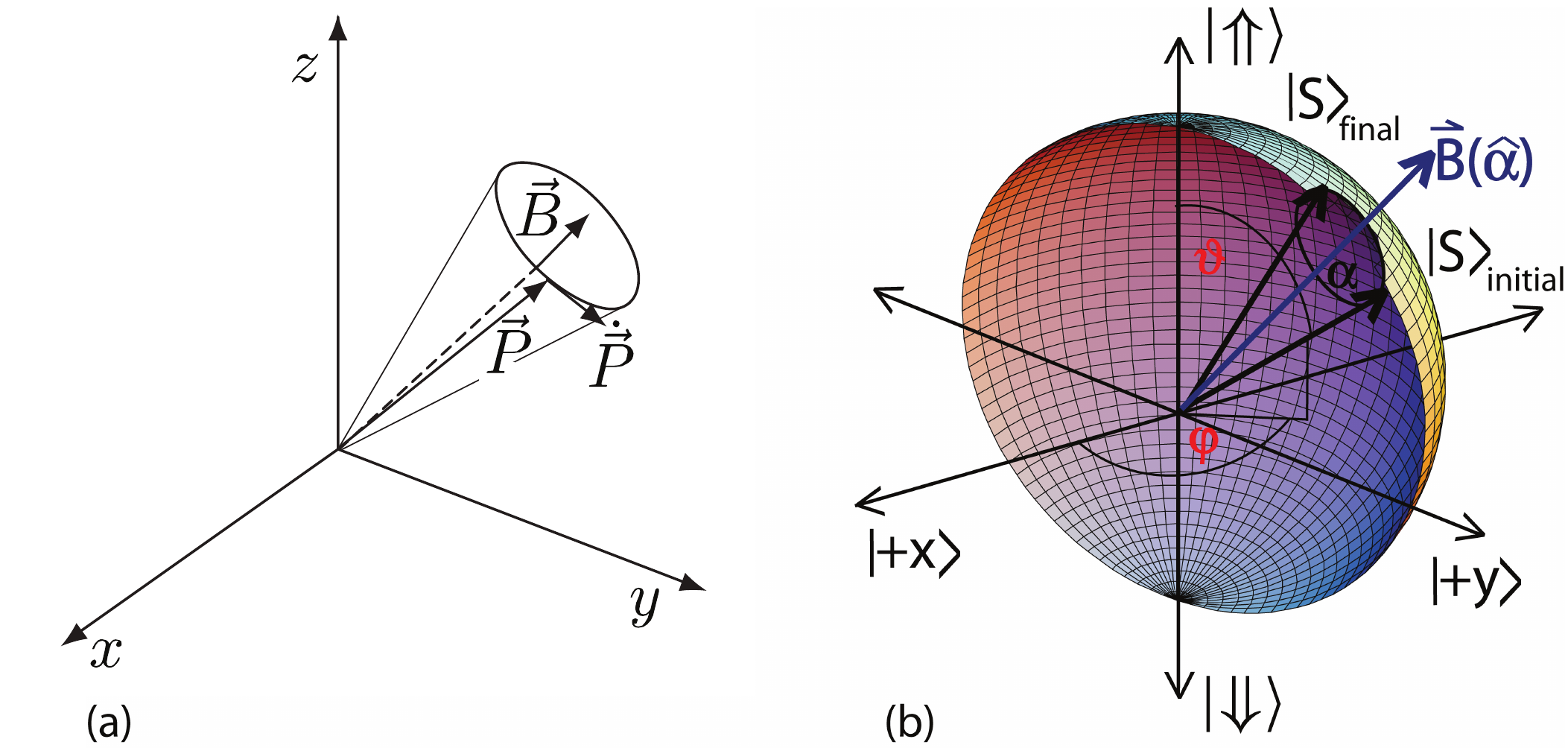}
\caption{(a) Motion of polarization vector in real space. (b) Bloch
sphere description of precession of an arbitrary spin state defined by polar angle
$\vartheta$ and azimuthal angle $\varphi$.}\label{fig:larmorall}
\end{center}
\end{figure}

\begin{figure}
\begin{center}
 \includegraphics[width=153mm]{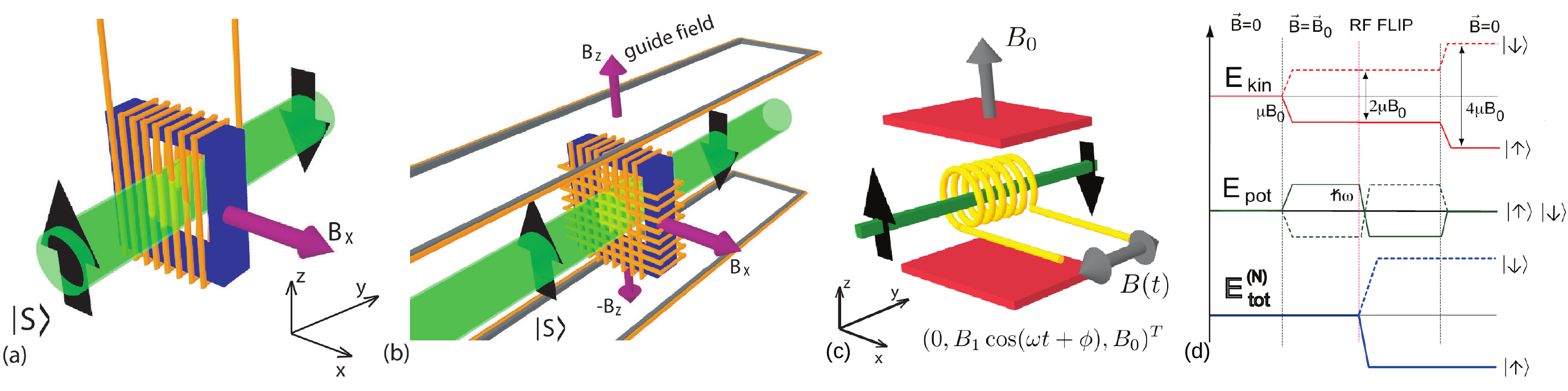}
\caption{(a) DC spin-flipper functional principle. (b) In practice, a second coil with its field pointing to the $-z$-direction (perpendicular to the original coil) is necessary to compensate the guide field. (c) Combination of static and oscillating magnetic field. (d) Energy scheme for the RF-flip process.}\label{fig:dcspinflipper}
\end{center}
\end{figure}

When a neutron beam is exposed to a stationary magnetic field, the motion of its polarization vector -- its vector components being the expectation values of the Pauli spin-matrices, $\vec P=\langle\Psi|\vec\sigma|\Psi\rangle$ -- is described by the Bloch-equation $d\vec P/dt =  \vec P \times\gamma\vec B$. That motion is called Larmor precession of the polarization vector about an axis defined by the magnetic field direction (cf. Fig.\,\ref{fig:larmorall}). $\vec P$ precesses at the so-called Larmor frequency $\omega_{\textrm{\scriptsize{L}}} = -\gamma|\vec B|$,
where $\gamma$ is the gyromagnetic ratio given by $\gamma=2\mu/\hbar$.
The Larmor-precession angle (rotation angle) solely depends on the magnitude of the applied magnetic field and the propagation time $\tau$ within the field and is given by  
\begin{equation}\label{3.4}
\alpha  = -\frac{2\mu}{\hbar}\int_0^\tau {B\,dt}=-\frac{2\mu}{\hbar}B\frac{L}{v},
\end{equation} 
where $L$ and $v$ are the length of the magnetic-field region traversed by the neutrons and the neutron velocity, respectively.
A direct-current (DC) spin-rotator -- essentially an aluminum frame with wire windings around it to form a coil of a couple of centimeter size -- and its working principle that is based on Larmor precession, are shown schematically in Fig.\,\ref{fig:dcspinflipper}\,(a,b).  

Using the formalism of quantum mechanics (QM), a spin rotation  through an angle $\alpha$ about an axis pointing in direction $\hat\alpha$ is described by the unitary transformation operator 
\begin{equation}\label{eq:unitary}
U(\vec\alpha)=\exp(-i\vec\sigma\vec\alpha/2) 
\end{equation}
that can be written as 
$
U(\vec\alpha)=\1\cos\alpha/2-i\vec\sigma\hat\alpha\sin\alpha/2
$. 
For instance, a superposition state $|\psi\rangle=1/\sqrt 2 (|\!\Uparrow\rangle+|\!\Downarrow\rangle)$ undergoing a rotation through the angle $\varphi$ about the $+z$-axis (defining the up-direction) transforms as 
$
U(\varphi\hat z)|\psi\rangle=1/\sqrt 2(e^{-i\varphi/2}|\!\Uparrow\rangle+e^{i\varphi/2}|\!\Downarrow\rangle)
$. 

\subsubsection{Neutrons in a time-dependent magnetic field: photon exchange}\label{Sec:Time}
 
A completely different physical situation arises when neutrons interact with a purely time-dependent magnetic field: Here, the total energy of the neutron is not conserved. Energy can be exchanged between neutrons and the radio-frequency (RF) field via photons of energy $\hbar\omega$. This behaviour is described by the dressed-particle formalism \cite{muskatPRL1987,summhammerPRA1993}. An oscillating RF-field and a static magnetic field -- a configuration used in nuclear magnetic resonance (NMR) -- is also capable of spin flipping. An oscillating RF field can be viewed as two counter-rotating fields. In the frame of one of the rotating components, the other is rotating at double-frequency and can be neglected (rotating-wave approximation; see, for instance, \cite{allenBook1987}). The static field component of magnitude $B_0$ is fully suppressed in case of frequency-resonance, i.e. for the oscillation frequency $\omega_{\rm res}=-\gamma B_0$.  If, in addition, the amplitude-resonance condition $B_1=\pi\hbar/2\tau\vert\mu\vert$ -- determining the amplitude of the rotating field $B_1$ -- is fulfilled, a spin flip occurs. A time-dependent phase shift emerges due to the RF-induced total-energy difference of the two spin eigenstates. That phase shift results in spin rotation even in field-free regions. Therefore it is referred to as zero-field precession in literature \cite{golubAJP1994}, which is exploited in zero-field spin-echo spectroscopy \cite{golubPLA1987}.
A consequence of the rotating-wave approximation is the so-called Bloch-Siegert shift, which gives rise to a correction term for the frequency-resonance now reading as $\omega_{\rm res}=2\vert\mu\vert B_0/\hbar[1+B_1^2/(16B_0^2)]$ \cite{blochPR1940}. 

The above explained combination of static and time-dependent magnetic fields is exploited in RF flippers, as depicted in Fig.\,\ref{fig:dcspinflipper}\,(c). Potential, kinetic and total neutron energies for RF-flipping are illustrated in Fig.\,\ref{fig:dcspinflipper}\,(d).

\subsubsection{Neutron Polarimeter}\label{sec:polarimeter}

\begin{figure}
\begin{center}
\includegraphics[width=4.5in]{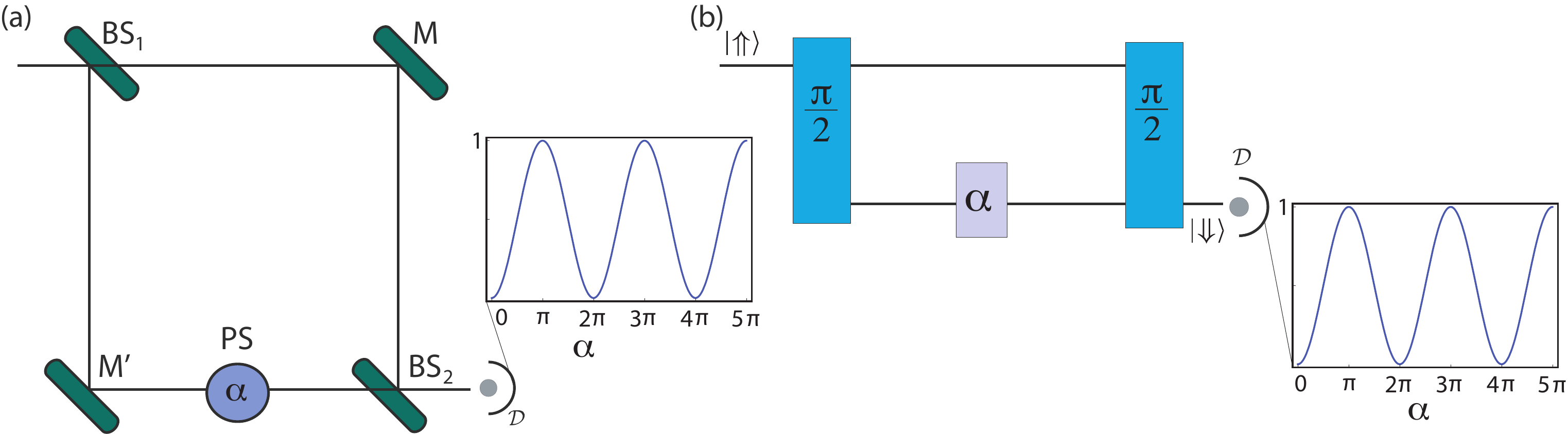}
\caption{(a) Scheme of a Mach-Zehnder IFM with two indistinguishable paths. (b) Scheme of the Ramsey IFM with two indistinguishable spin states. The conceptional analogy with the Mach-Zehnder IFM is evident.}
\label{fig:IFM_POL}
\end{center}
\end{figure}

\begin{figure}
\begin{center}
\includegraphics[width=4.9in]{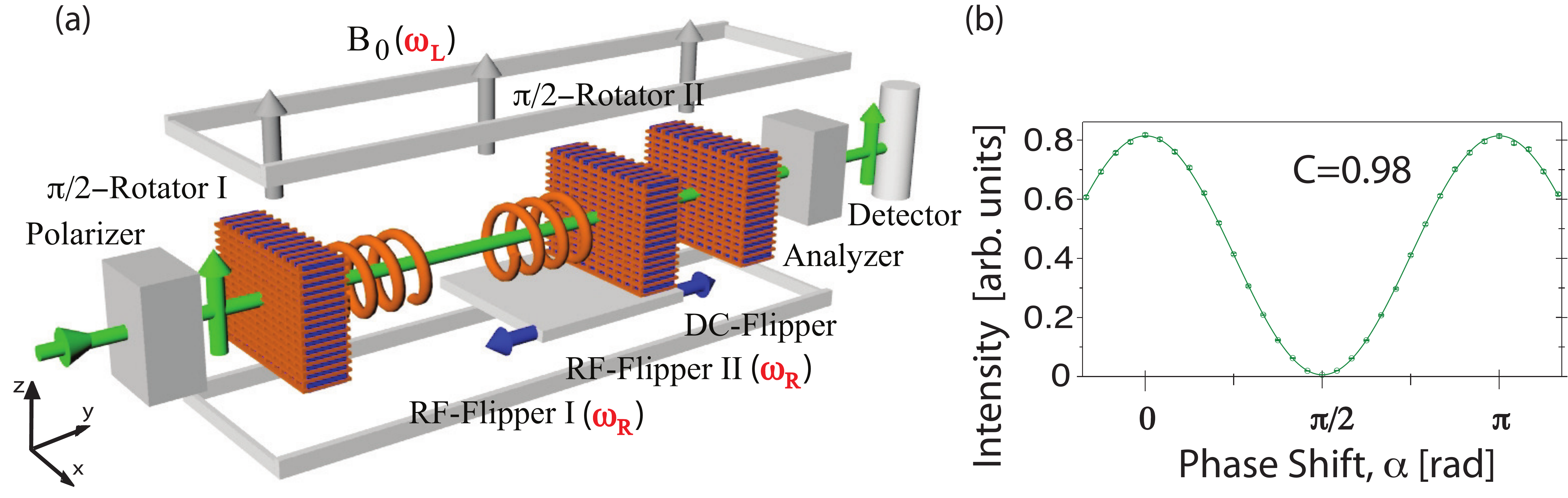}
\caption{(a) Experimental setup for separation of zero-field phase and Larmor-precession phase \cite{sponarPLA2008}. (b) Typical high-contrast (98\,\%) oscillations obtained in neutron polarimetry \cite{hasegawaNJP2012,sponarNJP2012}.}
 \label{fig:SetupZeroOsci}
\end{center}
\end{figure}

A combination of two $\pi/2$-pulses (triggering spin rotations) and a phase shift applied in between is generally referred to as Ramsey IFM in NMR \cite{hahnPR1950} and atomic physics \cite{chuNature1999}. In neutron optics, a similar scheme is usually called neutron polarimeter. An illustration of the polarimeter scheme in comparison to the IFM scheme is provided in Fig.\,\ref{fig:IFM_POL}. The first $\pi/2$-rotation (about the $+x$-direction, say) creates a coherent superposition of the orthogonal spin eigenstates by transforming the initial state $\ket{\!\Uparrow}$ to $1/\sqrt{2} (\ket{\!\Uparrow}+\ket{\!\Downarrow} )$. Before the second $\pi/2$-rotation probes it, a tunable phase shift $\alpha$ between the orthogonal spin eigenstates is induced (by, for example, a static magnetic field). Finally, the probability of finding the system in the state $\ket{\!\Uparrow}$ or $\ket{\!\Downarrow}$ is given by $
P_{\Uparrow,\Downarrow}=1/2(1\pm\sin\alpha)
$, predicting sinusoidal intensity oscillations [see, for instance, Fig.\,\ref{fig:SetupZeroOsci}\,(b)].

Neutron polarimetry has several advantages compared to perfect-crystal neutron interferometry. It is insensitive to ambient disturbances and therefore provides far better phase stability. Furthermore, efficiency of manipulations (including state splitting and recombination) are considerably high, typically up to 99\,\%. These benefits result in a better contrast compared to perfect-crystal interferometry (up to 98\,\%). In addition, perfect-crystal IFMs accept neutrons propagating in directions within an angular range of a few arc seconds, which leads to a significant loss of intensity. Polarimeters, however, can make use of a broader momentum distribution allowing for count rates that are higher by about one order of magnitude. 

Many polarimetric experiments described in this article were carried out at the
tangential beam tube of the TRIGA Mark II reactor at the research reactor facility (Atominstitut) of the Vienna University of Technology. There, a neutron beam is monochromatized by pyrolytic graphite-crystals selecting wavelengths between 
 1.7\,\AA\,and 2\,\AA\,(with spectral width $\Delta\lambda/\lambda\approx0.015$) and polarized up to $P_0 \approx 99\,\%$ by reflection from a bent Co-Ti supermirror. A polarizing supermirror is a multilayer structure consisting of alternating magnetic and non-magnetic media $A$ and $B$ with different coherent scattering lengths $b_{c(A,B)}$ and magnetic scattering length $p_A$. The combination is chosen such that its reflectivity -- proportional to $[N_A(b_{c(A)}\pm p_A)-N_Bb_{c(B)}]^2$ -- vanishes for one of the spin eigenstates. 
In addition, the thickness of the layers is chosen in such way that, for the reflected beams, constructive interference occurs according to the Bragg-condition \cite{willisBook2009}.  
 
A polarimeter combining static and time-dependent magnetic fields is depicted in Fig.\,\ref{fig:SetupZeroOsci}\,(a) \cite{sponarPLA2008}. A neutron beam propagating in $y$-direction and interacting with a static magnetic guide-field is described by the Hamiltonian $H=H_{\rm kin}+H_{\rm mag}=-\hbar^2 \vec\nabla^2/2m-\vec\mu \cdot\vec B_0(y)$, where the first term accounts for the kinetic energy of the neutron with its mass $m=1.674\times 10^{-27}$\,kg. The second term, already mentioned in Sec.\,\ref{sec:SpinManipulation}, leads to Zeeman-splitting of the kinetic energy of the spin eigenstates equal to $2\vert\mu\vert B_0$ [see Fig.\,\ref{fig:dcspinflipper}\,(d)]. 
A solution of the Schr\"{o}dinger equation is given by $\cos(\theta/2)\ket{\!\Uparrow}\ket{k_\Uparrow}+e^{i\phi}\sin(\theta/2)\ket{\!\Downarrow}\ket{k_\Downarrow}$, where $\ket{k_\Uparrow},\ket{k_\Downarrow}$ are the momentum eigenstates within the field $\vec B_0(y)$. $\theta$ and $\phi$ denote the polar and azimuthal angles determining the direction of the spin with respect to $\vec B_0(y)$. $k_{\Uparrow,\Downarrow}\approx k_0\mp\Delta k$ where $k_0$ is the momentum of the free particle and $\Delta k=m\mu\vert \vec B_0(y)\vert/\hbar^2k_0$ is the field-induced momentum shift due to Zeeman-splitting. A similar analysis can be done for interaction of neutrons with time-dependent fields \cite{hasegawaNJP2012}, for which  $\ket{k_\Uparrow},\ket{k_\Downarrow}$ would be substituted by $\ket{E_\Uparrow},\ket{E_\Downarrow}$, say.

\begin{figure}
\begin{center}
\includegraphics[width=5.1in]{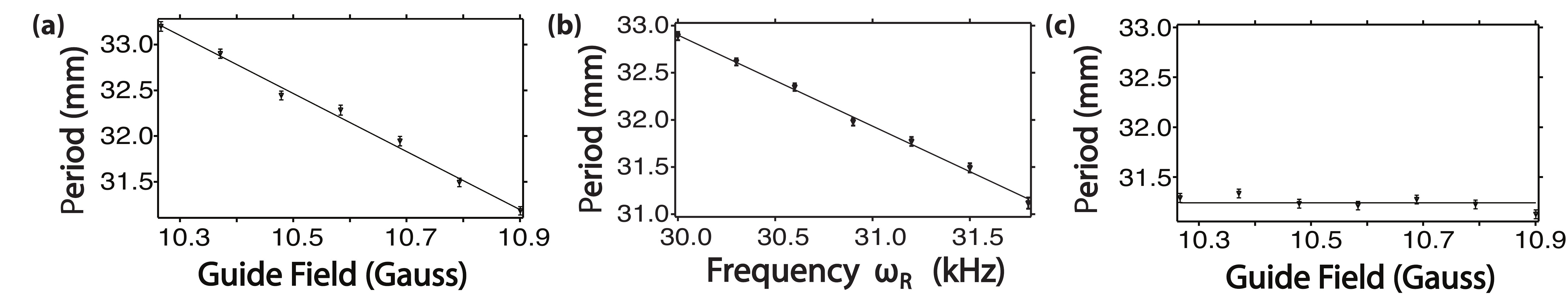}%
\caption{Dependence of the period of measured intensity oscillations on: (a) the strength of the guide field with both RF flippers off (pure Larmor precession), (b) the frequency of the RF flippers (pure zero-field precession), (c) the strength of the guide field with both RF flippers on (pure zero-field precession) \cite{sponarPLA2008}. \label{fig:Results_Zero}}
\end{center}
\end{figure}

For observation of pure Larmor precession with the experimental setup in Fig.\,\ref{fig:SetupZeroOsci}\,(a), both RF flippers were turned off and only the DC flipper and the two DC-$\pi/2$ spin-rotators were in operation. 
Then, the superposed states $\ket{\!\Uparrow}$ and $\ket{\!\Downarrow}$ acquire a pure Larmor-phase due to the guide field. Varying the position of the DC flipper, intensity oscillations were recorded.
The dependence of the period of these Larmor-precession-induced oscillations on the guide field is plotted in Fig.\,\ref{fig:Results_Zero}\,(a).
Characteristics of the zero-field precession were investigated by additionally turning on both RF flippers. 
In that case, the spin precession angle is expected to be a function of propagation time and RF-frequency, independent of the guide field. 
The linear frequency-dependence of the period can be seen in Fig.\,\ref{fig:Results_Zero}\,(b).
Furthermore, observation of pure zero-field phase was confirmed at constant frequency by varying the guide-field strength, confirming that no spin rotation due to Larmor precessions occurs. The respective results are plotted in Fig.\,\ref{fig:Results_Zero}\,(c) - a constant period, independent of the strength of the guide field.

\subsection{Very-cold and ultra-cold neutron optics}\label{sec:VCNUCN}

For most of the experiments described in this paper thermal neutrons were used. Slow neutrons, with wavelengths 
in the ranges 4\,\AA\,$<\lambda <$ 30\,\AA\, and 30\,\AA\,$\le\lambda <$\,100\,\AA\,, are often 
termed cold neutrons (CN) and very-cold neutrons (VCN), respectively. Usually, they are needed to reach particular values of the scattering vector for clarifying the internal structure or dynamics of certain samples and/or to maximize interaction time, especially in fundamental physics. However, the gain in interaction time often comes at the price of low intensity, simply because most neutron sources -- due to a moderator material at ambient temperature -- provide a Maxwell-Boltzmann spectrum with its intensity maximum at thermal wavelengths. CN and VCN can be produced by further moderation (cooling) of neutrons in a cold source. The latter essentially consists of a tank of liquid deuterium at about 20 K, for instance, close to the reactor core, in which thermal neutrons with $E=k_BT\approx 20$ meV collide with atoms and lose their energy until they are in thermal equilibrium with the D$_2$. 

At small-angle-neutron-scattering (SANS) instruments, the wavelength distribution $\Delta\lambda/\lambda$ of CN, as incident from a cold source and filtered by a velocity selector, is typically around 10\,\% (see, for instance, \cite{kohlbrecherJAC2000} or \cite{YellowBookILL2008}). 
The holographic-grating IFM tests described in Sec.\,\ref{subsec:HoloGratings} were carried out with such instruments.
CN were also used for the experiments described in Secs.\,\ref{subsec:DoubleResRFSPinFlippers}, \ref{subsec:AvarietyOfTopAndGeoPh}, \ref{subsec:GeoPhase}, \ref{subsec:GoosHanchen} and \ref{subsec:Whispering}.

At PF2 of the ILL \cite{YellowBookILL2008}, VCN with very broad wavelength distribution are available.
During travelling in a curved vertical guide connected to the D$_2$ cold source, neutrons are cooled further by their movement in the gravitational potential. Faster neutrons are filtered out because their angles of incidence are too small for reflection within the guide tube. The vertical guide tube leads to a turbine several meters above the cold source, where the tube is split and one part is used as VCN source. VCN are quite slow (about 100\,m/s for $\lambda=40$\,\AA), their interaction with the earth's gravitational field -- they fall down by a centimeter on a flight path of about 4.5\,m -- is easily observable. 
Using diffraction gratings, a moderately divergent incident beam can be used for interferometry \cite{vanDerZouwNIMA2000} and diffraction experiments with holographic gratings (see Sec.\,\ref{subsec:HoloGratings}).

The second part of the split guide at PF2 is used to feed the aforementioned turbine (the so-called `Steyerl-turbine' \cite{steyerlPLA1986}), which generates ultra-cold neutrons (UCN) by Doppler-shifting the energy of the incident VCN spectrum: The turbine contains a rotating wheel on the outer frame of which curved Ni-mirrors are mounted such that they move slower than the VCN. In particular, they move in parallel to the incident direction of the beam. The VCN lose energy on reflection from that Ni surfaces just like a tennis ball does on the racket when playing a drop shot. The resulting velocity of UCN is around 5\,m/s. At PF2, UCN are distributed to four beam ports to supply different experiments. The low kinetic energy of UCN allows to guide them with tubes made of materials with high Fermi-potential as, for instance, Cu, for which UCN are totally reflected for any angle of incidence. In particular, UCN can also be stored in bottles for up to their lifetime (almost 15\,min) to accurately measure the latter or provide an experimental limit to the neutron electric dipole moment \cite{abelePPNP2008,dubbersRMP2011}. The measurement of the neutron rest charge \cite{gaehlerPRD1982} is another example for UCN-application \cite{plonkaSpehrNIMA2010}. Within the frame of the present review, UCN play a role in investigations of the Berry phase (Sec.\,\ref{subsec:AvarietyOfTopAndGeoPh}) and its robustness under noisy spin-evolutions 
(Sec.\,\ref{subsec:RobustBerry}). 

\section{Historical Experiments}\label{sec:History}

\subsection{$4\pi$-symmetry of the spin-1/2 wave function}\label{sec:4pi}

The evolution and manipulation of a spin-1/2 system can be conventionally represented by
the two-component spinor formalism introduced by Pauli in 1927 \cite{pauliZFP1927}. As before, we use
the Pauli equation (or Pauli-Schr\"{o}dinger equation), i.e., the Schr\"{o}dinger-equation
for spin-1/2 particles, which considers the interaction of the particle's spin with the external magnetic field.
It poses the non-relativistic limit of the Dirac equation. The Pauli equation is given by
\begin{equation}\label{3.1}
\hat H\,\Psi \left( {\mathord{\buildrel{\lower3pt\hbox{$\scriptscriptstyle\rightharpoonup$}}
\over r} ,t} \right) = \left[ { - \frac{\hbar^2}{{2m}}\nabla ^2  - \mu \mathord{\buildrel{\lower3pt\hbox{$\scriptscriptstyle\rightharpoonup$}}
\over \sigma } \cdot\mathord{\buildrel{\lower3pt\hbox{$\scriptscriptstyle\rightharpoonup$}}
\over B} \left( {\mathord{\buildrel{\lower3pt\hbox{$\scriptscriptstyle\rightharpoonup$}}
\over r} ,t} \right)} \right]\Psi \left( {\mathord{\buildrel{\lower3pt\hbox{$\scriptscriptstyle\rightharpoonup$}}
\over r} ,t} \right) = i\hbar \frac{\partial }{{\partial t}}\Psi \left( {\mathord{\buildrel{\lower3pt\hbox{$\scriptscriptstyle\rightharpoonup$}}
\over r} ,t} \right).
\end{equation}
A solution of the above equation is denoted as
\begin{equation}\label{3.2}
\Psi \left( {\mathord{\buildrel{\lower3pt\hbox{$\scriptscriptstyle\rightharpoonup$}}
\over r} ,t} \right) 
= \left[ {\begin{array}{*{20}c}
   {\Psi _\Uparrow  \left( {\mathord{\buildrel{\lower3pt\hbox{$\scriptscriptstyle\rightharpoonup$}}
\over r} ,t} \right)}  \\
   {\Psi _\Downarrow  \left( {\mathord{\buildrel{\lower3pt\hbox{$\scriptscriptstyle\rightharpoonup$}}
\over r} ,t} \right)}  \\
\end{array}} \right] 
= f_\Uparrow \left( {\mathord{\buildrel{\lower3pt\hbox{$\scriptscriptstyle\rightharpoonup$}}
\over r} ,t} \right)\,e^{-i\Phi/2 }\cos \frac{\Theta }{2}\left|\Uparrow  \right\rangle  + f_\Downarrow  \left( {\mathord{\buildrel{\lower3pt\hbox{$\scriptscriptstyle\rightharpoonup$}}
\over r} ,t} \right)\,e^{i\Phi/2 } \sin \frac{\Theta }{2}\left|\Downarrow  \right\rangle
\end{equation}
with space-time dependent coefficients of the wave functions $f_{\Uparrow,\Downarrow}  \left( \vec{r},t \right)$,
polar/azimuthal angle $\Theta/\Phi$ of the spin vector, and the spin basis $\{|\!\Uparrow\rangle,|\!\Downarrow\rangle\}$ along the quantization axis.

Using Eq.\,(\ref{eq:unitary}), it is straightforward to see that while the polarization vector $\vec P$ returns back to the initial directions after
a $2\pi$-rotation, the wave function itself has $4\pi$-symmetry:
$
\Psi(\alpha=0)=-\Psi(\alpha=2\pi)=\Psi(\alpha=4\pi)$.
Physically, this relation indicates the $\exp(i\pi)=-1$ phase factor or, equivalently, a $\pi$ phase shift after a spin-1/2 system was affected by a $2\pi$-rotation.
It should be emphasized that the $4\pi$-symmetry of the neutron wave function appears equally for polarized and unpolarized beam experiments.

\begin{figure}
\begin{center} 
\scalebox{0.3}
{\includegraphics {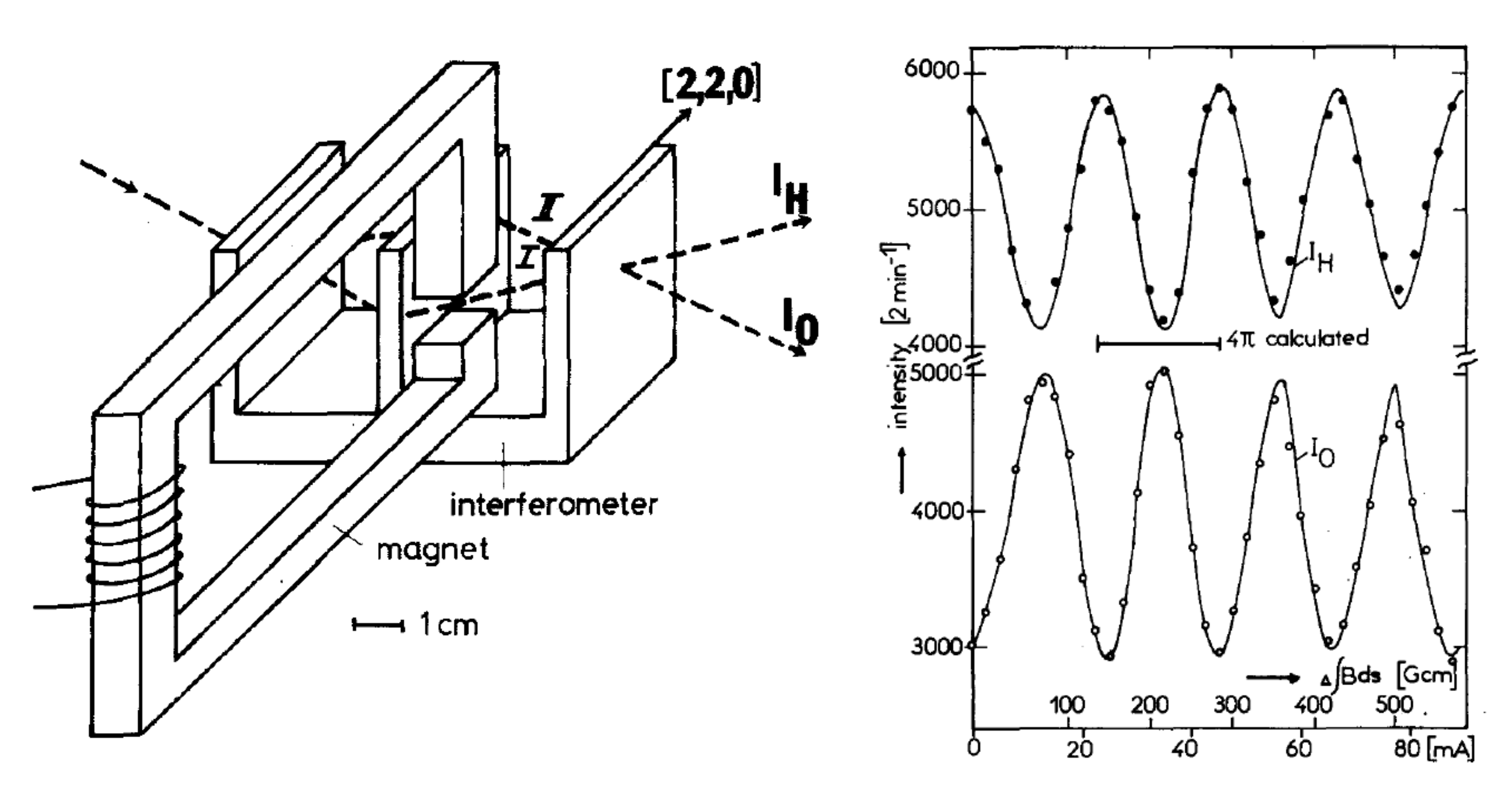}}
\caption{Experimental setup (left) and the results (right) of the demonstration of $4\pi$-symmetry of the fermionic
spinor wave function \cite{rauchPLA1975}. }
\label{fig:f_4pi} 
\end{center}
\end{figure}

The $4\pi$-symmetry was known at an early stage of the development of quantum theory.
Nevertheless, the $\exp(i\pi)$ phase factor was treated as inaccessible since in most experiments only the absolute square of the wave function is measured as intensity and the phenomenon is hidden in this kind of measurements.
In 1967, two publications \cite{aharonovPR1967,bernsteinPRL1967} appeared independently, which deal with the possibility of observing $4\pi$-symmetry of spin-$1/2$ particles
on the gedanken level. They showed $4\pi$-symmetry of the fermionic spinor wave function can even be observable
in a split beam experiment. They pointed out that, by using
one of the split beams in the IFM as a reference beam and utilizing the interference effect,
the phase factor indeed is observable in the shift of the interference fringes.

In general, when the neutron spin in one of the beams in the IFM is rotated by $\alpha$,
the intensity of the interfering beam in the forward directions becomes $I_0  = \left| {\Psi \left( {\alpha  = 0} \right) + \Psi \left( \alpha  \right)} \right|^2
\propto \left| \Psi  \right|^2 \left( {1 + \cos \alpha/2} \right)$. 
The first experimental demonstration of this phenomenon \cite{rauchPLA1975} followed directly the invention of
the silicon perfect-crystal neutron IFM. The result was confirmed by Werner's group as well \cite{wernerPRL1975}. The experimental setup and the results are depicted in Fig.\,\ref{fig:f_4pi}. An electromagnet was used to tune the strength of current supplying the magnetic field yoke. Intensity modulations
of 0- and H-beams were recorded as a function of current. The intensity oscillation period was determined by $(144\pm8)$\,Gcm,
which corresponds to the Larmor precession angle of $(704\pm38)\,^\circ$. The obtained results are in good agreement with theory. The results were improved using magnetized Mu-metal sheet, for which considerably large magnetic fields can be confined within the sample.
Other investigations of the $4\pi$-symmetry with neutrons using Fresnel diffraction at the ferromagnetic domains \cite{kleinPRL1976} or RF-flippers \cite{kraanEPL2004} have been reported.

\subsection{Gravity induced phases}\label{sec:gravity}

The neutron as a massive particle is affected by Newton's gravitational force as a consequence of classical mechanics. Parabolic trajectories of neutrons in the earth's gravitational field are observed, which again
confirms the equivalence of gravitational and inertial mass for the neutron \cite{koesterPRD1976}. In QM,
the consequences of the interaction appear not only in the trajectories of motion but also in the phase of the wave function as determined by the potential. The perfect-crystal neutron IFM enabled observations of the phases induced by the earth's gravitational potential \cite{colellaPRL1975}, earth's rotation (Sagnac effect) \cite{wernerPRL1979}, and motional effect on the wave function (Fizeau effect) \cite{arifPRA1989}. Here, the experiment performed by Colella, Overhauser, and Werner (COW) \cite{colellaPRL1975} is described. The peculiarity of this experiment lies in the fact that both gravity and
QM play a very important role due to the earth's gravitational acceleration \textit{g} and Planck's constant \textit{h}, that both enter the prediction of the phase shift in the experiments.

\begin{figure}
\begin{center} 
\scalebox{0.72}
{\includegraphics {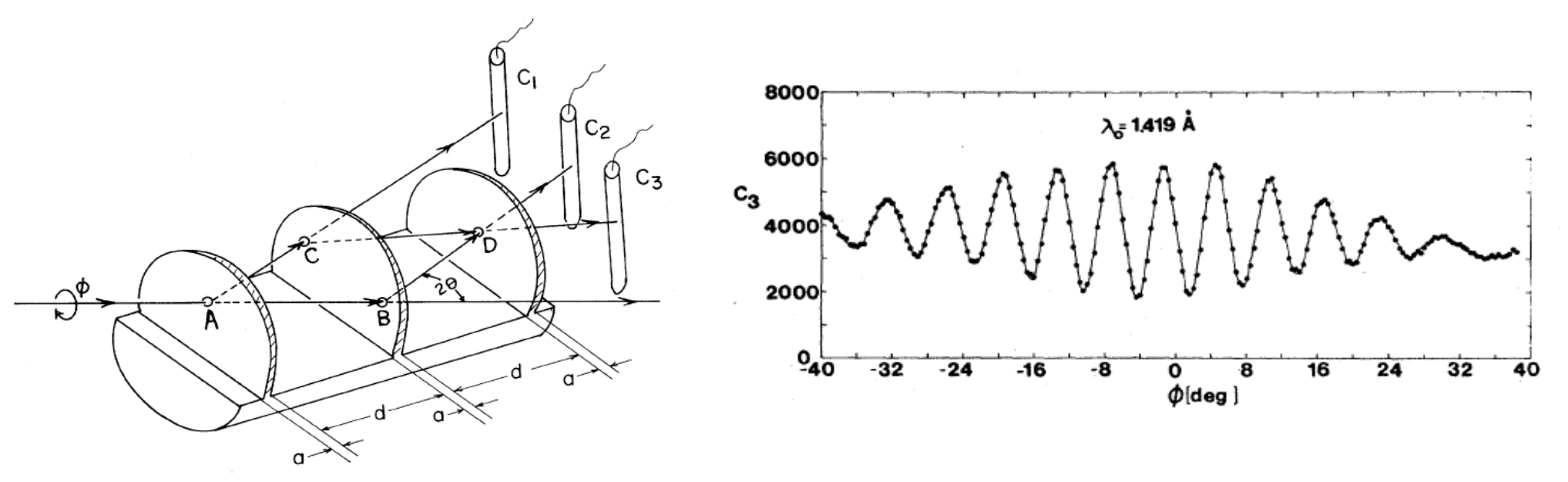}}
\caption{Left: Experimental setup. The neutron IFM is rotated about the axis of the incident beam. Right: The results of the experiment described in \cite{staudenmannPRA1980}. The intensity measured in the 0-detector (here, referred to as C$_3$) is plotted versus the IFM inclination angle $\phi$. Reprinted with permission from  \cite{colellaPRL1975,staudenmannPRA1980}. Copyright (1975,1980) by the American Physical Society.}
\label{fig:f_COW} 
\end{center}
\end{figure}

In the classical equation of motion, a particle with mass $m$ is affected by the earth's gravitational force
and is predicted to fall down according to
$
m\ddot{\vec{r}}=-\vec\nabla U_{grav}=-mg\hat{z}
$, with the gravitational potential $U_{grav}=mgz$ at a vertical distance $z$, close to the surface of the earth. This equation suggests that the mass term drops out
and that the equation of motion is independent of the mass of the particle. The situation in QM
is somewhat different: The Schr\"{o}dinger equation with the gravitational potential is written in the form
\begin{equation}\label{3.8}
\hat H\,\Psi \left( {\mathord{\buildrel{\lower3pt\hbox{$\scriptscriptstyle\rightharpoonup$}}
\over r} ,t} \right) = \left[ { - \frac{\hbar^2}{{2m}}\vec\nabla ^2  + U_{grav} } \right]\Psi \left( {\mathord{\buildrel{\lower3pt\hbox{$\scriptscriptstyle\rightharpoonup$}}
\over r} ,t} \right) = i\hbar \frac{\partial }{{\partial t}}\Psi \left( {\mathord{\buildrel{\lower3pt\hbox{$\scriptscriptstyle\rightharpoonup$}}
\over r} ,t} \right).
\end{equation}
Here the mass term \emph{m} does not cancel any more and the term $\hbar/m$ appears instead:
both the neutron mass \emph{m} and Planck's constant \emph{h} play a role. Here, we assume one of the beams propagating on a lower level than the other beam in the IFM. In this case, energy conservation
demands that the gravitational potential energy is transformed into the kinetic energy:
\begin{equation}\label{3.9}
\frac{\hbar^2 {k_0}^2}{2m}=\frac{\hbar^2 k^2}{2m}+mg \Delta H,
\end{equation}
where $k_0~(k)$ and $\Delta H$ are the wave vectors of the lower (upper) beam path and their difference in height, respectively. An approximation due to the small value of the gravitational potential
(typically $mg\Delta H \approx 1$\,neV) as compared to the kinetic energy of neutron (typically $E_{kin}=20$\,meV) is made for the phase shift due to the gravitational potential:
\begin{equation}\label{3.10}
\Delta \Phi_{grav}=\Delta k L=(k-k_0) L\approx -\frac{2 \pi \lambda m^2gL \Delta H}{h^2}
\end{equation}
with the path length L. It is worth noting here that, even though the trajectories due to
the gravitational force are practically the same for the lower and the higher beams, the difference of the potential
itself induces the phase shift, which is observable in a perfect-crystal neutron IFM.

The first experimental demonstration of the gravitationally induced phase shift was reported by COW in 1975 \cite{colellaPRL1975}.
The experimental setup is shown in Fig.\,\ref{fig:f_COW}\,(left). The IFM was rotated by $\phi$
about the axis of the incident beam. The wavelength was $\lambda=1.445$\,\AA, the path length $L\approx 4$\,cm and 
$\Delta H\approx 3\,\mbox{cm} \times \sin \phi $.
One period of the oscillation corresponds to about $\phi=6^\circ$. The agreement with theory was 90\,\%. The main deviation from theory was attributed to the bending of the IFM crystal during rotation.

\begin{figure}
\centering\includegraphics[width=5in]{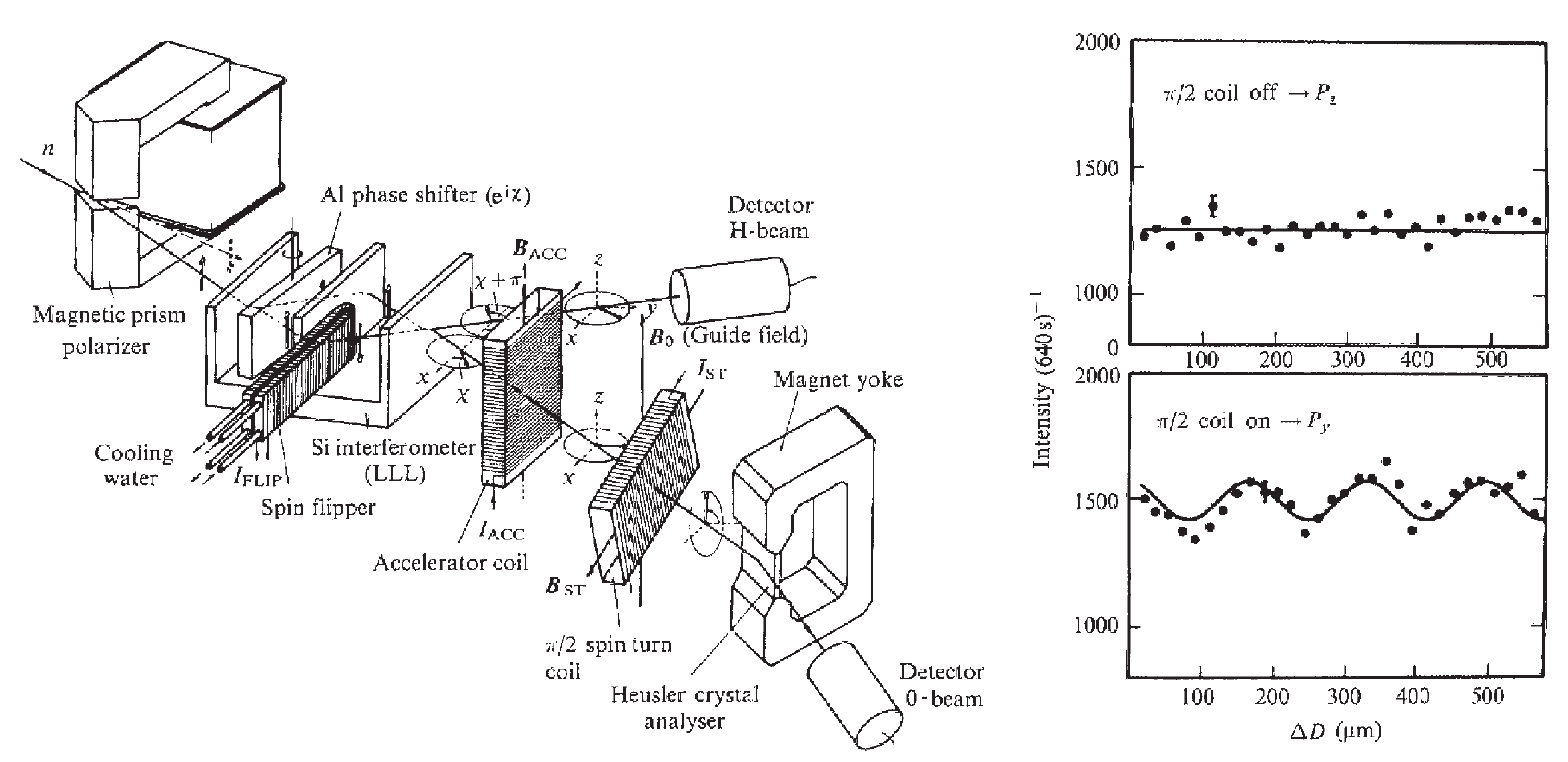}
\caption{
Experimental setup (left) and the results (right) of the demonstration of coherent spin superposition. Reprinted with permission from  \cite{summhammerPRA1983}. Copyright (1983) by the American Physical Society.}
\label{fig_DCSS}
\end{figure}

Later, a more-detailed investigation was carried out in which simultaneous effects of gravity, inertia and QM on the motion of neutrons were considered \cite{staudenmannPRA1980}. The interferogram, as shown in Fig.\,\ref{fig:f_COW}\,(right), exhibits clear sinusoidal intensity modulation as a function of $\phi$. This experiment provided convincing high-quality data deviating from theory by about 3\,\%. 
Afterwards, Bonse and Wroblewski reported an acceleration-induced quantum interference effect and pointed out
an additional influence on the IFM in a non-inertial frame due to dynamical diffraction \cite{bonsePRL1983,bonsePRD1984}:
They suggested a (downward) correction of the results in \cite{staudenmannPRA1980} of about 4\,\%. 
The discrepancy of about 1\,\% remained \cite{wernerPB1988}. 
A new measurement using a pair of almost harmonic wavelengths -- to allow monitoring the deformation of the IFM -- appeared \cite{littrellPRA1997}. 
In this experiment, the obtained values and the theoretical prediction still showed a discrepancy at the level of 1\,\%, the experimental error being only about 0.1\,\%. Another approach to measure precisely the gravitation-induced quantum phase with neutrons employed a grating IFM for VCN: Long-wavelength neutrons induce larger phase shifts [see Eq.\,(\ref{3.10})] and -- since gratings were thin, sputter-etched in quartz glass -- the grating-IFM was much less sensitive to bending during rotation.
The results of the measurements are consistent with theory but have a relatively large error of about 1\,\%, mainly due to the inaccurate measurement of the broad incident wavelength spectrum \cite{vanDerZouwNIMA2000}.
A completely different strategy is the use of neutron polarimetry, 
in particular, gravitational phase measurements with the spin-echo
spectrometer OffSpec at ISIS, Oxford, UK \cite{dalglieshPB2011}, where much longer path lengths and
a virtually white beam with high intensity are available.

Gravity-induced quantum phase was measured not only with neutrons but also with atoms: Kasevich and Chu have used
a fountain IFM for atoms \cite{kasevichPRL1991} to measure the gravitational acceleration of an atom.
They insist on high resolution of the $g$-measurement and report no significant discrepancy from theory \cite{kasevichAPB1992}.
About two decades later another paper appeared \cite{muellerNature2010} in which the authors consider
the atom IFM experiments as a measurement of gravitational redshift of a quantum clock operating 
with the Compton frequency $\omega_c={mc^2}/{\hbar}$. According to general relativity, a quantum clock 
runs slower by a factor of $1+{U}/{c^2}$ in higher gravitational potential. It was argued that, 
with a semiclassical non-relativistic analysis, atom interferometry exhibits extraordinary high accuracy 
in measurements of the gravitational redshift induced by the space-time curvature. 
This claim has triggered 
a stimulating debate \cite{wolfNature2010,wolfCQG2011,muellerNature2010b, hohenseePRL2011,schleichPRL2013,schleichNJP2013}.

Although there is a similarity between the
gravitation-induced phase measurements with atoms and neutron IFMs, subtle differences between
these devices are found which demand a special treatment of neutron interferometry \cite{greenbergerPRA2012}. Final agreement has not yet been reached.

\subsection{Spin superposition}

\begin{figure}
\centering\includegraphics[width=5.in]{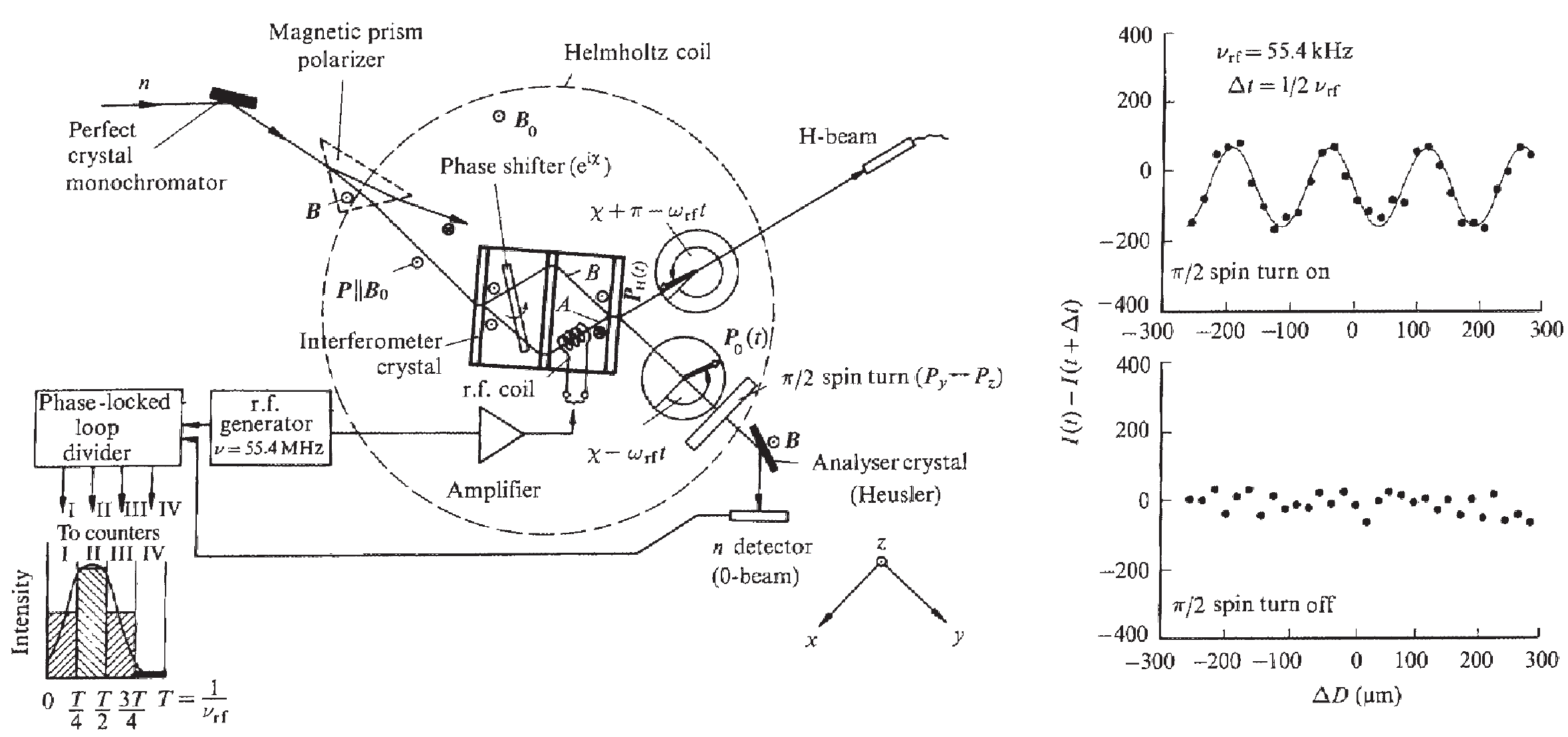}
\caption{
Experimental setup (left) and results (right) of the demonstration of time-dependent spin superposition. Typical results of the $y$- and $z$-component spin analysis are plotted as a function of the path difference $\Delta D$ (proportional to the phase shifter angle). Reprinted with permission from \cite{badurekPRL1983}. Copyright (1983) by the American Physical Society.}\label{fig_RFSS}
\end{figure}

The evolution of the spin vector obeys the Bloch equation (see 
Sec.\,\ref{Sec:TimeInDep}), which describes Larmor precession. This behaviour seems similar to that of angular momentum in classical physics. In addition to
the $4\pi$-symmetry of the spin-1/2 wave function, one sees a -- nowadays familiar -- feature in the superposition of
two spin eigenstates $|\!\Uparrow\rangle$ and 
$|\!\Downarrow\rangle$: it does not result in a (classical) mixture of
these states but in a new pure spin-state. In particular, quantum theory predicts, in this case,
that the final polarization vector lies in a plane perpendicular to the initial polarization axis and that
the azimuthal angle depends on the relative phase between the superposed spin states. It was Wigner who discussed the issue on the gedanken level \cite{wignerAJP1963}, followed by the actual observation using the neutron IFM.

\begin{figure}
\centering\includegraphics[width=5.in]{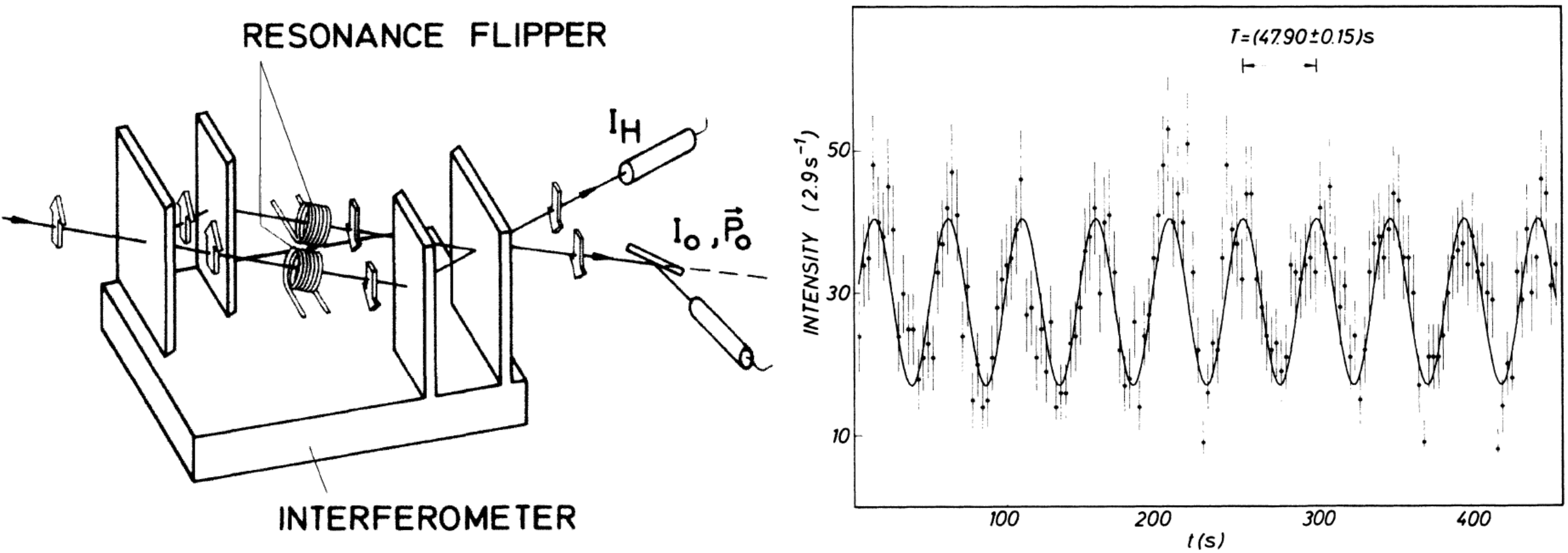}
\caption{
Experimental setup (left) and results (right) of the double-resonance IFM experiment. Reprinted with permission from \cite{badurekPRA1986}. Copyright (1986) by the American Physical Society.}
\label{fig_double1}
\end{figure}

Let us assume that the spin state of the incident beam 
is $|\!\Uparrow\rangle$, the beam is polarized to the $+z$-direction. This beam falls on
the IFM and is split into two beams. In one of the beam paths, a DC spin-flipper is inserted to
flip the neutron spin from $|\!\Uparrow\rangle$ 
to $|\!\Downarrow\rangle$. After the spin flipper and the phase shifter, the state in the IFM is written in the form
\begin{equation}\label{3.11}
|\Psi_0\rangle=\frac{1}{\sqrt 2}\left(|\!\Uparrow\rangle
+e^{i\chi}U(\pi\hat y)|\!\Uparrow\rangle\right)
=\frac{1}{\sqrt 2}\left(|\!\Uparrow\rangle
+e^{i\chi}|\!\Downarrow\rangle\right),
\end{equation}
where $\chi$ and $U(\pi\hat y)$ represent a relative phase between the two IFM paths and the $\pi$ spin-rotation
around the $y$-axis, respectively. 
The corresponding polarization vector lying in the $xy$-plane is given by $\vec{P_0}=[\cos\chi,\sin\chi,0]$ and its rotation can be revealed by spin analysis.
The experimental setup is depicted in Fig.\,\ref{fig_DCSS}\,(left).
A magnetic prism polarizer was used to obtain the polarized neutron beam. A phase shifter and a water-cooled DC spin-flipper were
inserted in the IFM. The 0-beam undergoes spin analysis by a combination of static $\pi/2$-rotation coil and a Heusler-crystal analyzer, while the intensity of the H-beam was
directly measured. Since $\vec{P_0}$ was rotating in the $xy$-plane, a sinusoidal intensity modulation was observed with the $\pi/2$-rotation turned on. Constant intensity was observed with the $\pi/2$-rotation turned off, as shown in Fig.\,\ref{fig_DCSS}\,(right). 
The results clearly demonstrate that the spin state of the 0-beam is a superposition of the spin states in left and right paths -- \emph{not} an incoherent mixture but a new pure state. An incoherent mixture cannot account for the observation of Larmor precession.

If an 
RF flipper (cf. Sec.\,\ref{Sec:Time}) is used instead of a DC flipper, a time-dependent effect comes in. The final superposition state is then written in the form
\mbox{$1/\sqrt 2(|\!\Uparrow\rangle+e^{i\chi}
e^{-i\omega_rt}|\!\Downarrow\rangle)$}  
with the frequency of the resonant RF operation $\omega_r$. The polarization vector is given 
by $\vec{P'_0}=[\cos(\chi+\omega_rt),\sin(\chi+\omega_rt),0]$. 
The equation suggests that the rotation of
the final polarization vector is time-dependent, describing a non-stationary 
interference effect, which can be measured by stroboscopic detection.

The experimental setup to observe time-dependent spin-superposition is depicted in Fig.\,\ref{fig_RFSS}\,(left). The RF spin-flip induced a shift of the total neutron energy, which is interpreted as a photon exchange of neutron and magnetic field during the spin flip. The spin-flipped spinor acquired a time-dependent phase factor
$e^{-i \omega_r t}$, resulting in the final time-dependent superposition state. In the experiment, the data-acquisition in the 0-detector was phase-locked to the signal generator of the RF spin-flipper. The data are shown in Fig.\,\ref{fig_RFSS}\,(right). The experiment clearly demonstrated that the coherent superposition of two orthogonal
spin states with different energies results in time-dependent rotation of the polarization vector.
Coherence properties of the neutron beam
are preserved after the energy exchange of neutrons traversing the RF field.
This has stimulated discussions on
the complementarity in a double-slit situation (see, for instance, \cite{feynmanLectures1966}). In particular, by detecting a missing/added photon in the RF field, it was argued that path detection and observation of interference fringes could be possible simultaneously \cite{dewdneyPLA1984}. It turned out, however, that
due to the extremely large number of photons in the RF field, detection of the exchange photon
is not feasible, not even in principle. The common consent is that the energy change itself does apparently not
represent a measuring process, the exchanged photon cannot be used to obtain path information.

\begin{figure}
\centering\includegraphics[width=4.8in]{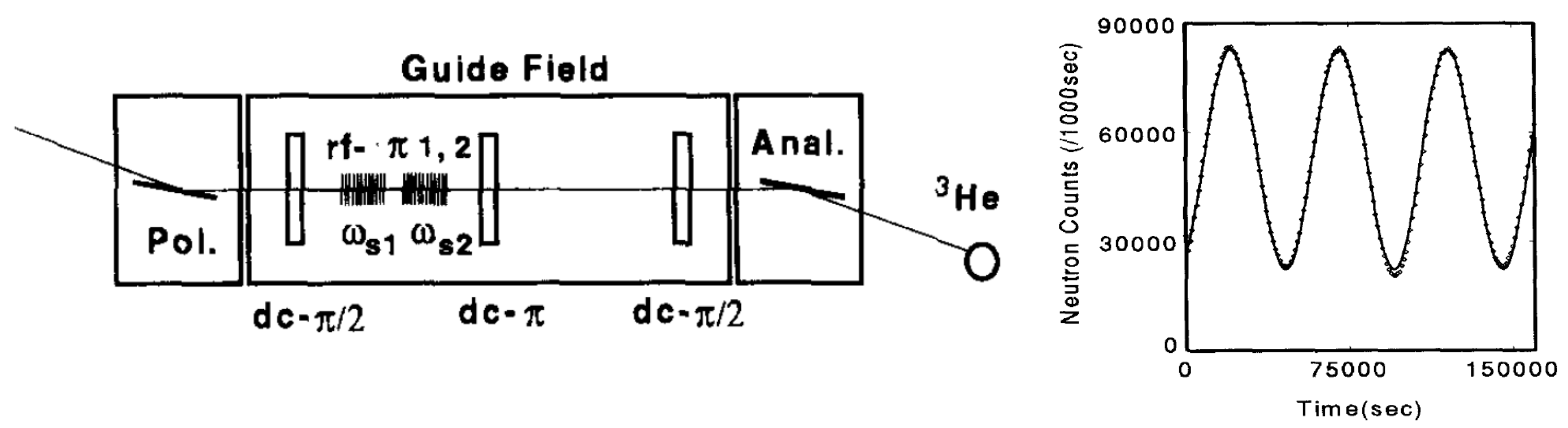}
\caption{
Experimental setup (left) and the results (right) of the double resonance experiment with cold neutrons \cite{yamazakiPB1998}. This experiment demonstrated intensity modulations with very long period, which resulted from the extremely small difference
of the neutron energy.
}
\label{fig_double2}
\end{figure}

\subsection{Double-resonance IFM experiment}
\label{subsec:DoubleResRFSPinFlippers}

After the performance of the spin-superposition experiment with an RF spin-flipper, the argument put forward by Dewdney \emph{et al.} \cite{dewdneyPLA1984} (see end of previous Section) inspired a neutron-IFM experiment with one RF coil in each path \cite{badurekPRA1986}.
The experimental setup is depicted in Fig.\,\ref{fig_double1}\,(left). Two independent RF spin-flippers, operating at
the frequencies $\omega ^{(1)}_r$ and $\omega ^{(2)}_r$, were inserted in each beam path of the IFM.
The intensity of the 0-beam is calculated to be
\begin{equation}\label{3.14}
I_{\mbox{\scriptsize{O}}}\propto\left|\frac{1}{\sqrt 2 }\left(e^{-i\omega_r^{(2)}t}|\!\Downarrow\rangle+e^{-i\omega_r^{(1)}t}|\!\Downarrow\rangle\right)\right|^2
=\frac{1}{2}\left(1+\cos\left[(\omega_r^{(2)}-\omega_r^{(1)})t\right]\right).
\end{equation}
The latter again suggests intensity modulations in time, when the resonant frequencies of the two coils are slightly detuned ($\omega_r^{(1)}\ne\omega_r^{(2)}$).
The result of the experiment is shown in Fig.\,\ref{fig_double1}\,(right). Frequencies were tuned to $\nu^{(1)}_r=71.89980$\,kHz
and $\nu^{(2)}_r=71.89978$\,kHz $(\Delta \nu =0.02$\,Hz). An intensity modulation with a period of $47.90 \pm 0.15$\,s was obtained. The energy difference of the neutron beams in the IFM after the spin flip
was $\Delta E=\hbar(\omega^{(1)}_r-\omega^{(2)}_r)=8.6\times10^{-17}$\,eV. The observation of interference confirmed again the fact that the coherence of the neutron beams is preserved in spite of the energy exchange. Therefore, it is clear that energy exchange does not present a path-measurement.

A group at the Kyoto University carried out another double-resonance experiment using a cold neutron beam
at the Kyoto University Research Reactor Institute (KURRI) and Japan Atomic Energy Research Institute (JAERI)
[now reorganized as Japan Atomic Energy Agency (JAEA)] \cite{ebisawaJPSJ1998,yamazakiPB1998}. The experimental setup is shown
in Fig.\,\ref{fig_double2}\,(left). The first DC-$\pi/2$ spin-rotator generated the superposition $|\Psi'\rangle=1/\sqrt 2(|\!\Uparrow\rangle+|\!\Downarrow\rangle)$. After going through two RF spin-flippers, the state evolved to
\begin{equation}
|\Psi'\rangle\propto1/\sqrt 2 \left(e^{-i(\omega_r^{(1)}
-\omega_r^{(2)})t}|\!\Uparrow\rangle+e^{-i(\omega_r^{(2)}-\omega_r^{(1)})t}|\!\Downarrow\rangle\right) 
\end{equation}
Two resonance coils were operated at very small frequency difference, $\Delta\nu=20\,\mu$Hz, which corresponds to
the tiny energy difference of $\Delta E=8.27\times 10^{-20}$\,eV. The final change of the polarization vector was
observed as intensity modulation by applying another $\pi/2$ spin-rotation, followed by spin analysis to
the $+z$-direction. Typical intensity modulations are depicted in Fig.\,\ref{fig_double2}\,(right). The extremely high energy-sensitivity of
this arrangement is worth mentioning. In addition, the observed oscillation period of $49904.9 \pm 27.2$\,s ($\approx14$ hours!)
was far longer than the coherence time of the neutron beam, which means that neutrons -- by no means -- felt the whole period of the magnetic-field beating. A valid interpretation is that each particle is affected only by the instantaneous magnetic field and the phase difference in the short passing-time, i.e. the interaction within the coherence time leads to the observed intensity modulation.

\begin{figure}
\centering\includegraphics[width=4.8in]{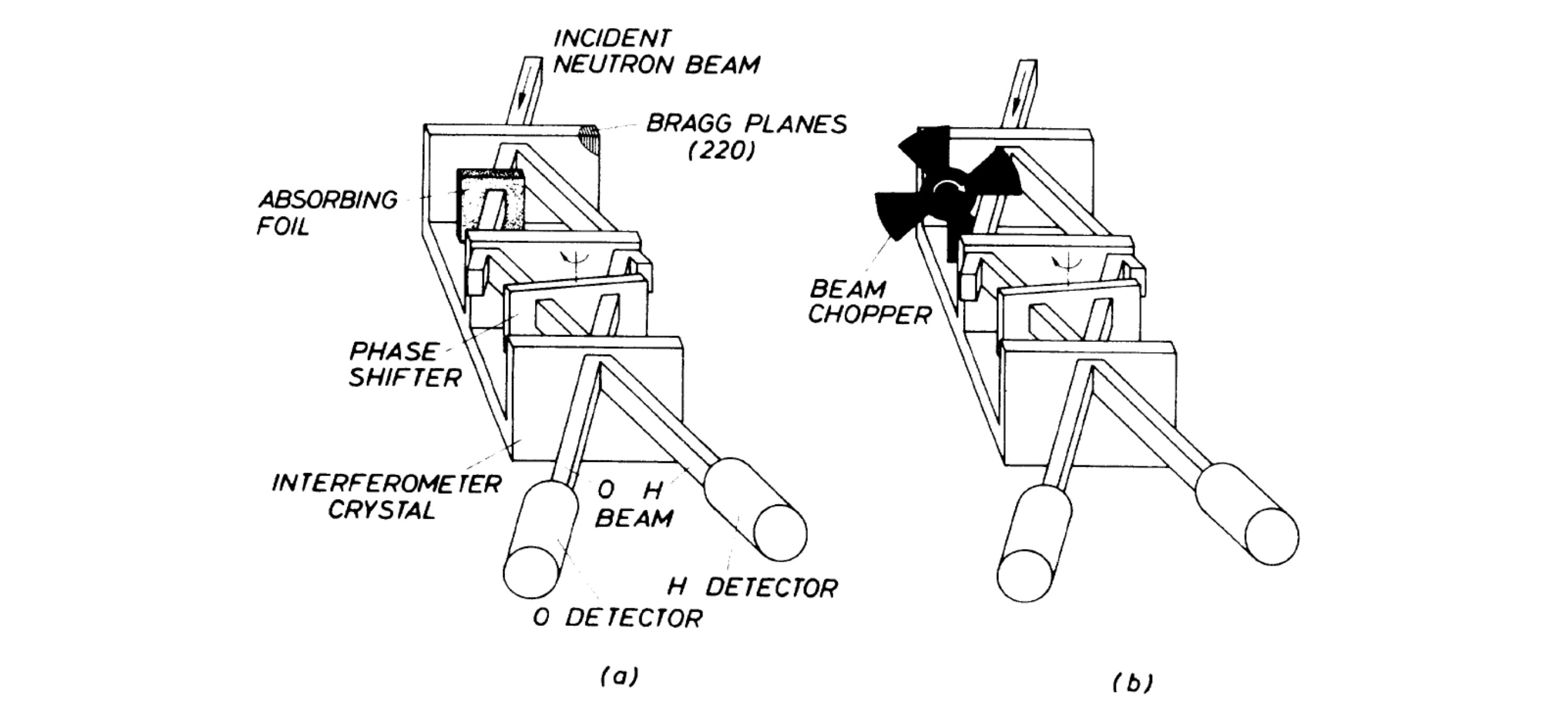}
\caption{
Experimental setup to observe the effects of stochastic and deterministic absorbers. (a) Stochastic absorber: A beam-attenuating absorber foil was inserted in one of the beam paths. (b) Deterministic absorber: A beam chopper was inserted in one of the beam paths. Reprinted with permission from \cite{summhammerPRA1987}. Copyright (1987) by the American Physical Society.
}
\label{fig_abs1}
\end{figure}

\subsection{Stochastic and deterministic absorption}

The IFM is a device where neutrons exhibit wave properties and, after traversing it, are detected 
as particles. Standard interpretations of QM consider this wave-particle duality of quantum "particles" 
a fundamental property in QM. To investigate this duality in more detail, neutron IFM experiments were carried out in which quantitative effects of beam attenuation in the IFM were studied \cite{summhammerPRA1987,rauchPRA1990}. 
Neutrons which are absorbed in one of the beam paths cannot contribute to the interference pattern measured 
behind the IFM. Quantum theory makes some remarkable predictions:
\emph{(i)} It makes a difference for the amplitude of the interference fringes whether neutrons are 
absorbed stochastically (without any chance to predict -- even in principle -- if neutrons will be 
absorbed or not) or deterministically (where it is known with certainty 
if neutrons will be absorbed or not in a certain instant of time). 
\emph{(ii)} Even when $99\,\%$ of neutrons in one of the beam paths are absorbed there is a case where the final interference
fringes show $10\,\%$ visibility. 

\begin{figure}
\centering\includegraphics[width=4.5in]{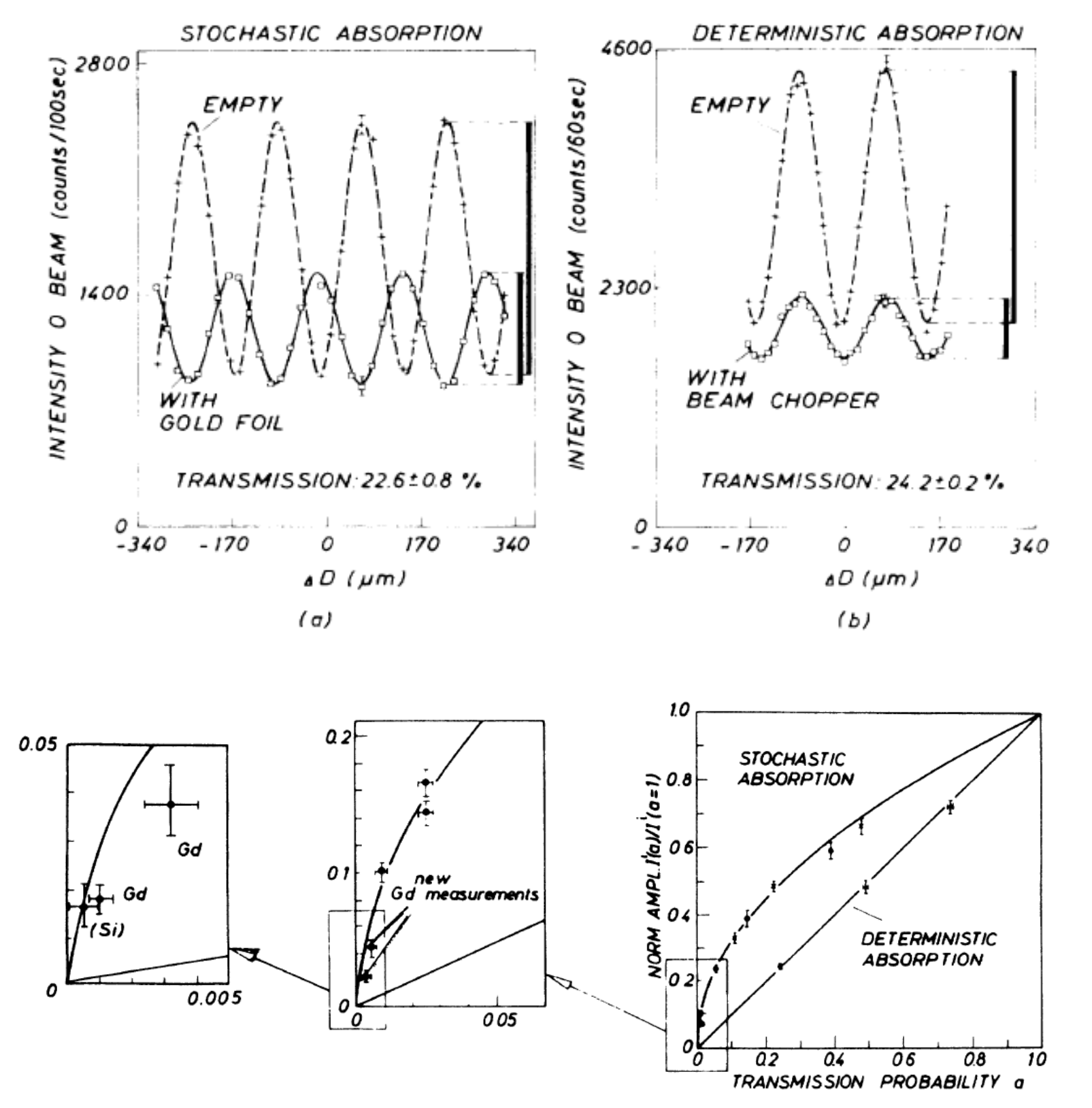}
\caption{
Experimental results for stochastic and deterministic absorption in the IFM.
Top: Interferograms for (a) stochastic and (b) deterministic absorption are plotted together with interferograms measured without absorbers. 
Bottom: The normalized amplitude of the interferogram is
plotted as the function of the transmissivity. Clear square-root/linear dependence is seen for stochastic/deterministic absorption. Reprinted with permission from \cite{summhammerPRA1987,rauchPRA1990}. Copyright (1987,1990) by the American Physical Society.   
}
\label{fig_abs2}
\end{figure}

The experimental setup to study the influence of stochastic and deterministic absorption on the interference fringes \cite{summhammerPRA1987} is shown in Fig.\,\ref{fig_abs1}. Two kinds of absorbers were involved in the experiments: 
absorber foil that absorbs neutrons stochastically according to its thickness and a beam chopper with its blades virtually opaque, i.e., a deterministic on/off-absorber. 
According to quantum theory, intensities for these stochastic and deterministic absorption cases
are written as 
\begin{eqnarray}\label{3.16}
I_{sto}\!\!\!\!&\propto&\!\!\!\!\left||\Psi_{\rm I}\rangle\right|^2 
\left(1 + T_{sto} + 2\sqrt{T_{sto}}\cos\chi\right)  \\
I_{det}\!\!\!\!&\propto&\!\!\!\!\left||\Psi_{\rm I}\rangle\right|^2 
\left(1 + T_{det} + 2T_{det}\cos\chi\right)
\end{eqnarray}
with the transmissivities $T_{sto}$ and $T_{det}$ representing beam attenuation by the absorber foil and the opening ratio
of the chopper, respectively. These equations suggest that the amplitude of the interference oscillations for the 
stochastic case is expected to exhibit a square-root dependence of the transmissivity $T_{sto}$, whereas that for the deterministic case is expected to be simply linear dependent on the transmissivity $T_{det}$. 
In particular, the former
gives the remarkable prediction \emph{(ii)}, mentioned above. Typical sinusoidal intensity modulations obtained with the transmissivity $T_{sto}\approx 23\,\%$ and $T_{det}\approx 24\,\%$ are depicted in Fig.\,\ref{fig_abs2}\,(top). As theory predicts, 
the contrast measured with the absorber foil was larger than that measured with the beam chopper even if $T_{sto}\approx T_{det}$. The amplitude of 
the interference fringes is plotted as a function of the transmissivity in Fig.\,\ref{fig_abs2}\,(bottom). Clear square-root and linear dependence is seen. Further studies allowed measurements
using absorbers with much lower transmissivity \cite{rauchPRA1990}. Experimental results of this measurements are shown in Fig.\,\ref{fig_abs2}\,(bottom, left and center). 
The values at very low transmissivity lie slightly
below the $\sqrt{T_{sto}}$ curve: in this low-contrast regime, other effects such as counting statistics become important, which can reduce the fringe contrast. Another experiment with x-rays studied 
the interference effect in the high absorption regime \cite{hasegawaPLA1994}. The latter experiment confirmed the square-root dependence 
even for low transmissivity.

By varying the reflectivity/transmissivity of the beam-splitting mirrors of the Mach-Zehnder IFM for 
visible light, similar phenomena were observed \cite{mittelstaedtFP1987}. The latter experiment was interpreted in terms of unsharp wave-particle behaviour \cite{wootersPRD1979,buschFP1987}. In contrast, the neutron IFM experiments were discussed in terms of
non-ideal measurements of the interference and the path \cite{muynckPRA1990}. 
Furthermore, reconsidering the effect of 
the beam-chopper, it turned out that the deterministic absorber generated not a pure state but a mixture of 
certain pure states. The chopper wheel generated a mixture in time of the states with
full and zero-contrast interference fringes. 
An experiment with a perfect-crystal IFM for X-rays was carried out,
where a beam-attenuating absorber was inserted \emph{partially} in one of the beams. This generated a mixture in space 
of beams with reduced and full intensity modulations \cite{hasegawaJJAP1991}. Here, the intermediate situation between 
the stochastic and the deterministic absorbers as well as an apparent destruction of the interference effect was observed.
Moreover, detailed studies of the combination of absorbers and the mixtures were carried out 
\cite{hasegawaZPB1994}. The phase difference of the intensity modulations plays an important role to induce an \emph{apparent} 
destruction of the interference effect. 

\begin{figure}
\begin{center} 
\scalebox{0.4}
{\includegraphics {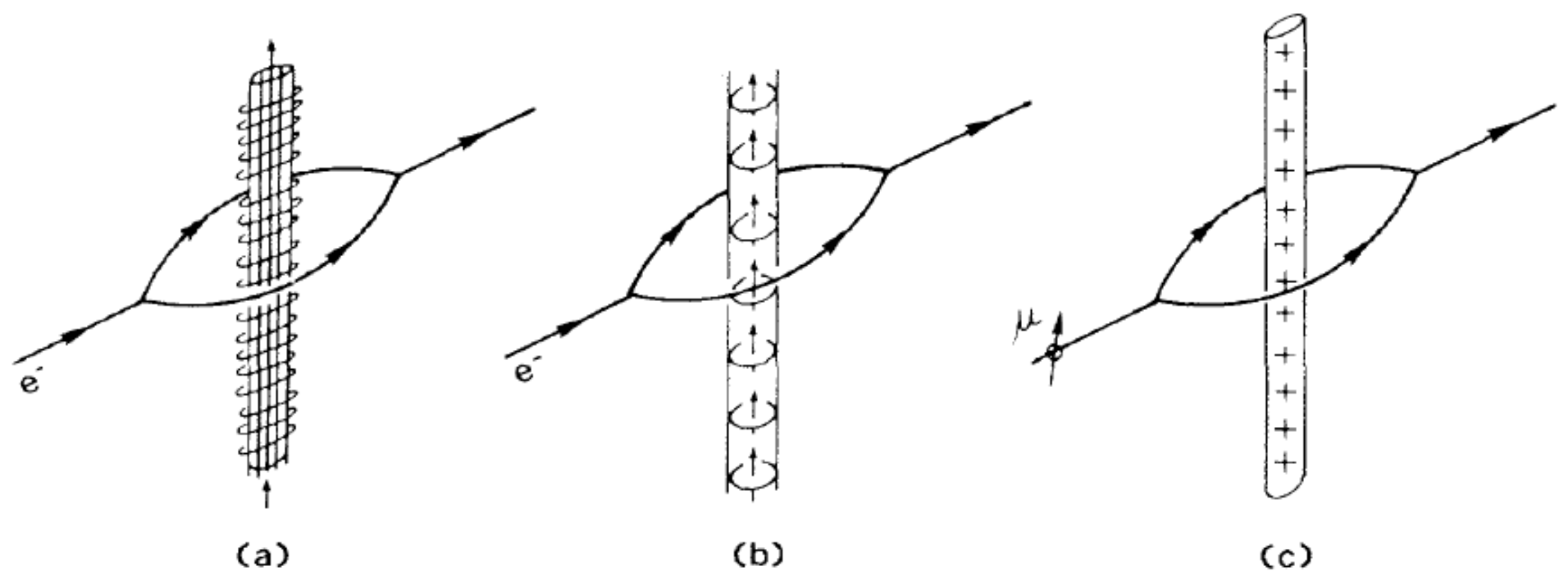}}
\caption{(a) The Aharonov-Bohm configuration in which the solenoid enclosed by the split electron beam can be seen as a chain of aligned magnetic dipoles (b), which -- for symmetry reasons -- suggests that a similar effect should exist (c) for a split beam of magnetic dipoles enclosing a line charge. Reprinted with permission from \cite{cimminoPRL1989}. Copyright (1989) by the American Physical Society.}
\label{fig:Cimmino1989a} 
\end{center}
\end{figure}

\subsection{Topological phases investigated with neutrons}\label{subsec:AvarietyOfTopAndGeoPh}

In classical electrodynamics, the measurable quantities -- the electromagnetic forces -- are calculated from electric and magnetic fields which can, in turn, be written in terms of so-called electromagnetic scalar- and vector-potentials. The potentials are usually seen as somewhat auxiliary quantities because they are not gauge invariant. However, in 1959 Aharonov and Bohm showed that in QM time-dependent scalar as well as time-independent vector potentials induce a measurable phase shift on the wave function of single electrons in a two-path IFM \cite{aharonovPR1959}. 
Interestingly, in the arrangement the electrons only travel in regions where all electromagnetic fields -- but not the potentials -- are zero, which demonstrates the previously unexpected physical significance of the electromagnetic potentials alone. For the vector-potential effect, the induced phase shift merely depends upon the magnetic flux enclosed by the IFM paths and not on the energy of the electrons. Due to the latter and the fact that the phase shift is independent of the particular geometry of the IFM paths as long as flux lines are encircled, the term \emph{topological phases} was coined. After decades of discussions about the existence of the Aharonov-Bohm phase (Aharonov-Bohm effect), conclusive evidence was given in an electron-holography experiment by Tonomura \emph{et al.} in 1986 \cite{tonomuraPRL1986}, in which the IFM-paths enclosed a torroidal ferromagnet ring. The electromagnetic stray fields were shielded by a superconducting- and a Copper-layer so that the electrons travelled in essentially field-free regions. 
Usually, the phases induced by electromagnetic scalar and vector potential are called scalar and vector Aharonov-Bohm phases (SAB- and VAB-phases or effects), respectively. 
 
\begin{figure}
\begin{center} 
\scalebox{0.3}
{\includegraphics {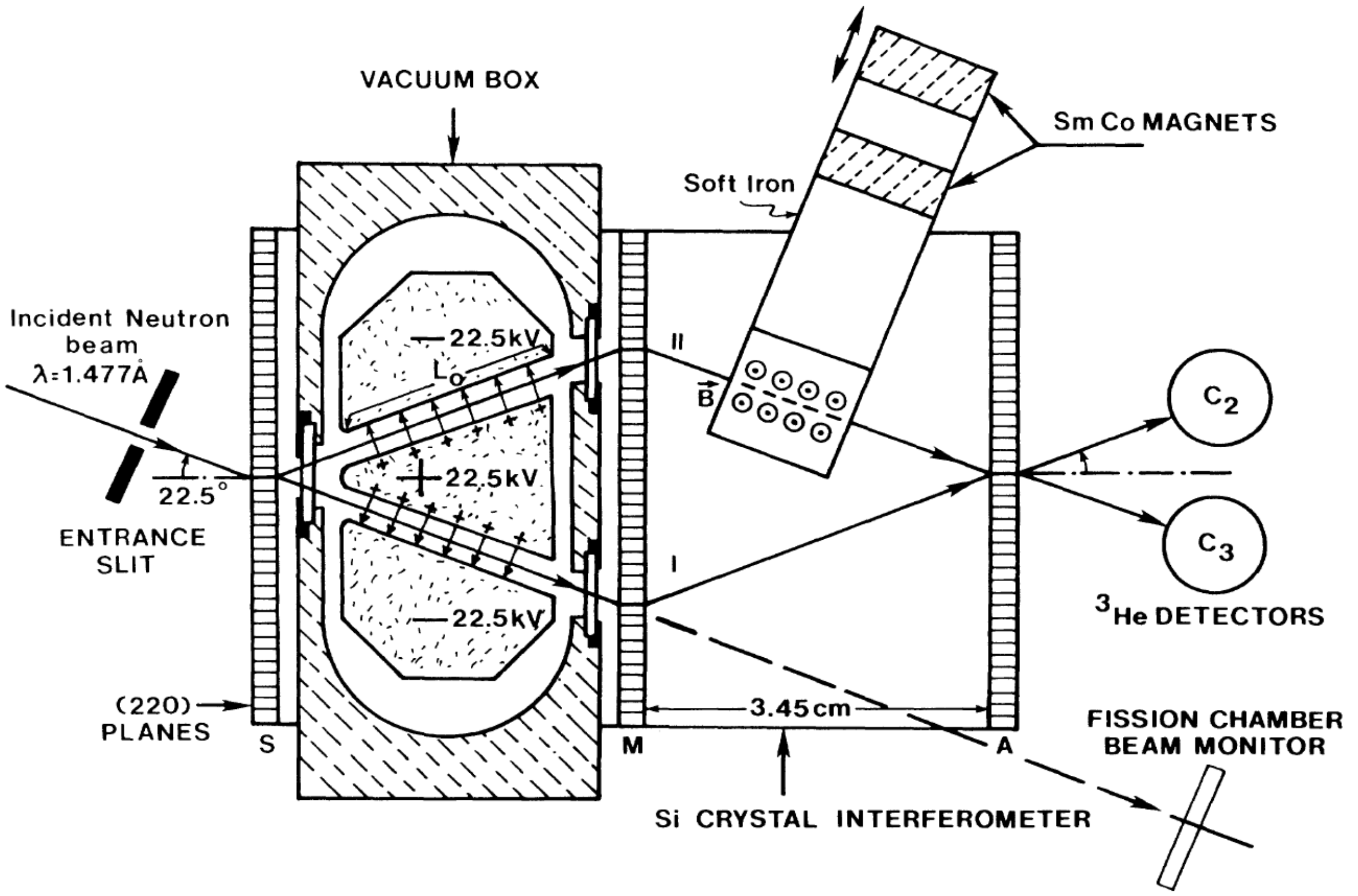}}
\caption{Experimental arrangement to measure the AC phase shift. Reprinted with permission from \cite{cimminoPRL1989}. Copyright (1989) by the American Physical Society.}
\label{fig:Cimmino1989b} 
\end{center}
\end{figure}

\subsubsection{Aharonov-Casher effect in neutron interferometry}

The VAB-arrangement can be envisaged as charged particles encircling a line of magnetic dipoles [see Fig.\,\ref{fig:Cimmino1989a}\,(a,b)].  
Therefore, also its counterpart -- neutral particles possessing magnetic dipole moment encircling a line charge -- should result in a measurable phase shift [see Fig.\,\ref{fig:Cimmino1989a}\,(c)], which was theoretically shown by Aharonov and Casher (AC) \cite{aharonovPRL1984} and could, indeed, be demonstrated experimentally with neutrons for the first time by Cimmino \emph{et al.} \cite{cimminoPRL1989}. 

In that experiment, a perfect-crystal neutron IFM was equipped with an electrode system that can be viewed as an array of line charges perpendicular to the IFM-path plane (here defined to be identical to the $xy$-plane, cf. Fig.\,\ref{fig:Cimmino1989b}). The applied voltage was 45\,kV at 0.154\,cm electrode-distance on a path-length of 2.53\,cm. The theoretical prediction for the AC-phase, arising due to an effective magnetic field $-\vec v/c\times\vec E$ is $\Delta\Phi_{\rm AC}=1.5\,\sigma\,$mrad, where $\sigma=\pm 1$, depending on the spin of the incident beam.
Even though the AC-phase depends on the spin direction, it was shown in \cite{cimminoPRL1989} that unpolarized neutrons can be used for the experiment if a suitable combination of gravitational (see Sec\,\ref{sec:gravity}) and magnetic (by a static magnetic field $B_z$, cf. Fig.\,\ref{fig:Cimmino1989b}) phase shifts  $\Delta\Phi_{\rm G}$ and $\Delta\Phi_{\rm M}$, respectively, are induced. Together, the AC-phase and $B_z$ result in the spin-dependent phase shifts $\pm\Delta\Phi_{\rm AC}$ and $\Delta\Phi_{\rm M}$, the former depending also on electrode-polarity, next to $\sigma$. 
For instance, the interference term of the prediction for the intensity in the $C_3$-detector can -- for electrode-polarity $\pm$ and an unpolarized incident beam -- be written as  
$\cos(\Delta\Phi_0+\Delta\Phi_{\rm G})\cos\left(|\Delta\Phi_{\rm M}|+|\Delta\Phi_{\rm AC}|\right)$, where $\Delta\Phi_0$ is an intrinsic IFM-phase. Now, tuning $\Delta\Phi_0+\Delta\Phi_{\rm G}$ to 0 (by inclining the IFM) and $\Delta\Phi_{\rm M}$ to $\pi/2$, the interference term becomes $\sin(\pm|\Delta\Phi_{\rm AC}|)\approx\pm|\Delta\Phi_{\rm AC}|$. Thus, with proper adjustment of $\Delta\Phi_{\rm G}$ and $\Delta\Phi_{\rm M}$, the count rate in the $C_3$ detector is linearly proportional to $|\Delta\Phi_{\rm AC}|$ for an unpolarized beam.

The AC-phase was measured to be $\Delta\Phi_{\rm AC}=2.19\pm 0.52$\,mrad in comparison to the theoretically expected value of 1.5\,mrad. The agreement with theory was improved in experiments with a Thallium fluoride molecular beam \cite{sangsterPRL1993}, a Calcium-atom Bord\'{e} IFM \cite{zeiskeAPB1995} and a Rubidium atom beam \cite{goerlitzPRA1995}. In the atom-IFM experiments, also the velocity-independence of the AC-phase could be demonstrated. However, the neutron-IFM experiment was the only one implementing the split-beam geometry, i.e., the correct topology. The atom IFMs used the interference between states of internal degrees of freedom instead and, therefore, resemble more the topology of the closely-related neutron spin-orbit coupling as demonstrated by Shull \cite{shullPRL1963}. The issue is -- together with suggestions for improvements of the neutron IFM experiment -- discussed in \cite{cimminoPhysB2006}.

\subsubsection{Scalar Aharonov-Bohm effect with neutrons} 

\begin{figure}
\begin{center} 
\scalebox{0.329}
{\includegraphics {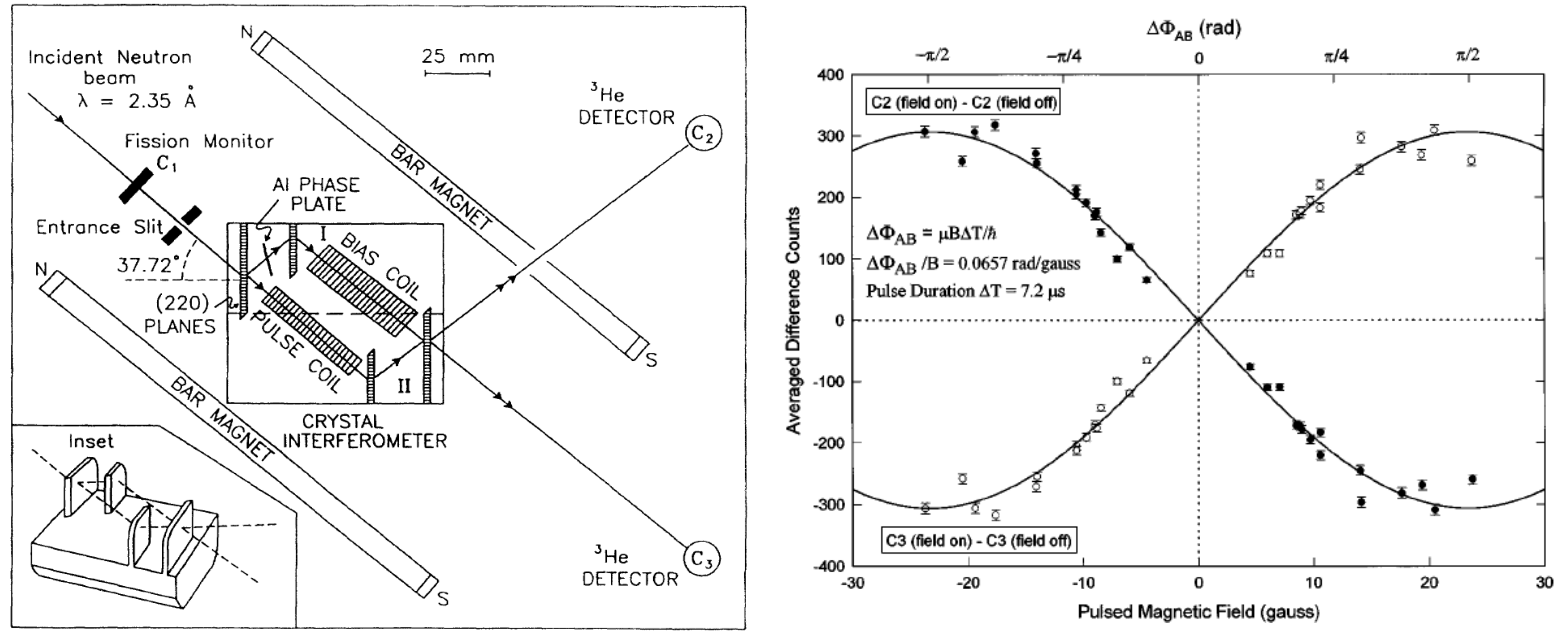}}
\caption{Left: Neutron-IFM setup to measure the SAB-phase with unpolarized neutrons. Right: Difference between the counts with pulsed magnetic-field on and off in the $C3$- and $C2$-detectors, measured with polarized neutrons. Reprinted with permission from \cite{allmanPRL1992,leePRL1998}. Copyright (1992,1998) by the American Physical Society.}
\label{fig:SABAllmanLee} 
\end{center}
\end{figure}

The SAB-phase for neutrons -- involving time-dependent magnetic fields -- has been discussed and tested by Allman \emph{et al.} \cite{allmanPRL1992}. 
An unpolarized neutron beam was subjected to a time-dependent scalar potential $V_{SAB}=-\mu B_2(t)$ in one IFM path, as shown in Fig.\,\ref{fig:SABAllmanLee}\,(left). Here, because of the unpolarized beam, a similar strategy as in the AC-experiment (see previous section) was pursued, but instead of an auxiliary gravitational phase shift a phase-shifter slab was employed. To ensure pure time-dependence of the potential for observation of the SAB-phase, the behaviour of $B_2(t)$ was logged. That way, neutron counts detected at certain field-configuration (`feeling' $B_2$ being turned on and off during their propagation within the coil) could be identified. The theoretical expectations were fully confirmed. 
The non-dispersive feature of the effect could not be investigated due to the rather narrow wavelength-acceptance of the IFM-crystal in Bragg position.

Since unpolarized neutrons were used for the experiment, the result launched a discussion about a possible classical torque and forces exerted on the neutrons that would render the suggested topological features of the SAB- and AC-effects inexistent \cite{peshkinPRL1992,pfeiferPRL1994,peshkinPRL1995}. A further neutron-IFM study with neutrons polarized in direction of the pulsed field \cite{leePRL1998} helped to somewhat settle the issue. Its results are shown in Fig.\,\ref{fig:SABAllmanLee}\,(right). Only recently, the discrepancy was seemingly resolved also from the theoretical point of view \cite{dulatPRL2012}. 

\begin{figure}
\begin{center} 
\scalebox{0.28}
{\includegraphics {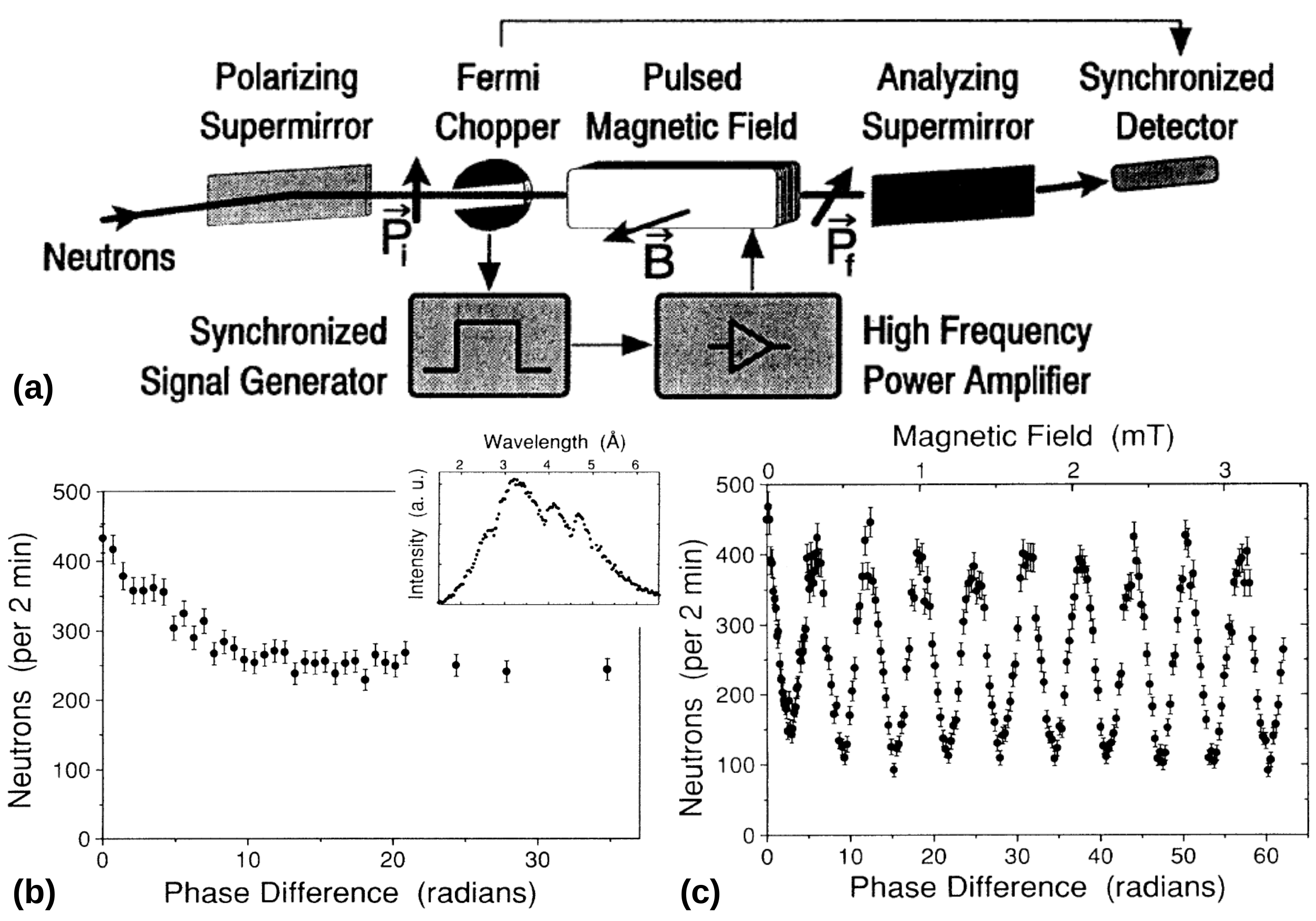}}
\caption{(a) Neutron-polarimeter setup to demonstrate the non-dispersive feature of the SAB effect.
(b) Count rate versus Larmor-phase (proportional to the static magnetic-field strength). Inset: Incident wavelength distribution. (c) Count rate versus SAB-phase (proportional to the pulsed magnetic-field strength). Reprinted with permission from \cite{badurekPRL1993}. Copyright (1993) by the American Physical Society.}
\label{fig:Badurek1993} 
\end{center}
\end{figure} 

In \cite{badurekPRL1993}, the non-dispersive feature of the neutron-SAB phase was demonstrated in a neutron-polarimeter experiment using a polarized beam with broad wavelength distribution [cf. Fig.\,\ref{fig:Badurek1993}\,(a),\,(b,\,inset)]. If the neutron beam -- prepared in a spin superposition -- traverses a strong \emph{static} magnetic field, the resulting dispersive Larmor-phase shift separates the wave packets associated to the up- and down-spin states (longitudinal Stern-Gerlach effect \cite{alefeldZPB1981}) by a distance larger than the longitudinal coherence length $\lambda^2/\Delta\lambda$ and an interference pattern cannot be observed [Fig.\,\ref{fig:Badurek1993}\,(b)]. However, this is not the case if the neutron wave packets travel trough a magnetic-field coil turned on only while the wave packets are inside and, therefore, do not experience any forces but only the potential $-\mu B(t)$ [Fig.\,\ref{fig:Badurek1993}\,(c)]. In the experiment, the neutron beam was pulsed by a chopper. The latter and the magnetic-field coil were phase-locked by a signal generator as illustrated in Fig.\,\ref{fig:Badurek1993}\,(a). An experiment demonstrating the non-dispersive feature was later also carried out in a VCN-IFM \cite{vanDerZouwNIMA2000}.
An excellent overview on the topic is given in Ref.\,\cite{wernerJPA2010}.

\subsection{Geometric phases}\label{subsec:GeoPhase} 

In 1984 Michael Berry realized that slow (so-called adiabatic) and cyclic evolutions of quantum systems comprise a so-far \lq{forgotten}\rq \,phase factor. Unlike the usual dynamical phase factor $\exp(-iHt/\hbar)$, it only depends on the solid angle $\Omega$ enclosed by the evolution path of a quantum state in parameter space as seen from the point of degeneracy \cite{berryPRSLA1984}. In particular, the Berry phase is equal to $-\Omega/2$ for two-level systems. A first experimental demonstration was soon accomplished 
using photons \cite{tomitaPRL1986}.

\subsubsection{The Berry phase tested with neutrons}\label{subsubsec:Berry}

\begin{figure}
\begin{center} 
\scalebox{0.5}
{\includegraphics {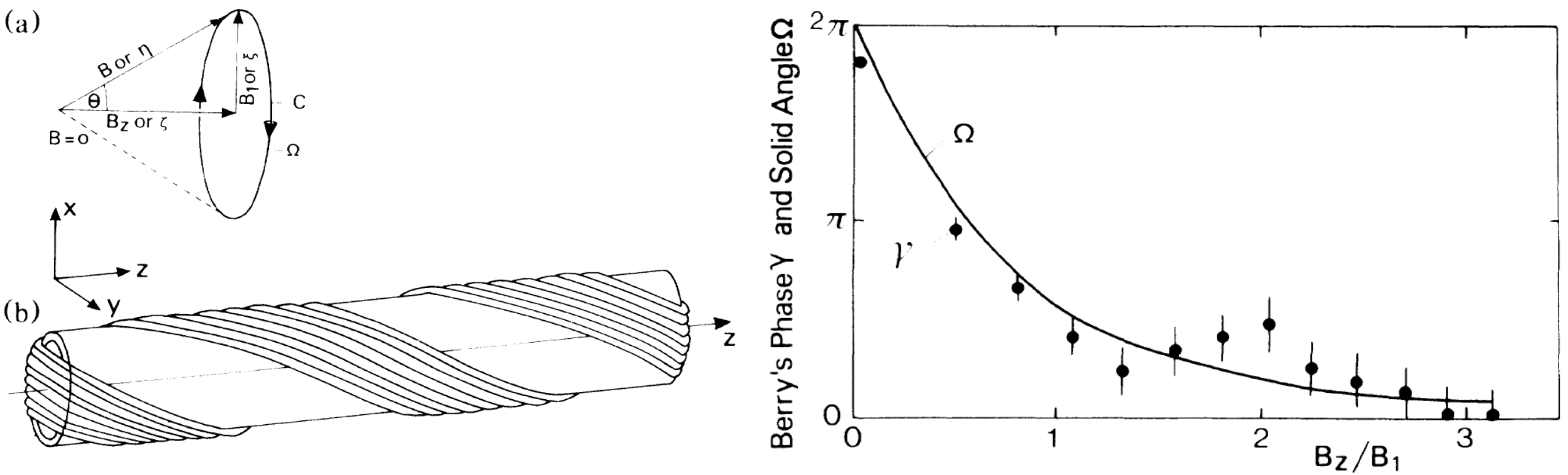}}
\caption{Left: (a) Field configuration. (b) Sketch of the helical coil to produce a twisted magnetic field $B$ by superposing the helical field $B_1$ and a solenoid field $B_z$ (parallel to neutron beam in $z$-direction, not shown).  
Right: Measured Berry phase versus field ratio $B_z/B_1$ that determines the solid angle as seen from the origin of the parameter space (at $B_x=B_y=B_z=0$). Reprinted with permission from \cite{bitterPRL1987}. Copyright (1987) by the American Physical Society.}
\label{fig:bitter1987} 
\end{center}
\end{figure}

A neutron spin-state can be taken around a circular path by a static magnetic field arranged along the neutron flight-path. The magnetic field would slowly change directions and take the spin along a circle and the accumulated phase can be measured behind the arrangement. Such a polarimeter experiment was indeed carried out by Bitter and Dubbers \cite{bitterPRL1987}. 
In that experiment, a helical coil was used to produce a (for a neutron velocity of about 500m/s) slowly varying magnetic field to induce an adiabatic evolution along a circular path $C$. By variation of the magnetic field-ratio $B_z/B_1$, the Berry phase was measured as a function of $\Omega$ (see Fig.\,\ref{fig:bitter1987}). 

Another early experimental test of the Berry phase was achieved with UCN by Richardson \emph{et al.} \cite{richardsonPRL1988}.
In contrast to \cite{bitterPRL1987}, here, the magnetic field direction was not varied in space but in time. The group used a well-shielded apparatus designed for establishing an experimental limit to the neutron electric-dipole moment, equipped with an additional coil to generate magnetic fields in arbitrary directions. 
In that experiment, it was also shown that the Berry phase is additive in the sense that multiple excursions along the same path add up to the multiple of the Berry phase induced by that path (cf. related discussions in Sec.\,\ref{sec:GeometricPhaseForMixedStates}). 

\subsubsection{Geometric phase arising from various types of quantum evolutions}\label{susubsec:PanchPhase}

Soon, it was realized that Berry's concept was closely related to Pancharatnam\rq{}s work \cite{pancharatnamPIAS1956,berryJMO1987}. The Pancharatnam phase is defined as the argument of a complex number 
$\langle \psi|U|\psi\rangle$, where 
$|\psi\rangle$ and $U|\psi\rangle$ are any non-orthogonal and, in general, non-collinear state vectors. 
Here, $U$ is an operator denoting a unitary evolution. The phase is measured by some sort of interferometry experiment, in which a state $|\psi\rangle$ is prepared, split up (not necessarily in space) and one part let evolve to 
$U|\psi\rangle$, which is finally brought to interference with $|\psi\rangle$. 
The measured signal is usually an intensity oscillation with fringe contrast 
$|\langle \psi|U|\psi\rangle|$ that appears due to application of an auxiliary phase shift. 
The obtained fringes are shifted by arg$\langle \psi|U|\psi\rangle$ in comparison to the fringes measured in a situation with $U=\1$ or, more generally, when $\langle \psi|U|\psi\rangle$ is real and positive. Only in cases in which $U$ takes the state along a great circle on the Bloch-sphere, the Pancharatnam phase is a purely geometric phase. In general, it comprises a dynamical- and a geometric-phase part. 

The theoretical concept of Berry was rapidly generalized to non-adiabatic \cite{aharonovPRL1987}
and non-cyclic \cite{samuelPRL1988} evolutions. 
In non-cyclic evolutions, the path on the geodesic is not closed. 
It turned out that a last geodesic part of the evolution path can be spared to still obtain the very same geometric phase as for a cyclic path that fully encloses $\Omega$. 

\begin{figure}
\begin{center} 
\scalebox{0.43}
{\includegraphics {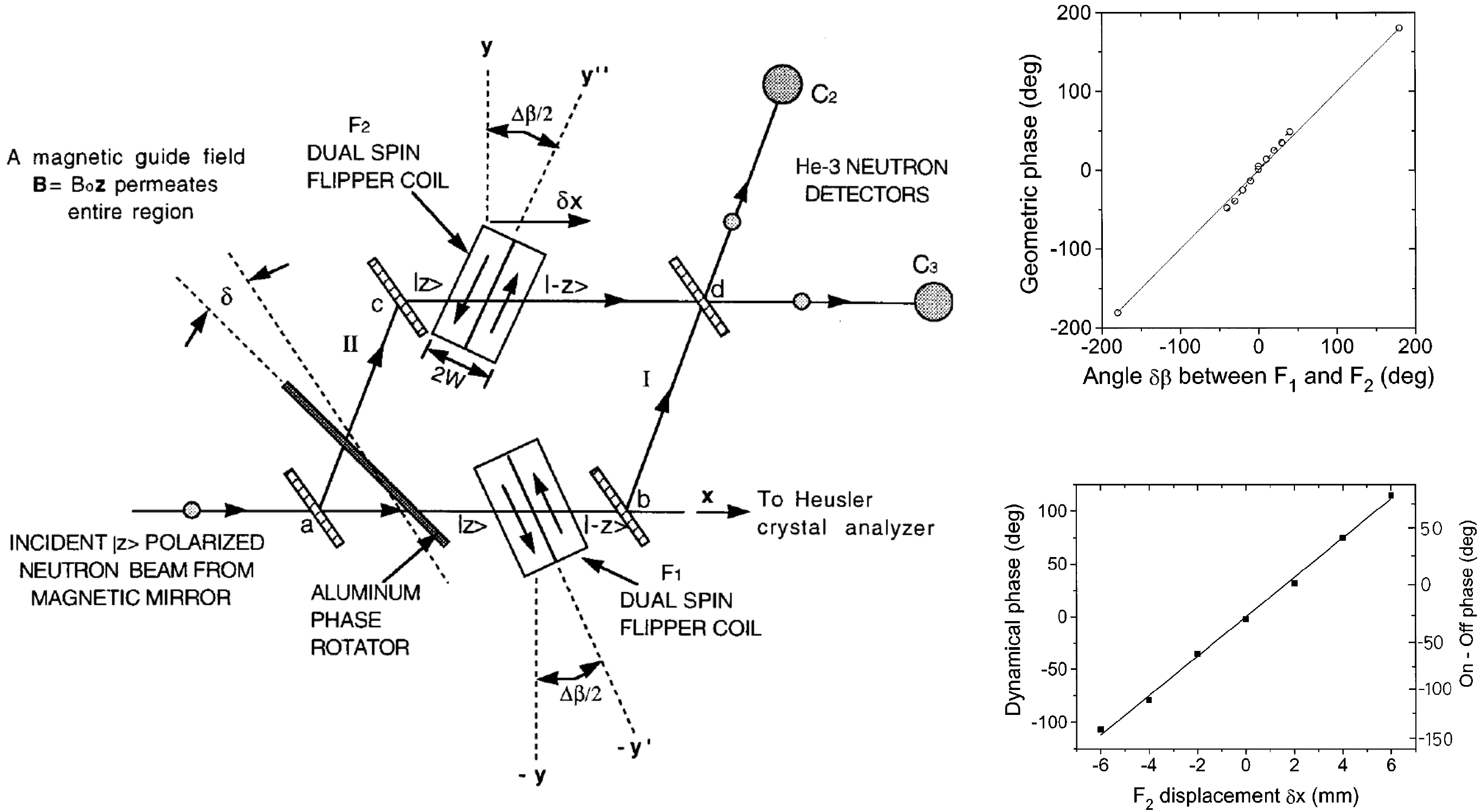}}
\caption{Left: Perfect-crystal neutron IFM setup to address geometric and dynamical phases separately. Right: Experimental data for rotation (above) and displacement (below). Reprinted with permission from \cite{waghPRL1997,allmanPRA1997}. Copyright (1997) by the American Physical Society.}
\label{fig:Wagh1997} 
\end{center}
\end{figure}

Relevant experimental data for non-adiabatic (but cyclic) evolutions was first obtained in perfect-crystal IFM experiments by Wagh \emph{et al.} \cite{waghPRL1997,allmanPRA1997}, in which the spin state was rotated from the north- to the south-pole of the Bloch sphere along different effective meridians in each IFM path (see Fig.\,\ref{fig:Wagh1997}). The meridians were separated by an azimuthal angle $\delta\beta$ that determined the enclosed solid angle $\Omega$. The spin state was rotated by DC flippers F$_1$ and F$_2$ in each path. $\delta\beta$ was set by rotation of the flippers about the vertical axis. The phases were obtained by comparison of the curves measured with F$_1$ and F$_2$ on and off. By variation of $\delta\beta$, only the geometric phase was varied, while displacement of F$_2$ along the IFM path led to varied dynamical phases. The results were later confirmed at improved accuracy using a similar method in a polarimeter experiment \cite{waghPLA2000}.  

\begin{figure}
\begin{center} 
\scalebox{0.33}
{\includegraphics {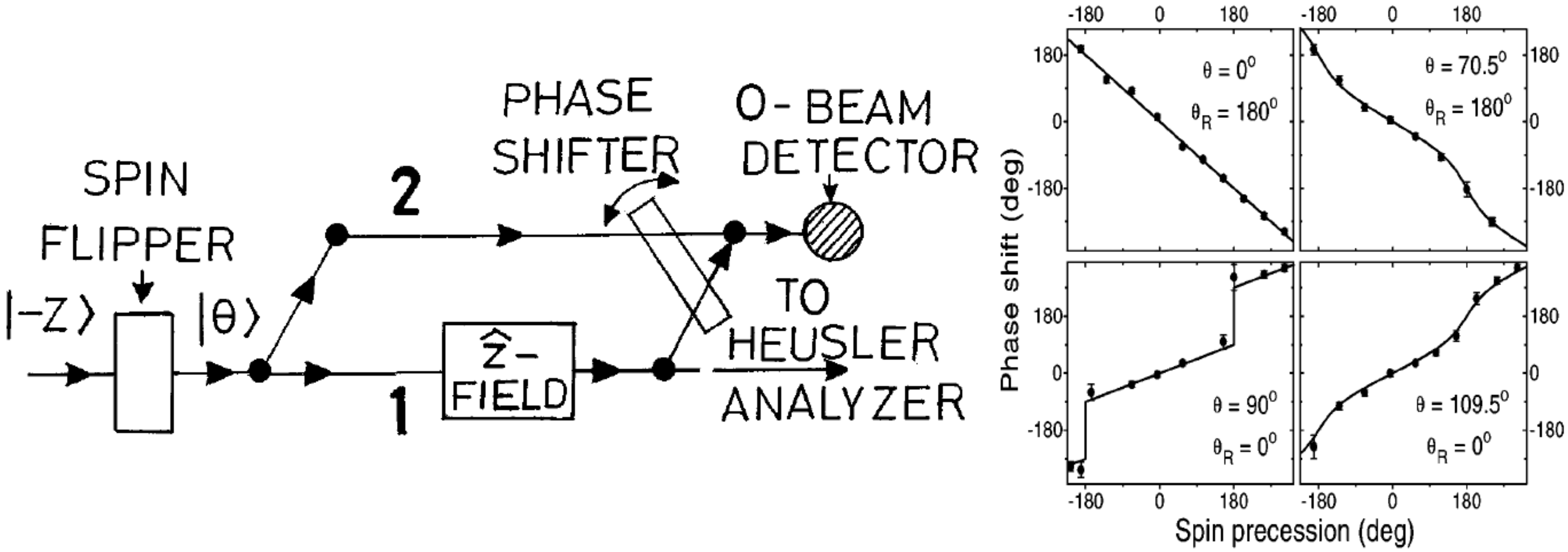}}
\caption{Left: Perfect-crystal IFM setup for measuring Pancharatnam phases in non-adiabatic and non-cyclic evolutions \cite{waghPLA1995}. Right: Theoretical expectations (solid lines) and measured phases (data points) for four different incident-to-reference state combinations ($\theta$ to $\theta_R$ combinations) versus $\phi_L$ induced by the $\hat z$-field in path 1. Reprinted with permission from \cite{waghPRL1998}. Copyright (1998) by the American Physical Society.}
\label{fig:Wagh1998} 
\end{center}
\end{figure}

In \cite{weinfurterPRL1990}, a neutron polarimeter experiment is described that was designed to measure the adiabatic and non-cyclic geometric phase. After some discussions about if the physical quantity addressed in that paper was a phase or merely a precession angle \cite{waghPLA1995b}, a perfect-crystal IFM experiment was carried out to measure the total Pancharatnam phase \cite{waghPLA1995,waghPRL1998} and, in particular, the non-adiabatic and non-cyclic geometric phase for the special case of spin evolutions along the equator of the Bloch sphere. The setup and the resulting data are shown in Fig.\,\ref{fig:Wagh1998}. A spin flipper prepared the incident polarized neutron beam in a state $\cos\theta/2|\!\Uparrow\rangle + \sin\theta/2|\!\Downarrow\rangle$. 
With an additional static magnetic field (aligned to $|\!\Uparrow\rangle$) in one IFM path, intensity oscillations were measured upon rotation of a phase-shifter slab for either incident state. The same was done for 
a reference incident state $\cos\theta_R/2|\!\Uparrow\rangle + \sin\theta_R/2|\!\Downarrow\rangle$. 
The phase shift of intensity oscillations to the reference-curve is the Pancharatnam phase measured in the experiment. The induced phase is equal to the non-cyclic geometric phase for $\theta=\pi/2$. A proposal for a polarized-neutron interferometry experiment in which the dynamical phase cancels out and only the non-cyclic geometric phase is measured was made in \cite{sjoeqvistPLA2001}.

Note that the geometric phase only depends on the evolution path of the system and not on dynamical properties such as neutron energy (It is, however, important to keep in mind that a certain experimental setting leads to a particular evolution-path only for a small wavelength-band.). It is, therefore, not surprising that topological and geometric phases are related. As already pointed out in \cite{berryPRSLA1984}, the VAB-phase can be interpreted as a special case of the Berry phase in which the adiabaticity constraint is lifted.

A closer look at more recent neutron-optics experiments related to geometric phases is taken in Sec.\,\ref{sec:TopologicalAndBerryPhases}.

\section{Quantum contextuality and entanglement studied with neutrons}\label{sec:contextuality}

\subsection{Entanglement in various quantum systems}

Led by his abhorrence of non-locality -- a feature at the very heart of the standard interpretation of QM -- Einstein believed that non-locality demonstrated QM to be incomplete. 
Together with his co-workers Podolsky and Rosen (EPR), Einstein argued that more complete, deterministic \emph{hidden} physic must underly QM. 
His believes are expressed  in the famous EPR-paper of 1935 \cite{einsteinPR1935}. Einstein concludes with the following sentences: 
\emph{``While we have thus shown that the wavefunction does not provide a complete description of the physical reality, we left open the question of whether or not such a description exists. We believe, however, that such a theory is possible''}. 
Such a theory should be supplemented by additional \emph{hidden-variables}, addressing objective properties (\emph{elements of reality}) of physical systems in order to restore causality and locality. 
In 1951, Bohm reformulated the EPR-argument for spin observables of two spatially separated entangled particles to illuminate the essential features of the EPR-scenario \cite{bohm1951}. 
In 1964, Bell -- initially a follower of Einstein's realistic view -- proved in his celebrated theorem that all hidden-variable theories which are based on the assumptions of locality and realism conflict with the predictions of QM \cite{bellPhysics1964}. 
Bell introduced inequalities which hold for the predictions of any local hidden-variable theory applied, but are violated by QM. Violation of a Bell-inequality proves the presence of entanglement (also known as \emph{EPR-correlation}), a term coined by Schr\"{o}dinger \cite{schroedingerNW1935}. 
Entanglement has become a key ingredient for quantum-communication and quantum-information science \cite{nielsenChuangBook2000}. Bell's theorem has finally ruined Einstein's dream of a realistic description of nature and laid the cornerstone for the present view of QM.

Five years after Bell's paper, Clauser, Horne, Shimony and Holt (CHSH) reformulated Bell's inequality pertinent to the first experiment aiming at a demonstration of quantum non-locality \cite{clauserPRL1969}. 
Polarization measurements of correlated photon pairs produced in an atomic cascade and the use of one-channel polarizers allowed for the first experimental violation of Bell's inequality in 1972 \cite{freedmanPRL1972}. 
With the use of two-channel polarizers, experiments similar to the scheme described by Bohm were performed \cite{aspectPRL1982,aspectPRL1982b}. 
Development of a new type of entangled-photon source using parametric down conversion led to violations of the CHSH-inequality in almost perfect accordance with the prediction of QM \cite{kwiatPRL1995,weihsPRL1998,tittelPRL1998}. 
To date, entanglement has been verified for a number of quantum systems such as $^9\rm{Be}^+$ions \cite{roweNature2001}, photon-ion hybrid systems \cite{moehringPRL2004}, protons \cite{sakaiPRL2006}, $\rm{Yb^+}$ ions \cite{matsukevichPRL2008}, and neutrons \cite{hasegawaNature2003}.

\begin{figure}
\begin{center}
 \includegraphics[width=120mm]{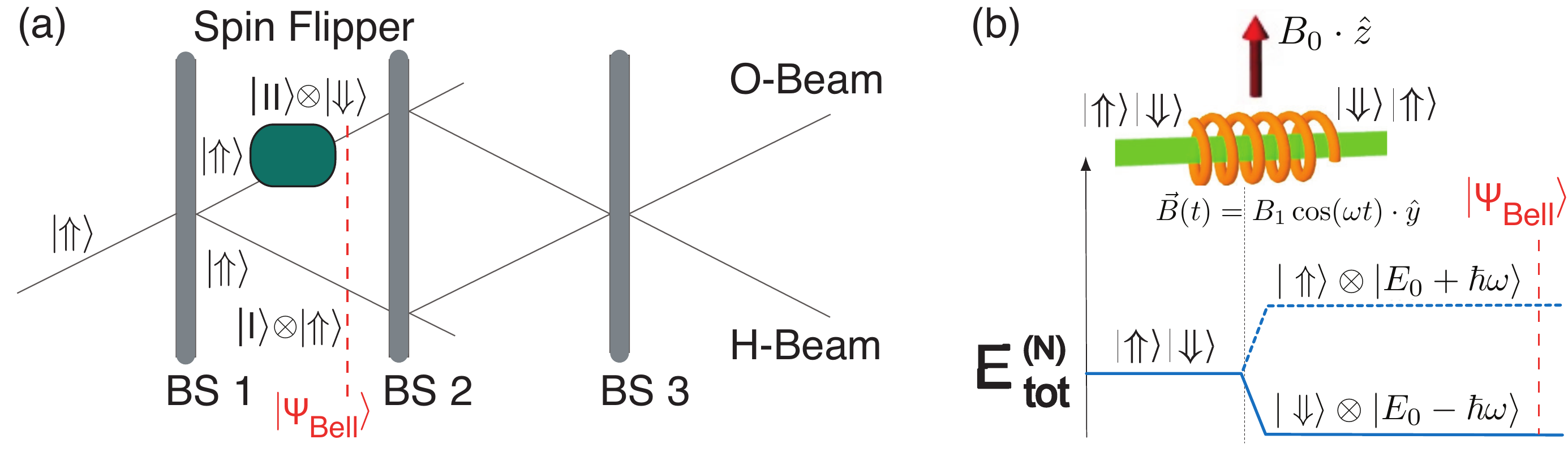}
\caption{(a) Principal of spin-path entanglement preparation in neutron interferometry. The spin in one path is flipped, thereby creating a Bell state. (b) Principle of spin-energy entanglement preparation within an RF-flipper. Both spin components are flipped due to photon absorption or emission, thereby preparing a Bell state.}\label{fig:IFM_POL_Ent_Prep}
\end{center}
\end{figure}

\subsubsection{Quantum non-locality and contextuality}

Bell's original inequality is based on the joint assumptions of locality
and realism. The corresponding class of hidden-variable theories are 
accordingly the local hidden-variable theories (LHVTs). A class of  theories
that maintains realism but abandons locality was proposed by Leggett in 2003 \cite{leggettFOP2003}: the non-local hidden-variable theories (NLHVTs). Leggett also proposed an incompatibility theorem proving the contradictoriness of this class of models with quantum predictions. An experimental falsification of the NLHVTs for entangled photons is reported in  \cite{groeblacherNature2007}.

Furthermore, it is possible to derive a Bell-like inequality by introducing the concept of
non-contextuality. 
Non-contextuality implies that the value of an observable is predefined and independent of the
experimental context, i.e. of previous or simultaneous measurements
of a commuting observable \cite{merminRMP1993}.
Non-contextuality is a more stringent demand than locality because it
requires mutual independence of the results for commuting
observables even if there is no space-like separation involved \cite{simonPRL2000}.
The corresponding class of realistic theories is called non-contextual hidden-variable theories NCHVTs.
Bell's locality is a special case of this non-contextual hidden-variable hypothesis.

Apart from Bell's theorem, there exists a second powerful argument against the possibility of extending QM into a more complete theory, namely the Kochen-Specker (KS) theorem \cite{kochenJMM1967}. 
While violations of Bell-inequalities discard LHVTs, the KS-theorem stresses the incompatibility of QM with NCHVTs. 
The theorem is based on the following two assumptions: 
\emph{(i)} value definiteness, i.e. observables $A$ and $B$ have predefined values $v(A)$ and $v(B)$; \emph{(ii)} non-contextuality, i.e. properties of the system exist independently of any measurement context, in particular, independently of other measurements of compatible observables performed simultaneously/before/after. 
According to these assumption, the relations $v(A+B)=v(A)+v(B)$ and $v(A \cdot B)= v(A)\cdot v(B)$ hold for compatible observables, which have a common eigenbasis. It has been proven mathematically that it is impossible to satisfy both relations for arbitrary pairs of compatible observables $A$ and $B$ within the framework of QM. 
Kochen and Specker's original proof involves 117 vectors in three dimensions. 
Simplified versions have been proposed by Peres \cite{peresPLA1990} and Mermin \cite{merminPRL1990,merminRMP1993}. The simplest proof of the KS-theorem was found by Cabello and uses only 18 vectors in four dimensions \cite{cabelloPLA1996}. 
Based on this proof, a state-dependent \cite{cabelloPRL2008} as well as state-independent \cite{cabelloPRL2008b} experimental test of the KS-theorem was proposed by Cabello. 
The former was carried out with photons
\cite{michlerPRL2000} and neutrons \cite{hasegawaPRL2006,bartosikPRL2009}, the latter using trapped ions \cite{kirchmairNature2009} and photons \cite{amselemPRL2009}. 

\begin{figure}
\begin{center}
 \includegraphics[width=70mm]{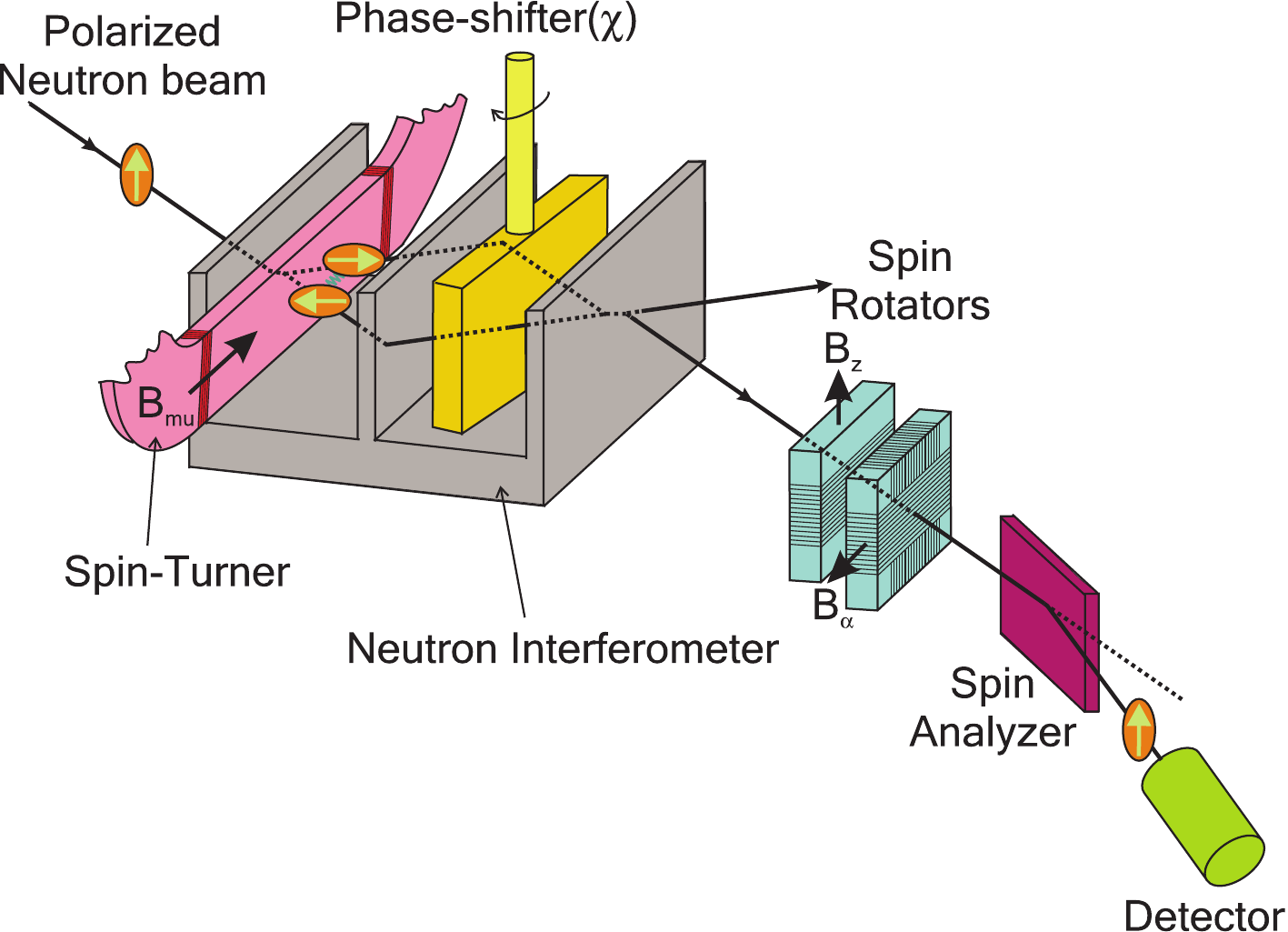}
\caption{Neutron interferometric setup using a Mu-metal spin-turner for the preparation of the spin-path entangled single neutron state \cite{hasegawaNature2003}.}\label{fig:Bell_Yuji}
\end{center}
\end{figure}

\subsubsection{Entanglement between particles and degrees of freedom}

In the case of neutrons, entanglement is achieved between different
degrees of freedom (\emph{intra-particle} entanglement) and not between individual particles (\emph{inter-particle} entanglement). Each individual degree of freedom (DOF) is described formally as a two-level system represented by state vectors in a two-dimensional complex Hilbert space  $\mathcal H_{\rm {DOF_i}}$. The overall system is described by the product Hilbert space given by $\mathcal H_{\rm {tot}}=\prod_i \mathcal H_{\rm {DOF_i}}$. Since the observables of a subspace commute with observables of a different subspace, the single-neutron system is suitable for studying NCHVTs with multiple DOF.

\subsection{Bi-partite entanglement: spin-path and spin-energy entanglement}\label{sec:Bi}

One example of intra-particle entanglement is an entangled state of the neutron spin- and path-DOF in neutron interferometry. The corresponding product Hilbert-space is  $\mathcal H=\mathcal H_{\rm {spin}}\otimes\mathcal H_{\rm{path}}$. $\mathcal H_{\rm {spin}}$ is spanned by spin-up and spin-down eigenstates, denoted as $\ket{\!\Uparrow}$ and $\ket{\!\Downarrow}$, referring to a quantization axis along a static magnetic field  (here usually pointing to the $+z$-direction). $\mathcal H_{\rm {path}}$ is spanned by the orthogonal states for paths $\vert\rm{I}\rangle$ and $\vert\rm{II}\rangle$ of the IFM. When the incident beam is polarized, the spin in one path of the IFM can be flipped and the neutron wavefunction exhibits entanglement between the spinor and the spatial part, which is schematically illustrated in Fig.\,\ref{fig:IFM_POL_Ent_Prep}\,(a). The corresponding state vector is a maximally entangled Bell state denoted as
$
\ket{\Psi_{\rm{Bell}}} =1/\sqrt{2}(
 \ket{\textrm{I}}\ket{\!\Uparrow}+
  \ket{\textrm{II}}\ket{\!\Downarrow})$.

\begin{figure}
\begin{center}
 \includegraphics[width=120mm]{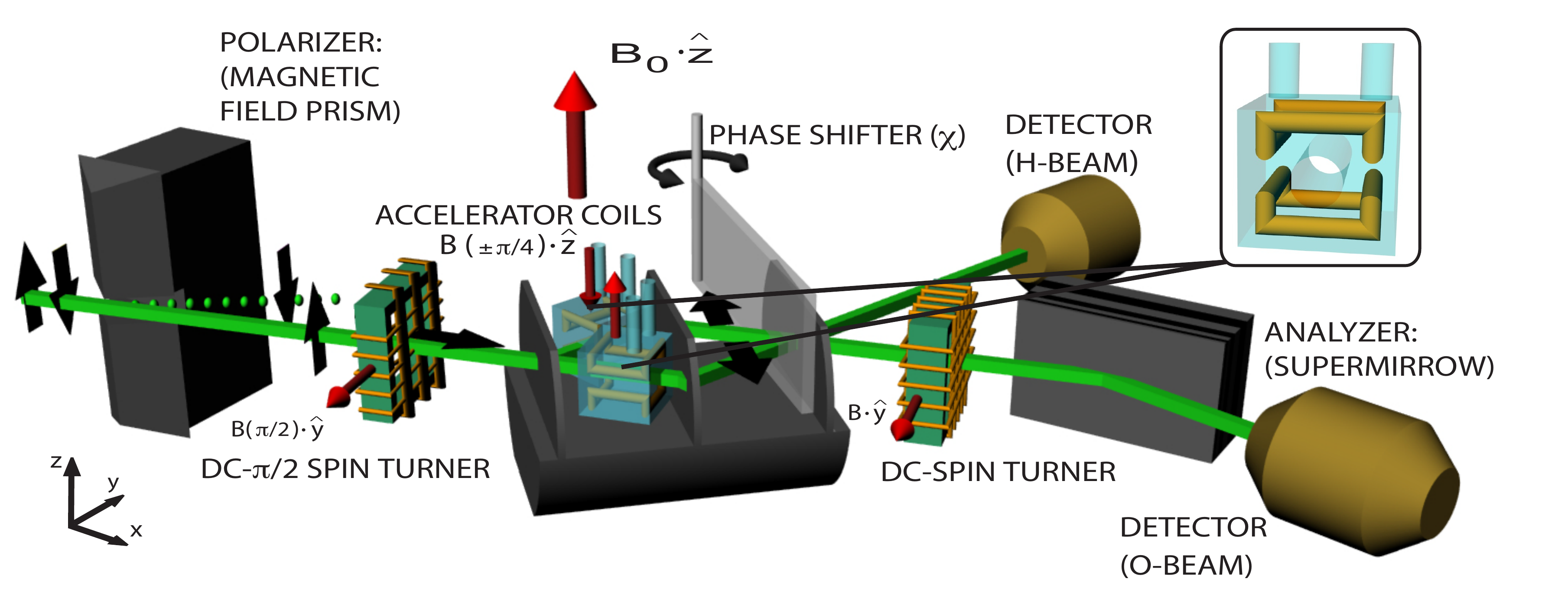}
\caption{Improved setup for demonstrating a violation of a Bell-like inequality for spin-path entanglement.}\label{fig:IFM_Bell_New}
\end{center}
\end{figure}

Another intra-particle type of entanglement is spin-energy entanglement. As discussed in Sec.\,\ref{Sec:Time}, when interacting with a time-dependent magnetic field the total energy of neutrons is no longer conserved. The total energy of the neutron decreases (or increases) by
$\hbar\omega$ during the interaction with the RF-flipper. This fact can be used to create a spin-energy entangled state expressed as  
$
\ket{\Psi_{\rm{Bell}}}=1/\sqrt 2(\ket{E_0+\hbar\omega} \ket{\!\Uparrow} + \ket{E_0-\hbar\omega}
\ket{\!\Downarrow})$.
The incoming spin-superposition can be created by applying a $\pi/2$ spin-rotation of an initially polarized beam. $\ket{E_0+\hbar\omega}$ and $\ket{E_0-\hbar\omega}$ are the energy eigenstates after interaction with a time-dependent magnetic field within an RF-flipper driven at frequency $\omega$. A graphical representation of spin-energy entanglement preparation is shown in Fig.\,\ref{fig:IFM_POL_Ent_Prep}\,(b).

\subsubsection{Violation of Bell-like inequality for single neutrons}\label{sec:Bell}

The first violation of a Bell-like inequality for a spin-path entangled state was achieved in 2003 
\cite{hasegawaNature2003}. 
The entanglement was realized
using a Mu-metal spin-turner consisting of a soft-magnetic
Mu-metal sheet with high permeability. 
In the experiment, both sub-beams traversed the Mu-metal. 
In one path, the initial spin was
turned from $\ket{\!\Uparrow}$ to $\ket{\!\Leftarrow}$, whereas in the
other path, due to different path lengths within the
soft-magnetic Mu-metal, the spin was turned from $\ket{\!\Uparrow}$ to
$\ket{\!\Rightarrow}$. 
Thus the initially prepared Bell state reads
$\ket{\Psi^{\mu}_{\rm{Bell}}}
=1/\sqrt 2(\ket{\!\Rightarrow}\ket{\textrm{I}}
+\ket{\!\Leftarrow}\ket{\textrm{II}})$. 
The expectation values for the joint spin-path measurements are given by 
$\bra{\Psi^{\mu}_{\rm{Bell}}}\hat P^{(\rm s)}_{\alpha\pm1}\hat P^{(\rm p)}_{\chi\pm1}\ket{\Psi^{\mu}_{\rm{Bell}}}
$, 
where $\hat P^{(\rm s)}_{\alpha\pm1}$ and $\hat P^{(\rm p)}_{\chi\pm1}$ are the projection operators to the states $1/\sqrt 2(\ket{\!\Uparrow}\pm e^{i\alpha}\ket{\!\Downarrow})$ and $1/\sqrt 2(\ket{\rm I}\pm e^{i\chi}\ket{\rm II})$, respectively. 
The required values for $\alpha$ and $\chi$ were tuned by spin rotators and a phase shifter, as depicted in Fig.\,\ref{fig:Bell_Yuji}. 
A maximum violation for the Bell-like inequality $-2\leq S\leq2$ of $S_{\rm th}=2\sqrt2$ is expected for $\alpha_1=0$, $\alpha_2=\pi/2$, $\chi_1=\pi/4$ and $\chi_2=-\pi/4$. 
In the experiment, the expectation values $E(\alpha_i,\chi_j)$ (with $i,j=1,2$) were determined by a combination of count rates with appropriate settings of $\alpha$ and $\chi$. 
The expectation values are expressed as
\begin{equation}\label{eq:expectExperimentCounts}
\!\!\!\!\!\!E(\alpha_i,\chi_j)=\frac{N(\alpha_i,\chi_j)+N(\alpha_i^\bot,\chi_j^\bot)-N(\alpha_i,\chi_j^\bot)-N(\alpha_i^\bot,\chi_j)}
{N(\alpha_i,\chi_j)+N(\alpha_i^\bot,\chi_j^\bot)+N(\alpha_i,\chi_j^\bot)+N(\alpha_i^\bot,\chi_j)},
\end{equation}
with $\alpha_i^\bot=\alpha_i+\pi$ and $\chi_j^\bot=\chi_j+\pi$. 
A final value of 
$S_{\rm exp}=2.051 \pm 0.019\nleq2$
was achieved, which violates the Bell-like inequality by almost three standard deviations.

\begin{figure}
\begin{center}
 \includegraphics[width=105mm]{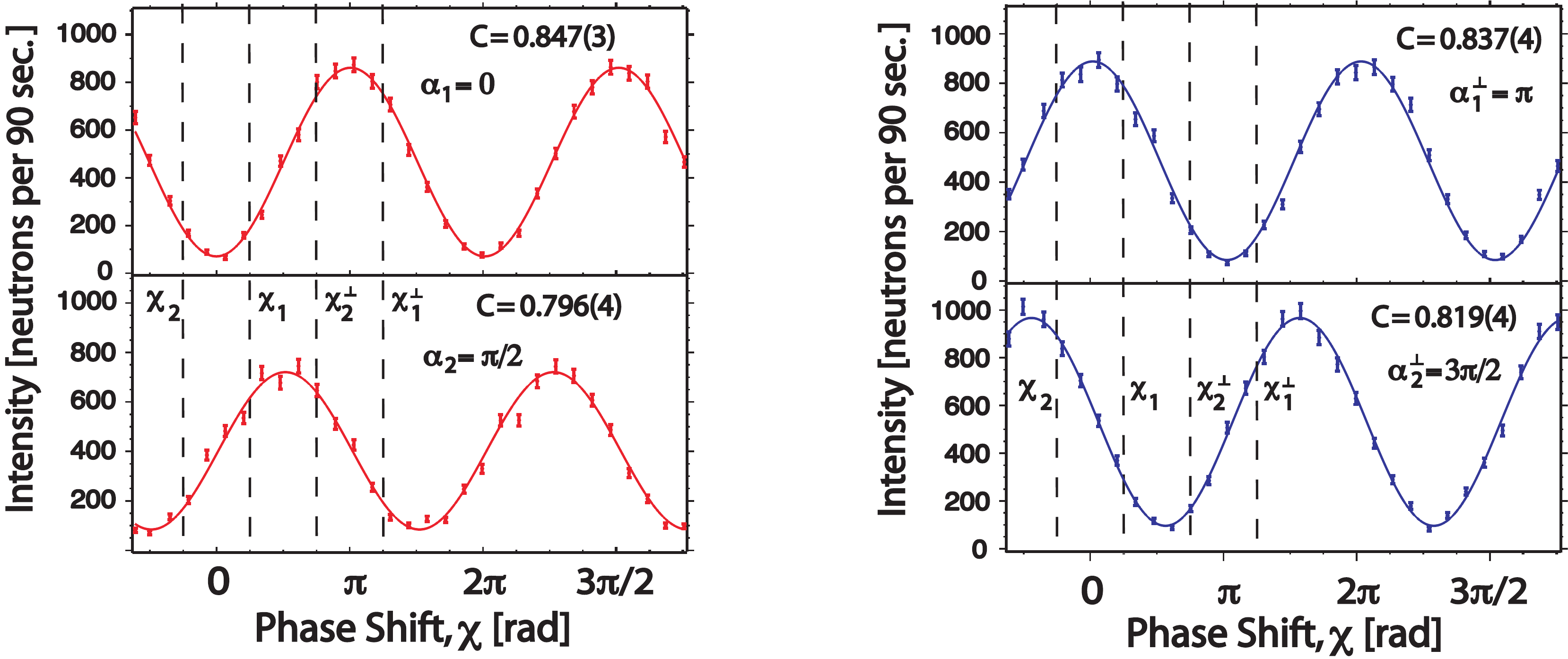}
\caption{Interference patterns for spin rotation-angle $\alpha=0$, $\pi/2$, $\pi$ and $3\pi/2$. Expectations values are calculated from intensities occurring at approroriate phase-shifter settings $\chi=\pi/4$, $-\pi/4$,  $7\pi/4$, and $5\pi/4$.}\label{fig:IFM_Bell_NewGraphes}
\end{center}
\end{figure}

This first experiment exhibited a violation of a Bell-like inequality. However the observed value was quite close to the classical border of 2. 
Thus, improvements of the setup were conceived. The Mu-metal sheet caused a considerable loss of interference contrast, therefore, it was replaced by two components: a DC-coil outside the IFM and an accelerator coil (a DC coil with magnetic field pointing in direction of the guide field to accelerate Larmor-precession within its field) in each arm of the IFM. 
Unlike in the previous experiment, the Bell-state preparation is split into two stages: 
\emph{(i)} The DC $\pi/2$ spin-turner rotates the spin into the $xy$-plane. The peculiarity of this coils lies in the fact that the horizontal windings are constructed using thin copper ribbons (instead of wire) to avoid small-angle scattering. 
\emph{(ii)} Behind the beamsplitter the sub-beams are exposed to accelerator coils rotating the spin by $\pm\pi/4$ in arm $\rm I$ and $\rm II$, respectively. 
The accelerator coils are aligned in Helmholtz-configuration. Their housings are temperature-controlled, water-filled boxes with tunnels, so that the beam can pass without material contact.
An illustration of the setup is shown in Fig.\,\ref{fig:IFM_Bell_New}. 

The measured expectation values, determined from the interference fringes shown in Fig.\,\ref{fig:IFM_Bell_NewGraphes}, are $E(0,\pi/4)=0.603(6)$,  $E(0,-\pi/4)=0.601(6)$, $E(\pi/2,\pi/4)=-0.526(6)$, and $E(\pi/2,-\pi/4)=0.635(7)$. These values lead to $S_{\rm exp}=2.365 \pm 0.013\nleq2$, which violates the Bell-like inequality by 28 standard deviations, clearly confirming the validity of the previous results.

\begin{figure}
\begin{center}
 \includegraphics[width=145mm]{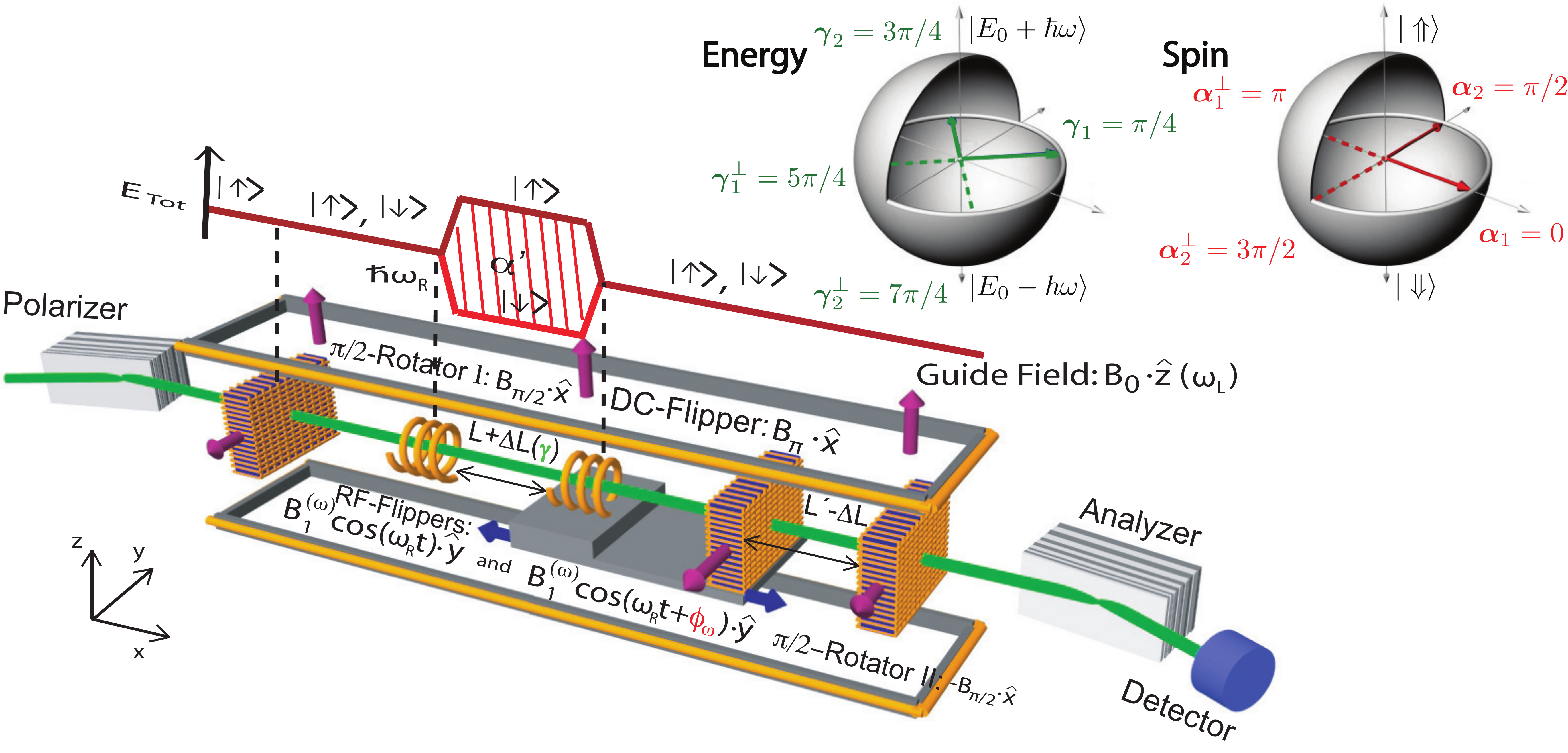}
\caption{Experimental apparatus for observation of spin-energy entanglement \cite{sponarPLA2010}.} 
\label{fig:Bell_Pol_Setup}
\end{center}
\end{figure}

In a polarimeter experiment \cite{sponarPLA2010} (see Fig.\,\ref{fig:Bell_Pol_Setup} for a sketch of the setup), violation of a Bell-like inequality for a spin-energy entangled single-neutron state was observed. The state preparation was the following: The first DC-coil, functioning as a $\pi/2$ spin-rotator prepared a coherent superposition of
the two orthogonal spin-eigenstates $\ket{\!\Uparrow}$ and
$\ket{\!\Downarrow}$. 
This incident state can be denoted as $\ket{\Psi_{S_x}}= 1/\sqrt{2}\, ( \ket{\!\Uparrow}+\ket{\!\Downarrow}) \ket{E_0 }$. The entanglement between spin- and energy-DOF was created exploiting the operation of a subsequent RF-flipper (see Sec.\,\ref{Sec:Time}). Interacting with a time-dependent magnetic field, the total energy of the neutron
is no longer conserved due to absorption and emission of photons of energy $\hbar\omega$, depending on the spin state. 
The RF-flipper was operating at the frequency $\omega/2\pi=32$\,kHz and, accordingly, the guide field was
tuned to $B_0\approx 1.1$\,mT. 
The entangled state vector can be represented as a Bell state
$\ket{\Psi_{\rm{Bell}} } = 1/\sqrt{2}( \ket{E_0+\hbar\omega} \ket{\!\Uparrow}+ \ket{E_0-\hbar\omega}
\ket{\!\Downarrow})$.
The directions for the Bell-measurement were set by adjusting the position of the second RF-flipper and by tuning the phase of its oscillating field for energy- and spin-subspace, respectively. Taking the second
RF-flipper (energy recombination) and the DC-flipper into account, this operation yielded the final state
$\ket{\Psi_{\rm{fin}} }=1/\sqrt{2}( e ^{-i\phi_\omega}\ket{\!\Uparrow}+ e ^{i\omega T}e^{i\phi_\omega}
\ket{\!\Downarrow}) \ket{E_0 }$.
Here, $\omega T=\gamma$ is the phase acquired in energy subspace, where $T$ is the propagation time at the distance $L+\Delta L$
between the two RF-flippers. $\phi_\omega$ is the tunable phase of the oscillating field of the second RF-flipper.

\begin{figure}
\begin{center}
 \includegraphics[width=100mm]{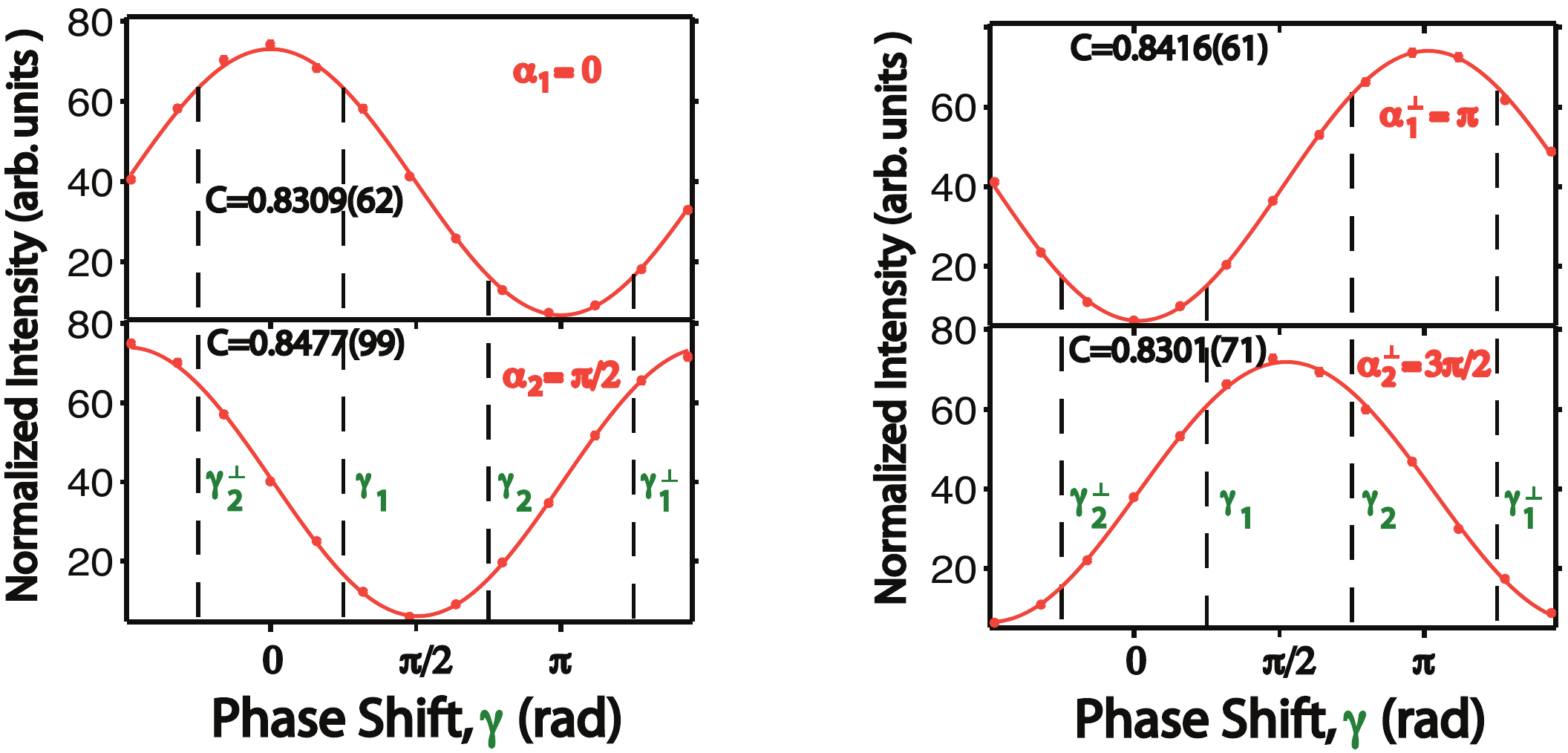}
\caption{Interference oscillations caused by variation of  $\gamma$. Dashed lines mark the $\gamma$-values for which a maximum violation of the CHSH-inequality is expected \cite{sponarPLA2010}.}\label{fig:Bell_Pol_Res}
\end{center}
\end{figure}

Intensity oscillations -- observed when the position of the translation stage (second RF-flipper) is varied ($\gamma$-scans) -- are plotted in Fig.\,\,\ref{fig:Bell_Pol_Res} for different settings of $\alpha$. 
The $\gamma$-scan for $\alpha_1=0$ was used to determine the positions of the translation stage corresponding to the values $\gamma_1=\pi/4, \gamma_2=-\pi/4$ $ (\gamma_1^\bot=5\pi/4$, $\gamma_2^\bot=3\pi/4$) which were, together with the spin phase
settings $\alpha_1=0$, $\alpha_2=\pi/2$ ($\alpha_1^\bot=\pi$, $\alpha_2^\bot=3\pi/2$), required for determining the $S$-value. A Bloch sphere description of these measurement directions is given in Fig.\,\ref{fig:Bell_Pol_Setup}. The final value
$S_{\rm exp}=2.333 \pm 0.002\nleq2$ was determined, which is notably above the value of
2, predicted by NCHVTs.

\subsubsection{Kochen-Specker Phenomena}

In our experimental realization, following the proposal in \cite{cabelloPRL2008}, the proof is based on the six observables $\sigma_x^{\rm s}$, $\sigma_x^{\rm p}$,  $\sigma_y^{\rm s}$, $\sigma_y^{\rm p}$, $\sigma_x^{\rm s}\sigma_y^{\rm p}$, and $\sigma_y^{\rm s}\sigma_x^{\rm p}$, (where s and p are abbreviations  for spin and path, respectively) and the following five QM-predictions for the maximally entangled state $\ket{\Psi}=1/\sqrt{2} \big(\ket{\!\Downarrow}\ket{\textrm{I}}
-\ket{\!\Uparrow}\ket{\textrm{II}}\big)$:

\begin{subequations}\label{eq:KS}
\begin{align}
 \sigma_x^{\rm s}\cdot\sigma_x^{\rm p}\ket{\Psi}=-\ket{\Psi}\\
\sigma_y^{\rm s}\cdot\sigma_y^{\rm p}\ket{\Psi}=-\ket{\Psi} \\
\sigma_x^{\rm s}\sigma_y^{\rm p}\cdot \sigma_x^{\rm s}\cdot\sigma_y^{\rm p}\ket{\Psi}=+\ket{\Psi} \\
\sigma_y^{\rm s}\sigma_x^{\rm p}\cdot \sigma_y^{\rm s}\cdot\sigma_x^{\rm p}\ket{\Psi}=+\ket{\Psi}\\
\sigma_x^{\rm s}\sigma_y^{\rm p}\cdot\sigma_y^{\rm s}\sigma_x^{\rm p}\ket{\Psi}=-\ket{\Psi}
\end{align}
\end{subequations}
In order to reproduce the predictions of QM within the framework of NCHVTs, predefined results have to be assigned to each of the six observables. Attempting to do so immediately leads to a contradiction to Eqs.\,(\ref{eq:KS}). An experimentally testable inequality can be derived from the linear combination of the five expectation values while taking into account that Eq.\,(\ref{eq:KS}c) and  Eq.\,(\ref{eq:KS}d) hold for any NCHVT due to their state independence (see \cite{bartosikPRL2009} for details). Thus, any NCHVT must satisfy the following reduced inequality:
\begin{equation}\label{eq:KSred}
\langle \sigma_x^{\rm s}\cdot\sigma_x^{\rm p}\rangle-\langle \sigma_y^{\rm s}\cdot\sigma_y^{\rm p}\rangle-\langle\sigma_x^{\rm s}\sigma_y^{\rm p}\cdot\sigma_y^{\rm s}\sigma_x^{\rm p}\rangle\leq 1,
 \end{equation}
whereas QM predicts a value of 3. Hence, a violation of inequality Eq.\,(\ref{eq:KSred}) reveals quantum-contextuality.

\begin{figure}
\begin{center}
 \includegraphics[width=135mm]{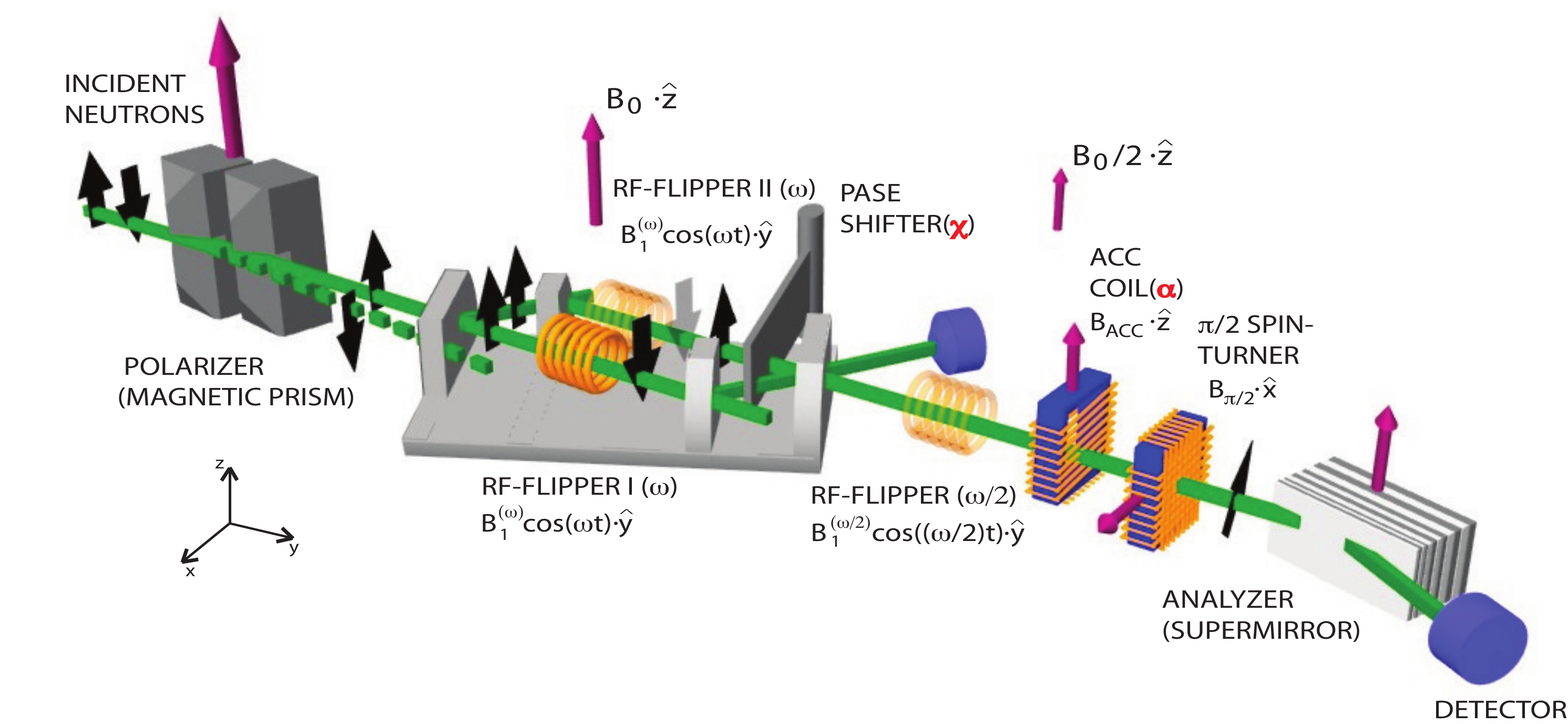}
\caption{Experimental apparatus for studying the KS-theorem following the proposal in \cite{cabelloPRL2008,bartosikPRL2009}.} 
\label{fig:KochenSpecker_Pol_Setup}
\end{center}
\end{figure}

The first and the second term in inequality Eq.\,(\ref{eq:KSred}) were measured in the usual manner using the setup depicted in Fig.\,\ref{fig:KochenSpecker_Pol_Setup}. For the path observable, the phase shifter was adjusted to $\chi=0$ and $\pi$ in order to measure $\sigma_x^{\rm p}$  ($\chi=\pi/2,3\pi/2$ for $\sigma_y^{\rm p}$, second term). The spin analysis in the $xy$-plane was accomplished by the combination of the Larmor accelerator, inducing Larmor-phases $\alpha=0,\pi$ ($\alpha=\pi/2,3\pi/2$ for $\sigma_y^{\rm s}$, second term), together with the $\pi/2$ spin-turner and the supermirror.

The third term in Eq.\,(\ref{eq:KSred}) required a simultaneous measurement of $\sigma_x^{\rm s}\sigma_y^{\rm p}$ and $\sigma_y^{\rm s}\sigma_x^{\rm p}$. This was achieved via a Bell-state discrimination: The two operators have the four common Bell-like eigenstates $\ket{\varphi_\pm}=1/\sqrt2(\ket{\!\Downarrow}\ket{\rm I}\pm i \ket{\!\Uparrow}\ket{\rm II})$ and $\ket{\phi}_\pm=1/\sqrt2(\ket{\!\Uparrow}\ket{\rm I}\pm i \ket{\!\Downarrow}\ket{\rm II})$. Consequently, the corresponding eigenvalue equations are $\sigma_x^{\rm s}\sigma_y^{\rm p}\ket{\varphi_\pm}=\pm\ket{\varphi_\pm}$, $\sigma_x^{\rm s}\sigma_y^{\rm p}\ket{\phi_\pm}=\pm\ket{\phi_\pm}$,  $\sigma_y^{\rm s}\sigma_x^{\rm p}\ket{\varphi_\pm}=\mp\ket{\varphi_\pm}$, and  $\sigma_y^{\rm s}\sigma_x^{\rm p}\ket{\phi_\pm}=\pm\ket{\phi_\pm}$. Hence, the outcome $-1$ and $+1$ for the product measurement of $\sigma_x^{\rm s}\sigma_y^{\rm p}\cdot\sigma_y^{\rm s}\sigma_x^{\rm p}$ are obtained for  $\ket{\varphi_\pm}$ and $\ket{\phi_\pm}$, respectively. In the setup this was realized by tuning on the second RF flipper in path II of the IFM, thereby transforming the state $\ket{\Psi}$ to $1/\sqrt{2} \big(\ket{\!\Downarrow}\ket{\textrm{I}}-\ket{\!\Downarrow}\ket{\textrm{II}}\big)$. Then, the states $\ket{\varphi_\pm}$ were found for phase-shifter settings $\chi=\pm\pi/2$ if the DC spin-turner was adjusted to induce a $\pi$-flip. $\ket{\phi_\pm}$ were obtained at the same phase-shifter position with the DC spin-turner switched off. The final value of $2.291\pm0.008\nleq1$, obtained from Eq.\,(\ref{eq:KSred}), is fully in favour of QM and clearly confirms the conflict with NCHVTs.

\subsubsection{Falsification of Leggett's model}

As already discussed in Section \ref{sec:Bell}, Bell proved in his celebrated theorem \cite{bellPhysics1964} that all HVTs which are based on the joint assumption of locality and realism conflict with certain predictions of QM. Taking this one step further, the question arises whether it is realism or locality that is responsible for this particular behaviour. By this means, Leggett proposed a class of realistic theories which abandons reliance on locality in 2003 \cite{leggettFOP2003}.
 
In a first experimental demonstration using entangled photons \cite{groeblacherNature2007}, rotational symmetry of
the correlation functions in each measurement plane was assumed since the
original inequality requires infinitely many measurement settings.
In a subsequent experiment \cite{paterekPRL2007}, this assumption was no longer needed. 
A different approach to applying a finite number of measurement
settings was accomplished in \cite{branciardPRL2007}. However, until 2012 Leggett-models had been examined experimentally only with photons.

In this Section, an experiment with neutrons analogous to a test of Leggett's non-local realistic model for
entangled pairs of particles is described \cite{hasegawaNJP2012}. Here, non-local correlations are replaced by correlations
between commuting (compatible) observables to study a contextual realistic model.

\begin{figure}
  \includegraphics[height=.173\textheight]{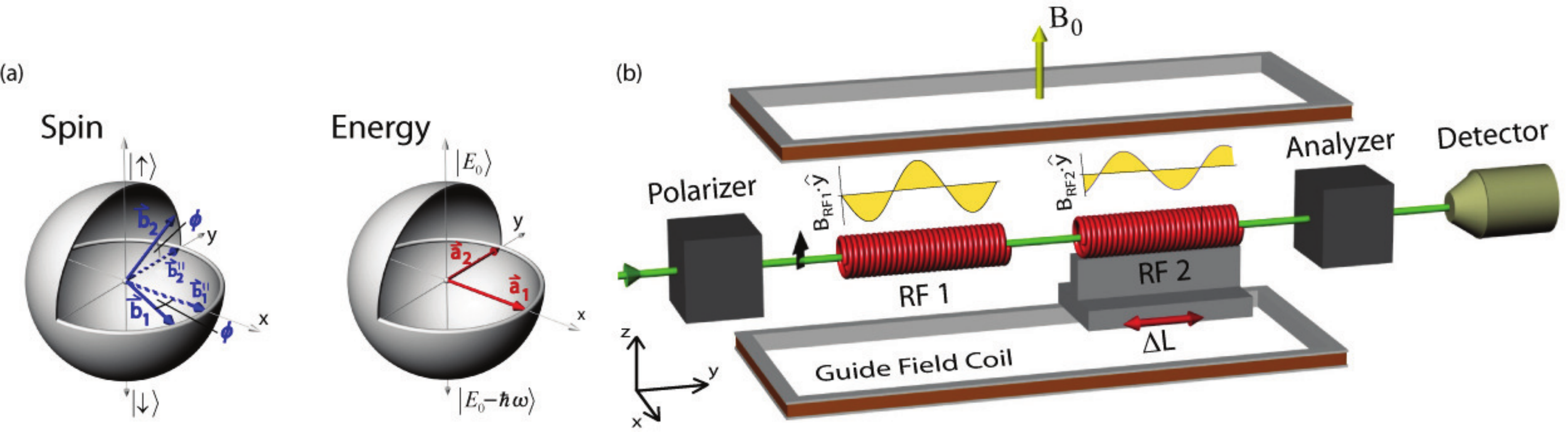}
  \caption{(a) Bloch sphere description of the spin- and energy-observables for falsification of Leggett's model. Measurement direction $\vec b_2$ lies outside
the equatorial plane. (b) Neutron polarimeter setup consisting of
two RF spin-rotator coils which were used for state preparation and
adjustment of the measurement direction for spin and energy \cite{hasegawaNJP2012}.} 
\label{fig:LeggettSetup2}
\end{figure}

For a polarimetric test, the criteria of the first experimental study
by Gr\"oblacher et al. \cite{groeblacherNature2007} were applied, using the following
assumptions for two commuting observables $A$ and $B$ for
two-dimensional quantum systems: 
\emph{(i)} All values of measurements are
predetermined (realism).
\emph{(ii)} States are a statistical mixture of
subensembles having definite polarization. 
\emph{(iii)} The expectation
values taken for each subensemble obey cosine dependence. 
While
assumption \emph{(i)} and \emph{(ii)} are common to experimental tests of
NCTs, assumption \emph{(iii)} is the peculiarity of this
model. 
Here, the outcome of $B[A]$ depends on the measurement
settings of $A[B]$. Assuming full rotational symmetry, an
inequality similar to the one in \cite{groeblacherNature2007} can be applied to our
test of a contextual model. 
The corresponding Leggett-like inequality is given by
\begin{equation}\label{eq:Legg}
S_{\rm Legg}\equiv \bigl \vert E_1\big(\vec
a_1;\phi\big)+E_1\big(\vec a_1;0\big) \bigr \vert + \bigl \vert
E_2\big(\vec a_2;\phi\big)+E_2\big(\vec a_2;0\big) \bigr \vert \leq
4-\frac{4}{\pi}\bigl \vert \sin\frac{\phi}{2}\bigr\vert.
\end{equation}
Here, $E_j\big(\vec a_j;\phi\big)$ with $j = 1, 2$, denote expectation values of joint correlation measurements at settings
$\vec a_j$ and $\vec b_j$ with relative angle $\phi$, and
expectation values $E_j\big(\vec a_j;0\big)$ represent
correlation measurements between $\vec a_j$ and $\vec b_j^\|$. The measurement directions $\vec a_1$, $\vec a_2$ and $\vec b_1$ are
assumed to lie in a single plane, whereas $\vec b_2$ is found in a
perpendicular plane, as depicted in Fig.\,\ref{fig:LeggettSetup2}\,(a). QM predicts
$E_j(\vec a_j;\phi)=-\cos\phi$ for an individual joint expectation value and therefore
$S_{\rm Legg}=2\vert1+\cos\phi\vert$. Thus a maximum violation is expected at
$\phi_{\rm max}\approx 0.1\,\pi$.
 
Our experiment exploited the joint expectation value measurements of two
commuting observables given by $A^{\rm spin}$ for the neutron spin
and $B^{\rm \,energy}$ for the total neutron energy \cite{hasegawaNJP2012}. 
A maximally
entangled Bell-like state $ \vert\Psi_{\rm N}^{\rm
Bell}\rangle=1/\sqrt 2( \vert\!\Uparrow\rangle\vert
E_0\rangle-\vert\!\Downarrow\rangle\vert E_0-\hbar \omega\rangle )$
was prepared by applying a $\pi/2$ spin-rotation within the first
RF-coil [see also Fig.\,\ref{fig:LeggettSetup2}\,(b)]. 
The measurement
directions for the spin-DOF, i.e., polar angle $\alpha$
and azimuthal angle $\beta$, were adjusted by amplitude and phase of
the oscillating magnetic field in RF\,2, respectively. 
The polar-angle setting was $\pi/2$ for the measurement directions in the equatorial plane and $\pi/2-\phi$ for the direction $\vec b_2$ (outside the equatorial plane). 
The relative
phase $\gamma$ between the energy eigenstates was induced by accurate displacement of the position of RF\,2.

\begin{figure}
\begin{center}
  \includegraphics[height=.22\textheight]{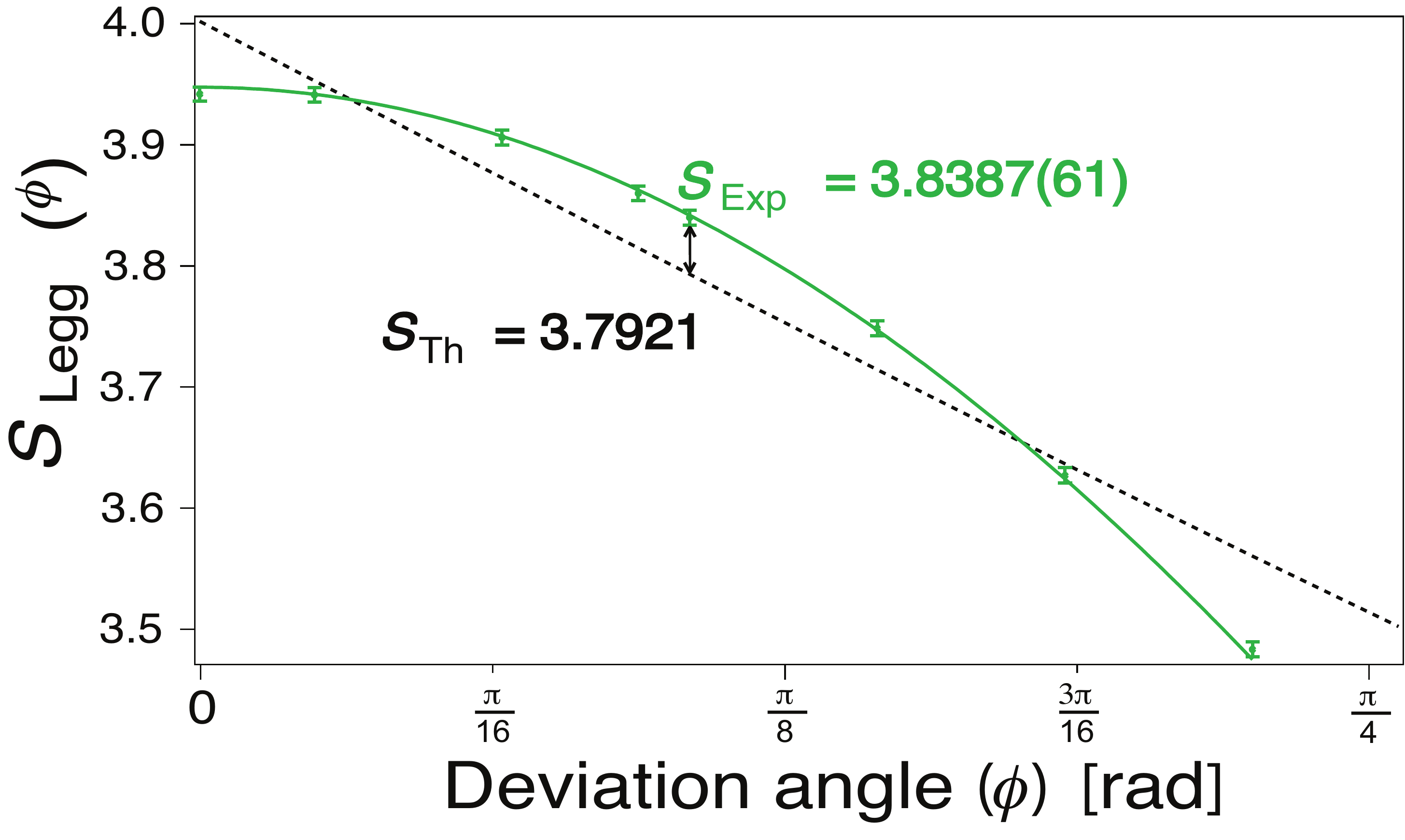}
  \caption{$S$-values as a function of the deviation angle $\phi$ for a contextual realistic model. The $S_{\rm Exp}$-value is clearly larger than the boundary \cite{hasegawaNJP2012}.\label{fig:LeggettSValue}}
\end{center}
\end{figure}

For a test of our Leggett-like contextual realistic model, a
mean contrast of C = 98.5\,\% was achieved. 
The four recorded
expectation values for directions $\vec a_1[\pi/2,0]$, $\vec a_2[\pi/2,\pi/2]$, $\vec b_1[\pi/2,-\phi]$, and $\vec b_2[\pi/2-\phi,\pi/2]$ resulted in a maximal value $S_{\rm Exp}=3.8387(61)$
at $\phi=0.14\,\pi$, which exceeds the  boundary 3.7921 by more than
7.6 standard deviations. A plot of the $S$-value for 8 settings of the deviation angle $\phi$  between 0 and 0.226\,$\pi$ is given in Fig.\,\ref{fig:LeggettSValue}.

\subsection{Tri-partite entanglement: spin-path-energy, spin-energy-momentum entanglement}

Not a statistical violation, but a contradiction between quantum
mechanics and LHVTs was found by Greenberger, Horne and
Zeilinger (GHZ) in 1989 for (at least) tripartite entanglement
\cite{greenbergerProceed1989,greenbergerAJP1990}. 
To date, several experimental realizations using multipartite entanglement have been achieved. 
Among them are experiments with polarized photons \cite{bouwmeesterPRL1999,zhaoNature2004,waltherNature2005,luNatPhys2007}, atoms \cite{leibfriedNature2005} and trapped ions \cite{haeffnerNature2005}.  

The GHZ-argument is independent of
the Bell-approach, thereby demonstrating in a non-statistic manner that QM and local realism are incompatible. 
The GHZ state for a general tripartite-entangled system, is an element of the product Hilbert space
$\mathcal H^{\rm total}=\mathcal H_2^A\otimes\mathcal H_2^B\otimes\mathcal H_2^C$ given, for instance, by the three-qubit state vector
$ \ket{\psi_{\rm{GHZ}}}=1/\sqrt2\left(\ket{\!\Uparrow^A}\ket{\!\Uparrow^B}\ket{\!\Uparrow^C}+\ket{\!\Downarrow^A}\ket{\!\Downarrow^B}\ket{\!\Downarrow^C}\right)$. 
Three measurements along two $y$-directions and one $x$-directions are performed with expectation values denoted as $E(\sigma_x^A,\sigma_y^B,\sigma_y^C)$, $E(\sigma_y^A,\sigma_x^B,\sigma_y^C)$ and $E(\sigma_x^A,\sigma_y^B,\sigma_x^C\big)$ where, for example, $E(\sigma_x^A,\sigma_y^B,\sigma_y^C)= \bra{\psi_{\rm{GHZ}}}\sigma_x^A\,\sigma_y^B\,\sigma_y^C \ket{\psi_{\rm{GHZ}}}$. 
A unique property of this system is that the result of the
$x$-measurement of one system can be predicted with certainty if the
results of the other two measurements -- for example the $y$-measurements of the other systems -- are known. 

From the point of view of a local realistic theory, this behaviour
can be reproduced simply by assigning predefined values to the
individual measurements $s_i^j$, with $i=x,y$ and $j=A,B,C$. 
For example, $s^A_y$ is the predefined
result of the $\sigma_x^A$ measurement, which can only be +1 or -1. 
Whatever combination is chosen, the prediction of QM will only be reproduced for three expectation values and, therefore, a contradictory result for the remaining expectation value emerges. 
However, since perfect correlations (or anticorrelations) cannot be observed in real experiments, an inequality is necessary in order to demonstrate the peculiar properties of the triply-entangled GHZ state. 

The GHZ-argument was analyzed in
detail by Mermin in \cite{merminPRL1990}, where an inequality is
derived for a state of $n$ spin-1/2 particles. 
That inequality is violated by
QM by an amount that increases exponentially with $n$. For a tripartite entangled GHZ state, the limit for a sum of four expectation values manifests as an experimentally testable figure of merit. The sum of expectation values -- usually referred to as $M$ -- is defined as 
\begin{equation}
 M=E(\sigma_x^A,\sigma_x^B,\sigma_x^C)-E(\sigma_x^A,\sigma_y^B,\sigma_y^C)-E(\sigma_y^A,\sigma_x^B,\sigma_y^C)-E(\sigma_y^A,\sigma_y^B,\sigma_x^C).
\end{equation}
NCHVTs set a limit for the maximum possible value, namely $\vert M \vert \leq 2$. In contrast, QM predicts an upper bound of 4. Thus, any measured value of $M$ that is larger than 2 decides in favour of quantum contextuality. 

The Pauli operators can be decomposed as
$
\sigma_x^{(i)}= \hat P^{(i)}(0)- \hat P^{(i)}(\pi)
$
and
$
\sigma_y^{(i)}= \hat P^{(i)}(\pi/2)- \hat P^{(i)}(3\pi/2)
$,
with $\hat P^{(i)} $ being the
projection operators onto an up-down superposition on the equatorial
plane, where the azimuthal angle is defined by a relative phase between the orthogonal eigenstates of the respective sub-system (DOF).
 
\subsubsection{Interferometer setup}

\begin{figure}
\begin{center}
 \includegraphics[width=135mm]{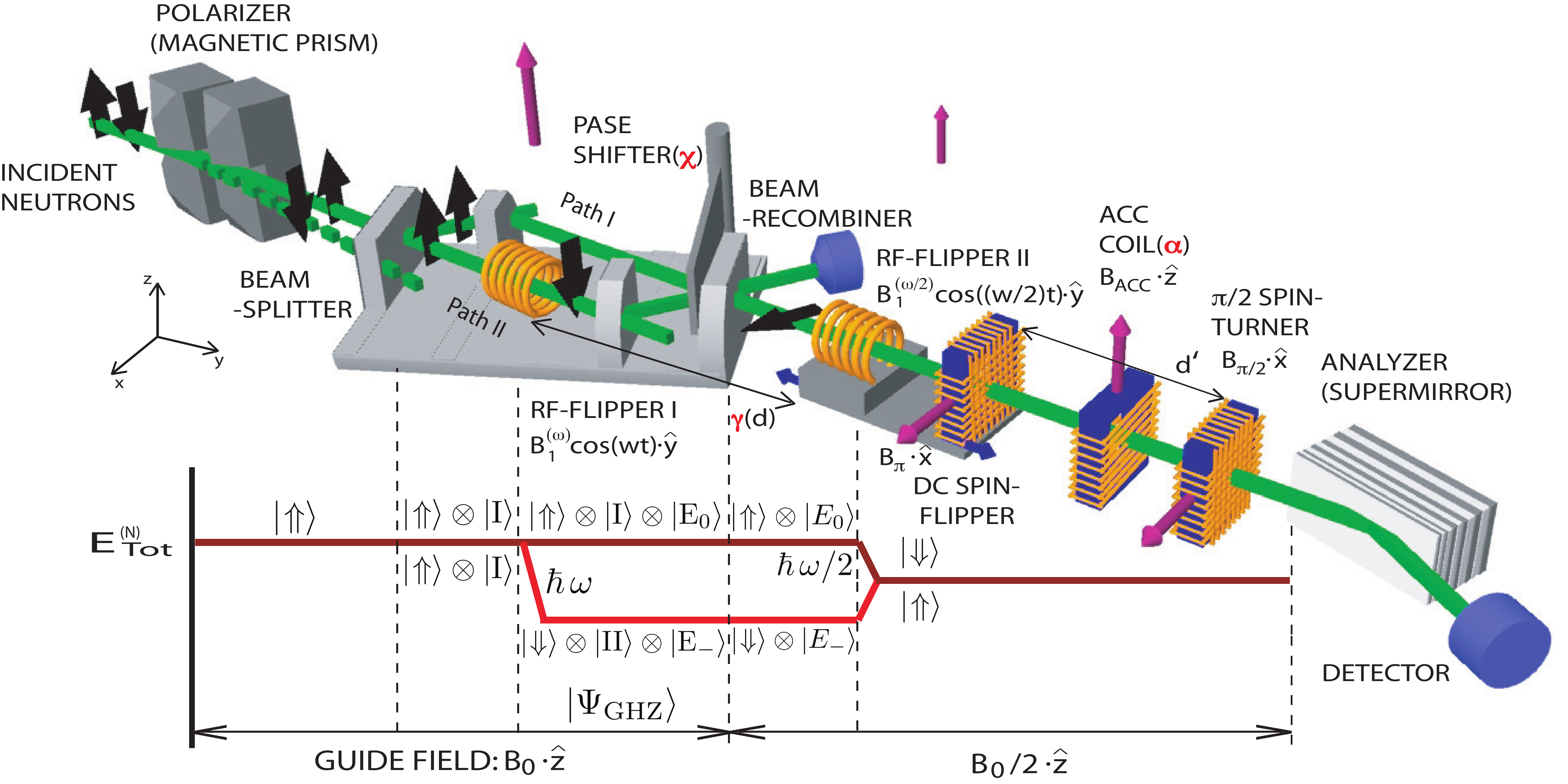}
\caption{Schematic view of the experimental setup for
stationary observation of interference between two RF-fields and its energy diagram \cite{hasegawaPRA2010}. After the first RF flipper the system is prepared in a GHZ state.}\label{fig:GHZ_IFM_Setup}
\end{center}
\end{figure}

As seen in Sec. \ref{sec:Bi}, bi-partite entanglement in an interferometric setup is achieved between spin- and path-DOF. In the polarimetric version, spin and energy are utilized. Combining these techniques allows for preparation of a tri-partite entangled state. Using a single RF flipper in one arm of the IFM, thereby manipulating the total energy, provides realization of triple-entanglement between the path-, spin- and energy-DOF. 

The state vectors of the
oscillating fields in the RF flippers are represented by coherent states
$\ket{\alpha}$, which are eigenstates of creation and annihilation operators $a^\dagger$ and $a$. The
eigenvalues of coherent states are complex numbers, so one can write
$
a\ket{\alpha}=\alpha\ket{\alpha}=\vert\alpha\vert
e^{i\phi}\ket{\alpha}\textrm{ with } \vert\alpha\vert=\sqrt{N}.
$
Hence, one can define a total state
vector including not only the neutron system
$
\ket{\Psi_{\textrm{N}}}= e ^{-(i/\hbar) E_0t}\ket{\psi_{\textrm{\,N}}}$, but also the two quantized oscillating
magnetic fields:
$
\ket{\Psi_\textrm{tot}}=\ket{\alpha_\omega}
\ket{\alpha_{\omega/2}}\ket{\Psi_{\textrm{N}}}.
$
The effect of one RF-field at frequency $\omega$ on $\ket{\psi_{\textrm{\,N}}}=\ket{\!\Uparrow}$ is:
\begin{eqnarray}
\left[\mu B^{(\omega)}_1(\textbf{r})/\sqrt{N_{\omega}}(
a_{\omega}^\dagger\sigma_+ + a_{\omega}\sigma_+)\right]\left(e ^{-(i/\hbar) E_0t}\ket{\alpha_\omega}\ket{\!\Uparrow}\right)\nonumber\\
=
e^{i \phi_\omega}e^{-(i/\hbar) (E_0-\hbar\omega)t}\ket{\alpha_\omega}\ket{\!\Downarrow}=\ket{\alpha_\omega} \ket{\!\Downarrow}\ket{E_0-\hbar\omega},
\end{eqnarray}
describing a spin flip due to emission of a photon of energy $\hbar\omega$ and
a phase factor $e^{i\phi_\omega(t)}$ from the coherent state of the
oscillating field.
Here, $e^{-(i/\hbar)(E_0-\hbar\omega) t}$
is associated with a corresponding
state vector $\ket{E_0-\hbar\omega}$ in analogy to an atomic two-level system, where a certain energy level is
associated with the exited state $\ket{e}$ and $\ket{E_0}$ with the ground state $\ket{g}$.
Thus, a third DOF becomes accessible in neutron interferometry. 
Due to its relatively simple preparation within a magnetic resonance field, the neutron total-energy DOF seems to be an almost ideal for candidate for multi-entanglement preparation.

\begin{figure}
\begin{center}
 \includegraphics[width=152mm]{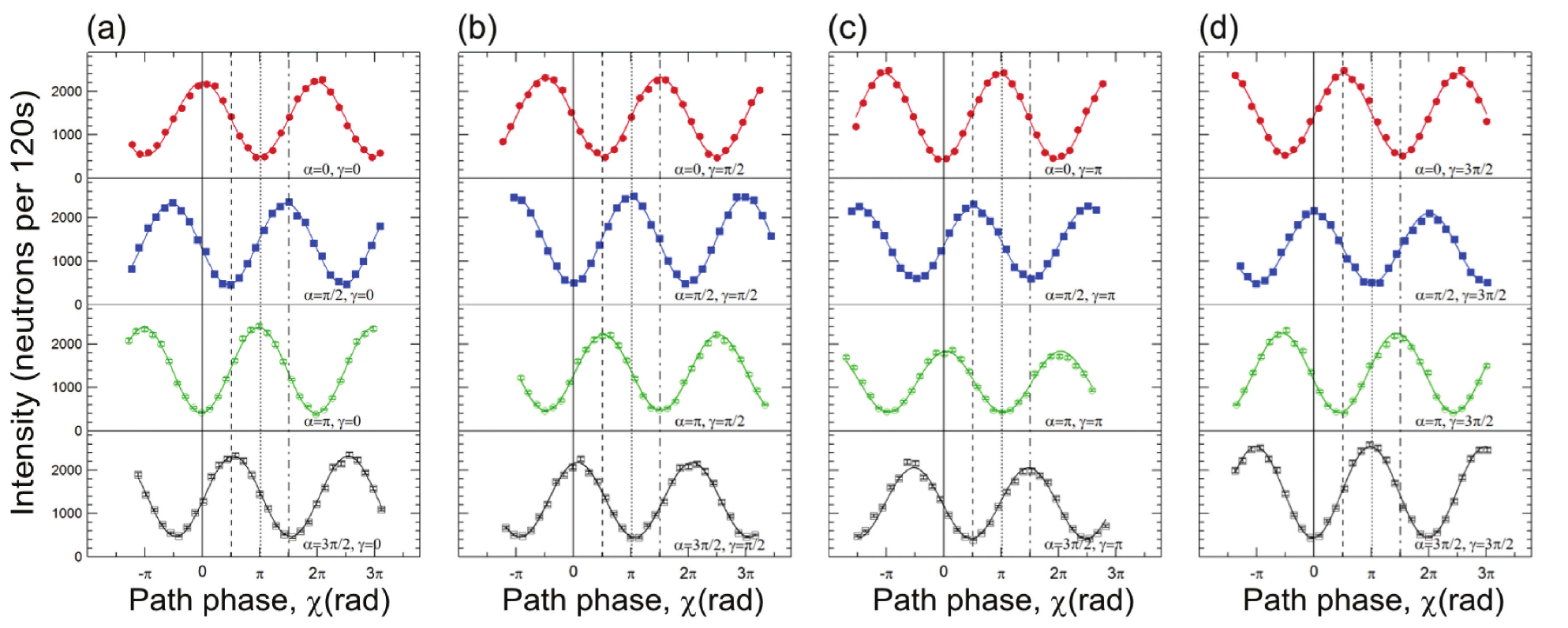}
\caption{Interference patterns obtained by varying the path phase $\chi$ \cite{hasegawaPRA2010}. }\label{fig:GHZ_IFM_Results}
\end{center}
\end{figure}

When operating an RF-flipper inside the IFM, a technical problem arises. The created entangled state is written as 
\begin{equation}
\ket{\Psi(t)}\propto
\ket{\alpha_\omega}\ket{\alpha_{\omega/2}}\frac{1}{\sqrt{2}}\left(\ket{\textrm{I}}\ket{E_0}\ket{\!\Uparrow}
 +e^{i\chi}\ket{\textrm{II}}\ket{E_0-\hbar\omega}e ^{i\phi_\omega
}\ket{\!\Downarrow}\right). 
\end{equation} 
The polarization vector of this state -- $\bra{\Psi_{\rm N}}\vec\sigma\ket{\Psi_{\rm N}}=(\cos (\chi-\omega t),\sin
(\chi-\omega t),0)$ -- is not stationary and measurements are best carried out with phase-locked detection systems \cite{badurekPRL1983}.
However, the energy difference between the orthogonal spin states can
be compensated before detection by inserting a second RF-flipper operating at a frequency $\omega/2$ behind the third
plate of the IFM \cite{sponarPRA2008}. 
This
flipper compensates the energy difference between the two spin components by absorption/emission of photons with energy $E=\hbar\omega/2$. Hence, a
combination of two different guide fields -- providing the requested
static magnetic fields to fulfill the frequency-resonance for $\omega$ (inside the IFM) and $\omega/2$ (after the IFM) -- is
required. An illustration of the setup is depicted in
Fig.\,\ref{fig:GHZ_IFM_Setup}, together with the corresponding energy diagram \cite{hasegawaPRA2010}. 

The neutron part of
the multi-entangled state-vector above, namely
$
\ket{\Psi^{\rm GHZ}_{\textrm{N}}}\propto
\left(\ket{\textrm{I}}\ket{E_0}\ket{\!\Uparrow}+
\ket{\textrm{II}}\ket{E_0-\hbar\omega}\ket{\!\Downarrow}
\right),
$
represents a spin-path-energy entangled state of GHZ-type.  
The respective phases are denoted as $\alpha$, $\chi$ and $\gamma$, respectively. The first RF-flipper induces the energy difference $\hbar\omega$, which is balanced by the second RF-flipper by choosing a frequency of $\omega/2$, resulting in the zero-field phase-difference $\gamma=\omega T$ (as described in Sec.\,\ref{sec:polarimeter}). Here, $T$ is the propagation time between the two RF-flippers at distance $d$. Displacement of the second RF-flipper is the crucial point in this experiment, since by increasing the distance between the RF-flippers not only the zero field phase $\gamma=\omega T$ is changed, but also the Larmor-precession angle within the static guide field $B_0$, which induces an additional undesired Larmor-phase contribution $\alpha'=\omega_{\rm{L(B_0)}}T$. However, the aim is to address zero-field precession independently from Larmor-precession. Compensation of undesired Larmor-phases is achieved by an auxiliary DC-flipper which is mounted
on the same translation stage as the second RF-flipper (see Fig.\,\ref{fig:GHZ_IFM_Setup}). Thus, in contrast to the zero-field phase, the Larmor-precession angle remains constant while $\gamma$ is varied. 

The phases $\alpha$ and $\gamma$ were tuned to 0, $\pi/2$, $\pi$ and $3\pi/2$ in order to accomplish projective measurements associated to $\hat P^j(0)$, $\hat P^j(\pi/2)$, $\hat P^j(\pi)$ and $\hat P^j(3\pi/2)$, with $j=$ spin, path and energy. The results are shown in Fig.\,\ref{fig:GHZ_IFM_Results}. The dashed lines denote values required for the determination of $M$. The average contrast of the oscillations were just below 70\,\%, which was clearly above the threshold visibility of 50\,\%, required for a violation of the Mermin-like inequality.
Measured intensity oscillations were fitted to sinusoidal
curves by applying a least squares method. The four
expectation values were extracted from the fit curves. In total, four sets of thirty-two oscillations were measured to reduce statistical errors. A final value of 
$M = 2.558 \pm 0.004\nleq2$,
was observed, exhibiting a clear violation of the non-contextual limit.

\subsubsection{Polarimeter setup}

\begin{figure}
\begin{center}
\includegraphics[width=153mm]{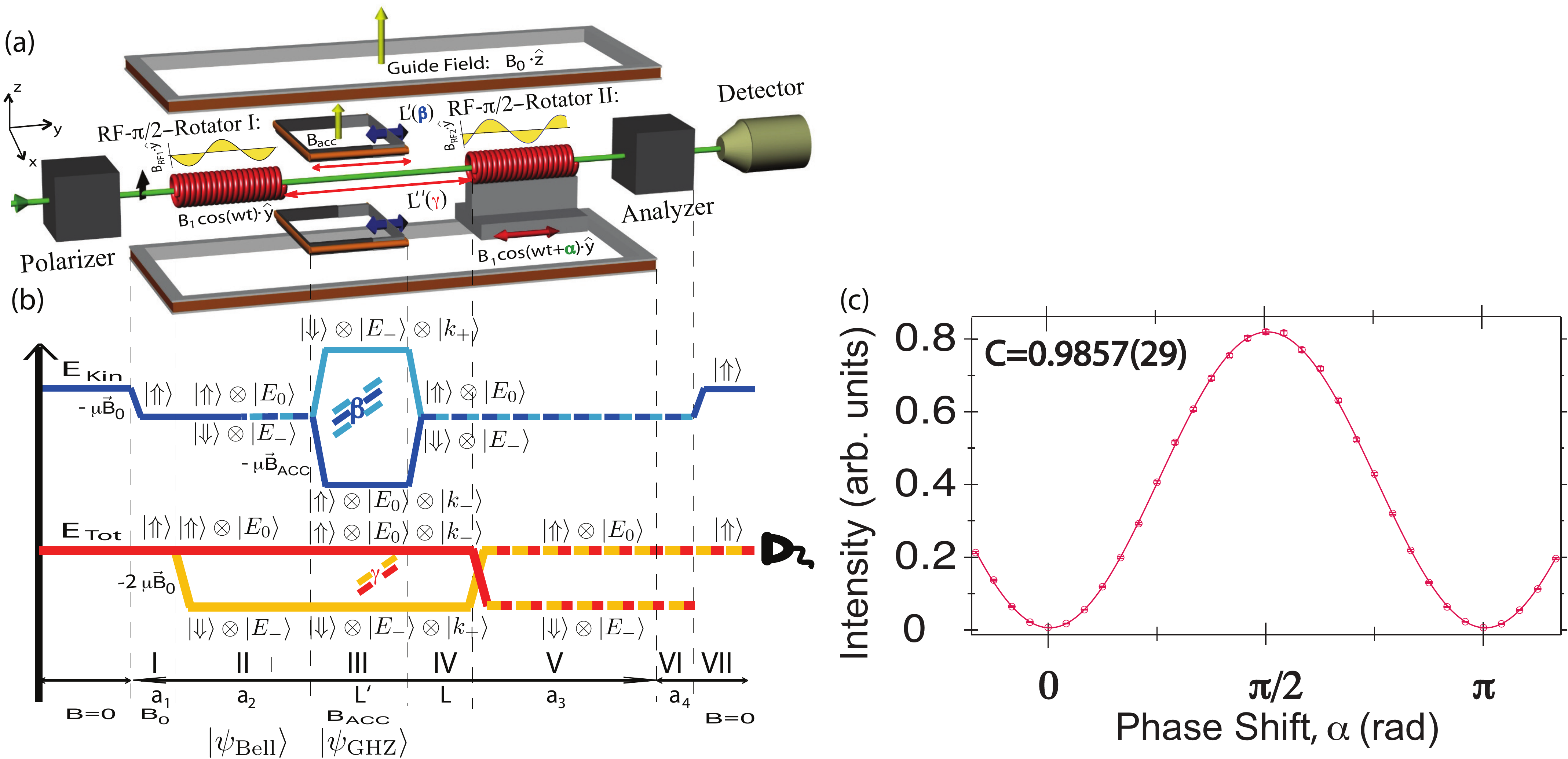}%
\caption{(a) Experimental
apparatus for observation of stronger-than-classic correlation
between the spin-, momentum- and energy-DOF. (b) Energy level diagram for momentum and
total energy (c) High-contrast intensity oscillations taken from \cite{sponarNJP2012}.
\label{fig:GHZ_Pol_Setup_Osci}
}
\end{center}
\end{figure}

In our polarimeter experiment \cite{sponarNJP2012}, tripartite entanglement was created between spin-, momentum-, and total-energy-DOF: $\mathcal
H=\mathcal H_{\rm {spin}}\otimes\mathcal
H_{\rm{momentum}}\otimes\mathcal H_{\rm{energy}}$. 
Momentum Hilbert-space is spanned by momentum eigenstates $\ket{ k_+}$ and $\ket{k_-}$ prepared by Zeeman splitting in a static magnetic field
$B_{\rm{acc}}$.

In the experiment, the spin-energy entanglement was achieved by a single RF-$\pi/2$
spin-rotator, as seen in Fig.\,\ref{fig:GHZ_Pol_Setup_Osci}.
The oscillating field was adjusted such that it
induces a spin flip with a probability of $1/2$ (region II in Fig.\,\ref{fig:GHZ_Pol_Setup_Osci}). Therefore, only the flipped spin component was affected by
the energy manipulation. It yielded an entangled state vector which
is represented as a Bell-like state
$\ket{\psi}
=1/\sqrt 2\left( \ket{E_0}
\ket{\!\Uparrow}+\ket{E_-}\ket{\!\Downarrow}\right)$, 
where $E_-=E_0-\hbar\omega$.
$E_0$ and $E_-$ are considered to be a two-state system with its state vectors spanning the Hilbert Space $\mathcal H_{\rm energy}$. 
The total system is now described by an entangled
neutron state of GHZ-type given by
$
\ket{\psi_{\rm{GHZ}}}=\frac{1}{\sqrt{2}}\left(
\ket{\!\Uparrow}\ket{k_-}\ket{E_0}+
\ket{\!\Downarrow}\ket{k_+}\ket{E_0-\hbar\omega}\right).
$
\begin{figure} 
\begin{center}
\scalebox{0.22}{\includegraphics{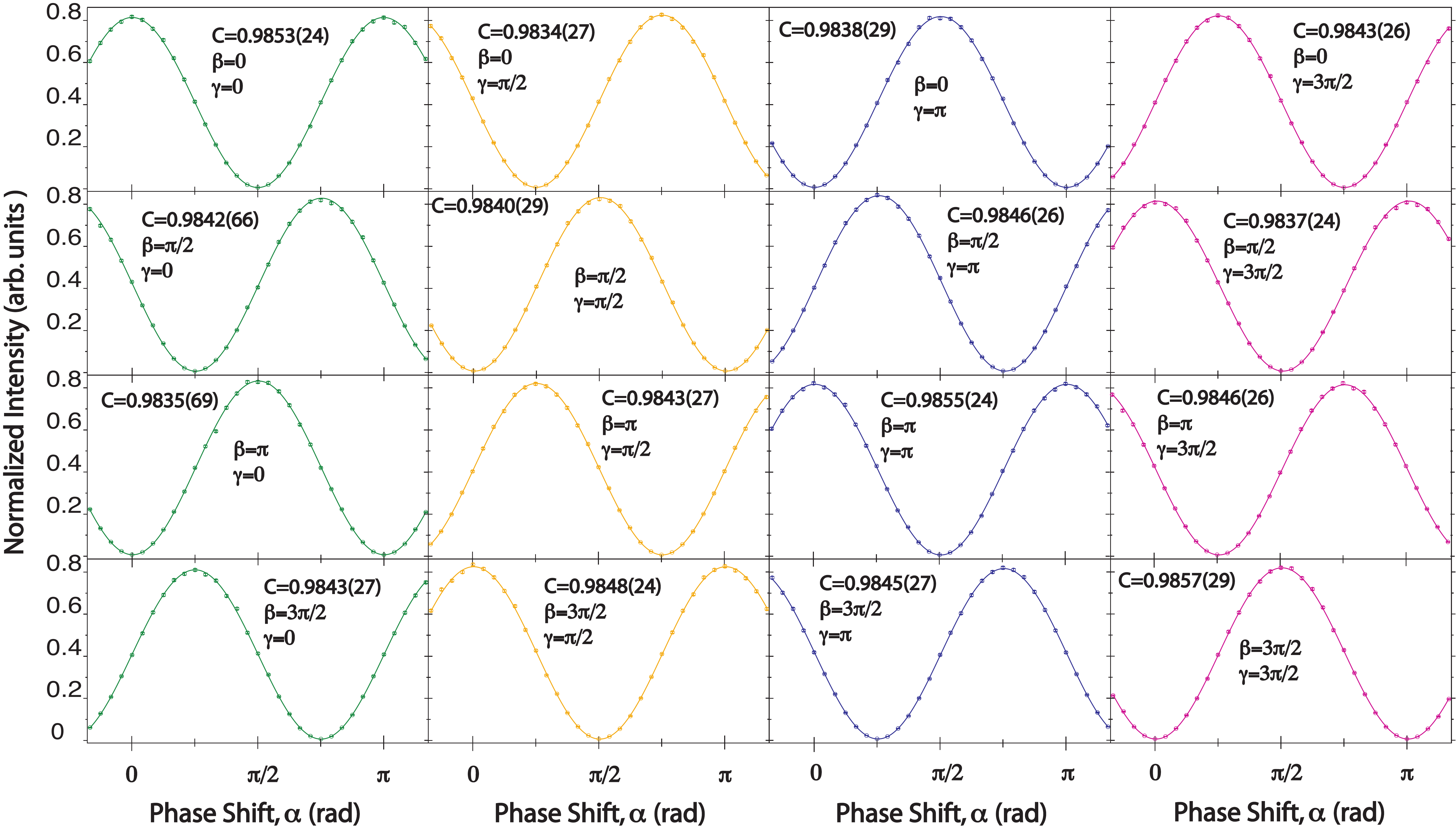}} \caption {High-contrast intensity
oscillation measured by varying the spin phase $\alpha$ for different
setting of the momentum phase $\beta$ and energy phase $\gamma$ \cite{sponarNJP2012}.}
\label{fig:GHZ_Pol_Results} 
\end{center}
\end{figure}

In the polarimetric setup in Fig.\,\ref{fig:GHZ_Pol_Setup_Osci}, the measurement apparatus consisted of the second
RF spin-rotator and the accelerator Field $B_{\rm{acc}}$ combined with the supermirror analyzer. The spin-phase measurement direction $\alpha$ was tuned by adjusting the phase of the
oscillating field within the second RF spin-rotator. 
As all spin states
to be analyzed lie in the $xy$-plane,
another $\pi/2$ spinor-rotation was performed. 
The measurement direction of the momentum phase $\beta$ was tuned by the
propagation time within the accelerator coil. The acquired phase in
momentum space was given by $\beta=\int B_{\rm acc}\,ds$. 
For convenience, the strength of the magnetic field was varied instead of
the length, in
practice. 

The measurement direction of the energy phase $\gamma$ was tuned by the
position of the second RF-$\pi/2$ spin-rotator, mounted on a
motorized translation stage, thereby varying the distance between
the two RF-$\pi/2$ spin-rotators. 
A change of the position of RF2 by
$\triangle L$, induced an undesired additional relative Larmor-phase
between the two spin eigenstates due to precession within
the guide field.
So, to achieve tuning of the energy phase $\gamma$ only, the phase of the oscillating magnetic field in
RF1 was adjusted to compensate for the Larmor phase.
With the momentum phase $\beta$ and the energy phase $\gamma$
tuned to the values 0, $\pi/2$, $\pi$ and $3\pi/2$, sixteen spin-phase $\alpha$ scans were carried out for a determination of $M$ yielding a value 
$M_{\rm exp} = 3.936 \pm 0.002\nleq2$. 
The results are shown in Fig.\,\ref{fig:GHZ_Pol_Results}.
The deviation of less than 2\,\% from the theoretical maximum $M_{\rm th}=4$ is worth noting here.

\section{Some recent neutron experiments on geometric phases}\label{sec:TopologicalAndBerryPhases}

The experiments on topological and geometric phases as described in Secs.\,\ref{subsec:AvarietyOfTopAndGeoPh} and \ref{subsec:GeoPhase} provided beautiful confirmations of the theoretical predictions made concerning topological and geometric phases. 
There is ongoing development regarding this exciting topic that also leads to important neutron experiments. Such experiments are discussed in the following Sections.

\subsection{Geometric phase in coupled interferometer loops}\label{sec:CoupledIFMLoops}

\begin{figure}
\begin{center} 
\scalebox{0.55}
{\includegraphics {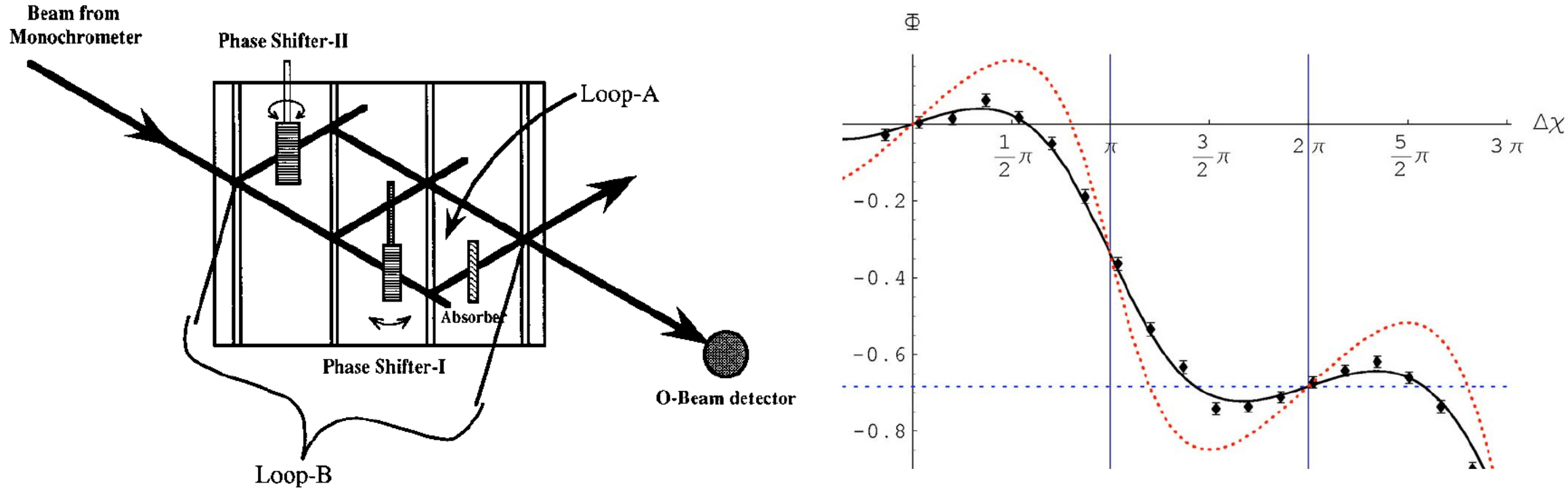}}
\caption{Left: Experimental setup used to measure the geometric phase accumulated by an evolution on the Bloch-sphere of the two orthogonal path states in loop A \cite{hasegawaPRA1996}. Right: Measured phase (data points) for non-cyclic path-state evolution compared with theory (solid line). The dotted line is a theory curve assuming perfect visibility \cite{filippPRA2005}.}
\label{fig:CoupledIFMLoops} 
\end{center}
\end{figure}

Since also the evolution of non-normalized states appearing from non-unitary evolutions can be depicted on projective Hilbert space \cite{mukundaAOP1993}, it is possible to assign a geometric phase also to the orthogonal states in two IFM paths as was pointed out and confirmed experimentally in \cite{hasegawaPRA1996}. Here, the incident neutron beam was split to create the IFM-loop B [see Fig.\,\ref{fig:CoupledIFMLoops}\,(left)]. A two-level system, made-up by two orthogonal path-states, was established in the lower IFM path by splitting the corresponding beam once more to form loop A. A Pancharatnam-phase was induced by realizing a suitable evolution on the Bloch-sphere associated to that two-level system (cf. Fig.\,\ref{fig:PathBlochSphere}). To accomplish that evolution, an arrangement of phase-shifters and absorbers was used to achieve a certain path on that Bloch-sphere. In particular, the superposition of path-states right after the beamsplitter for loop A is represented by a point on the equator of that Bloch-sphere. The azimuthal angle of the evolution path was set by phase-shifter-I, while the polar angle was set by the absorber as shown in Fig.\,\ref{fig:CoupledIFMLoops}\,(left). Phase shifter-II was needed to observe the interference (intensity oscillations) between evolved state and reference state. These oscillations were shifted by different Pancharatnam phases induced in loop A. 
 
The analogy is best understood by comparing Figs.\,\ref{fig:CoupledIFMLoops}\,(left) to \ref{fig:Wagh1998}\,(left). In contrast to the former, the (spin) two-level-system in Fig.\,\ref{fig:Wagh1998} was prepared before the first beamsplitter by polarizer and spin rotation coil. Its evolution on the spin Bloch-sphere was accomplished by the so-called $\hat z$-field (magnetic field) in path 1 of the IFM and the resulting state was compared to the reference state. In that case, only two IFM paths were needed, because one had two orthogonal spin-states to undergo the evolution in IFM-path 1 and the reference state in path 2. Apart from technical details, the analogy is valid. However, after serious criticism \cite{waghPRA1999}, an alternative experimental concept was suggested in \cite{sjoeqvistPRA2001} and the discussion was finally settled by repeating the neutron IFM experiment realizing also non-cyclic path state-evolutions and comparing the measurement results to exact calculations of the surface area enclosed by the intended path on the Bloch-sphere \cite{filippPRA2005}, as shown in Figs.\,\ref{fig:CoupledIFMLoops}\,(right).

\subsection{Off-diagonal geometric phase}

\begin{figure}
\begin{center} 
\scalebox{0.4}
{\includegraphics {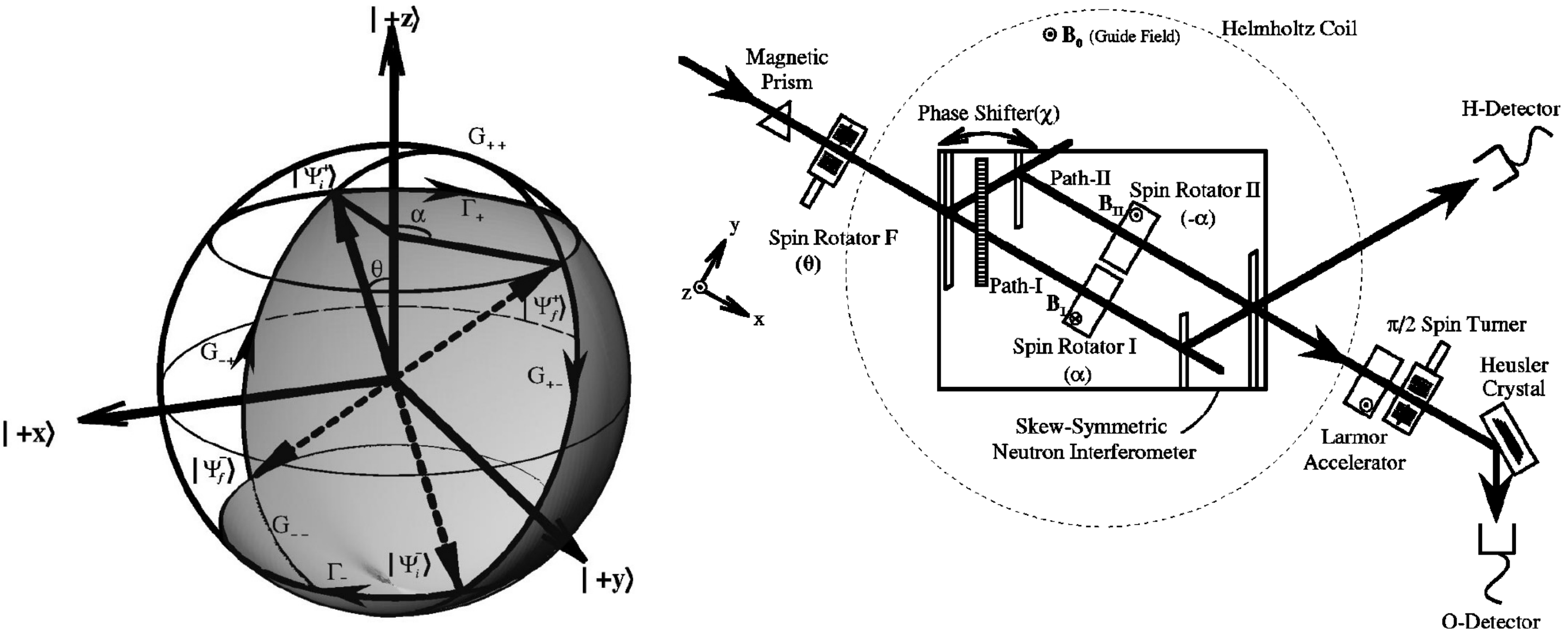}}
\caption{Left: Evolutions of the orthogonal states 
$|\Psi^+_i\rangle$ and $|\Psi^-_i\rangle$ on the Bloch sphere. The shaded area corresponds to an enclosed solid angle of $2\pi$, i.e. the off-diagonal geometric phase of $-\pi$ \cite{hasegawaPRL2001,hasegawaPRA2002}. Right: Experimental setup to measure off-diagonal geometric phase and Pancharatnam phase simultaneously in 0- and H- beams, respectively.}
\label{fig:Hasegawa2002} 
\end{center}
\end{figure}

The Pancharatnam phase (see Sec.\,\ref{susubsec:PanchPhase}) is well-defined except for orthogonal $|\psi\rangle$ and 
$U|\psi\rangle$, for which $\langle \psi|U|\psi\rangle=0$.
In \cite{maniniPRL2000}, Manini and Pistolesi describe a geometric phase of pairs of \emph{different} non-degenerate eigenstates 
$|\psi_i\rangle$ and $|\psi_j\rangle$ ($i\ne j$) of a Hamiltonian that undergo the \emph{same} adiabatic unitary parallel transport, i.e. the complex argument of products like $\langle\psi_m|U|\psi_n\rangle\langle\psi_n|U|\psi_m\rangle$. Such a phase is well-defined also for 
$|\psi_i\rangle \perp U|\psi_i\rangle$. The adiabaticity constraint was later removed in \cite{mukundaPRA2001}.

In Fig.\,\ref{fig:Hasegawa2002}\,(left), such an evolution is depicted on the Bloch-sphere for the mutually orthogonal states $|\Psi^+_i\rangle$ and $|\Psi^-_i\rangle$. Both states evolve according to the unitary operator $U$ along the paths $\Gamma^+$ and $\Gamma^-$ and are projected to the states $|\Psi^+_f\rangle=|\Psi^-_i\rangle$ and  $|\Psi^-_f\rangle=|\Psi^+_i\rangle$ via paths G$_{+-}$ and G$_{-+}$, respectively (Here, $i$ and $f$ refer to initial and final states.). Such a situation was mimicked in the experiment described in Refs.\,\cite{hasegawaPRL2001,hasegawaPRA2002}. In the neutron-IFM setup as sketched in Fig.\,\ref{fig:Hasegawa2002}\,(right), after being prepared by magnetic prisms in the up-state, the spin was rotated by an angle $\theta$ by the spin rotator F to obtain $|\Psi^+\rangle$. 
$U$ and $U^\dagger$ denote spin rotations by the angles $\alpha$ and $-\alpha$ around the $z$-axis, implemented by an equal pair of spin rotators in IFM paths I and II, respectively. Before the 0-detector the spin state was rotated to an appropriate direction by two more magnetic field coils in order to be analyzed by a Heusler-crystal, in particular, the 0-beam spin-state was projected to the $|\Psi^-\rangle$-direction for given $|\Psi^+\rangle$.
The measured intensity is expected to be 
\begin{eqnarray}\label{eq:offDiagObeam} 
I_0\propto |e^{i\chi}|\Psi_{\mbox{\scriptsize{I}}}\rangle
\!+\!|\Psi_{\mbox{\scriptsize{II}}}\rangle|^2\propto 1\!+\!|\langle\Psi^+|U|\Psi^-\rangle\langle\Psi^-|U|\Psi^+\rangle|\cos\big[\chi\!-\!\mbox{arg}\langle\Psi^+|U|\Psi^-\rangle\langle\Psi^-|U|\Psi^+\rangle\big]
\end{eqnarray} 
for the 0-beam, where $|\Psi_{\mbox{\scriptsize{I}}}\rangle=|\Psi^-\rangle\langle\Psi^-|U^\dagger|\Psi^+\rangle$ and $|\Psi_{\mbox{\scriptsize{II}}}\rangle=|\Psi^-\rangle\langle\Psi^-|U|\Psi^+\rangle$ with $|\Psi^+\rangle=\cos\theta/2|\!\Uparrow\rangle+\sin\theta/2|\!\Downarrow\rangle$ and $|\Psi^-\rangle\perp|\Psi^+\rangle$. Eq.\,(\ref{eq:offDiagObeam}) states that the measured intensity oscillations are expected to be shifted by precisely the off-diagonal geometric phase as defined above although, in that experiment $U$ did not act on $|\Psi^+\rangle$ and $|\Psi^-\rangle$ simultaneously. For the H-beam, the projection to $|\Psi^-\rangle$ was omitted and one expects
\begin{eqnarray}
I_H\propto |e^{i\chi}U^\dagger|\Psi^+\rangle+U|\Psi^+\rangle|^2
=2+2|\langle\Psi^+|U^2|\Psi^+\rangle|\cos\big[\chi-\mbox{arg}\langle\Psi^+|U^2|\Psi^+\rangle\big].
\end{eqnarray}

\begin{figure}
\begin{center} 
\scalebox{0.30}
{\includegraphics {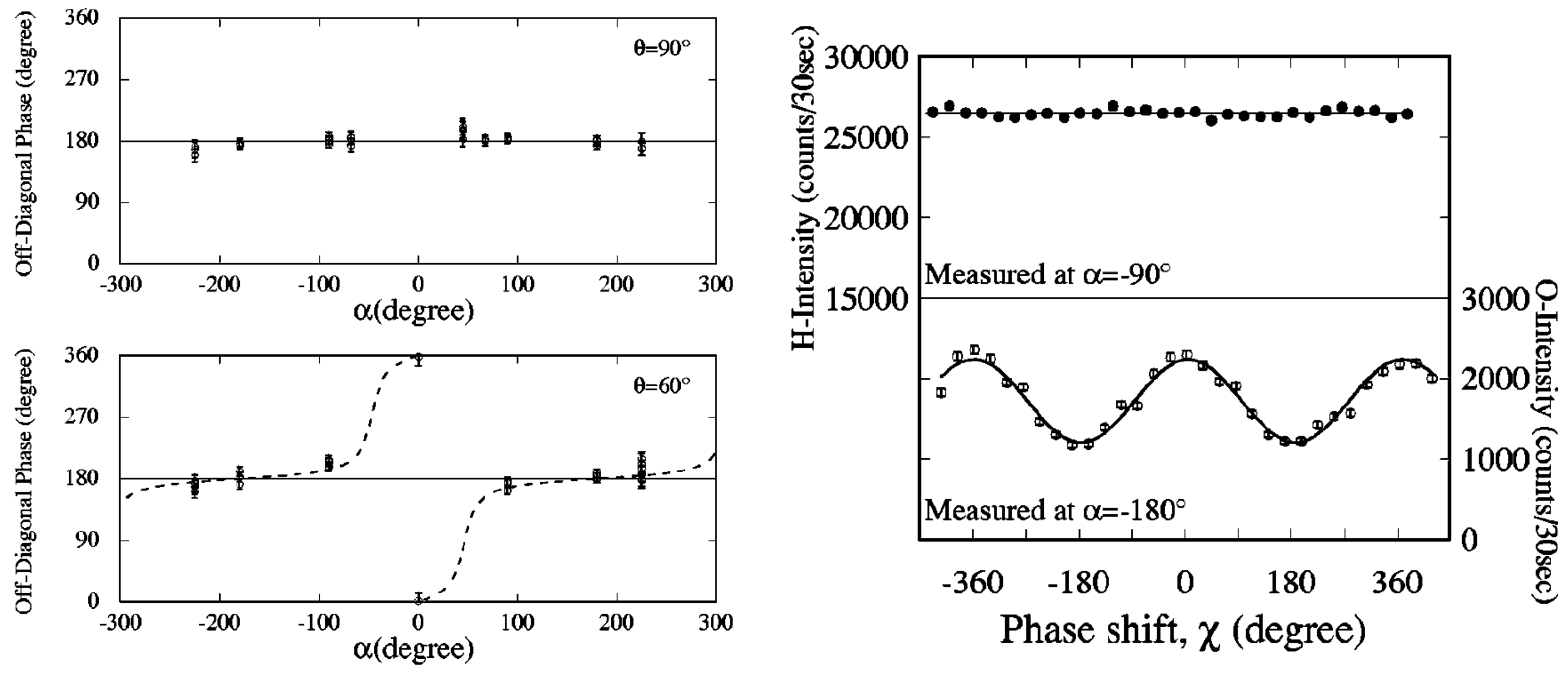}}
\caption{Left: Off-diagonal geometric phases as measured for various $\theta$ and $\alpha$ with imperfect incident polarization taken into account (dashed lines). Right: In case of $\theta=90^\circ$ and $\alpha=\pm 90^\circ$ the non-cyclic Pancharatnam phase is undefined (above), while the off-diagonal geometric phase (below) can still be observed \cite{hasegawaPRL2001,hasegawaPRA2002}.}
\label{fig:Hasegawa2002b} 
\end{center}
\end{figure}

Thus, it turned out that by measuring the H-beam intensity one could observe the diagonal non-cyclic Pancharatnam phase resulting from applying $U$ twice to the state $|\Psi^+\rangle$.
The associated path on the Bloch sphere (Fig.\,\ref{fig:Hasegawa2002}) corresponds to a rotation through $2\alpha$. The measurement results agree well with the prediction if the limited polarization of the incident beam is taken into account, as can be seen in Fig.\,\ref{fig:Hasegawa2002b}\,(left).

Note that the experiment can also be interpreted in terms of the sign change of the spinor for $2\pi$-rotations (see Sec.\,\ref{sec:4pi}) and, interestingly, as demonstration of an aspect of quantum erasure \cite{scullyPRA1982,scullyNature1991}: For $\theta=90^\circ$ and $\alpha=90^\circ$, the visibility of the H-beam disappeared because the spin states in paths I and II were orthogonal for that configuration, which labelled the paths and allows one to gain which-path information. Thereby, interference was destroyed, but could be regained by erasing the which-path information after beam-recombination if the final projection direction was changed, for instance. In \cite{hasegawaPRA2002}, which-path information was not erased but it is rather the degree of labelling that was changed by varying $\alpha$. Note that it is exactly the settings for full path-labelling for which the non-cyclic Pancharatnam phase is undefined [cf. Fig.\,\ref{fig:Hasegawa2002b}\,(right)]. 

\subsection{Geometric phase for mixed states}\label{sec:GeometricPhaseForMixedStates} 

In addition to an early approach by Uhlmann
\cite{uhlmannLettMathPhys1991}, a new concept of phase for mixed 
input states based on an interferometric point of view was developed \cite{sjoeqvistPRL2000}. 
Here, each eigenvector of the initial
density matrix acquires a phase independently. 
The total mixed-state phase was defined as the weighted average of the 
individual phase factors. 
In experiments, the system is always in a mixed state to a certain extent. Therefore, the concept of mixed-state phase is of great significance for certain
experimental situations in which pure-state theories may imply strong 
idealizations. Theoretical predictions have 
been tested by Du
\emph{et al.} \cite{duPRL2003} and Ericsson \emph{et al.}
\cite{ericssonPRL2005} using NMR and single-photon
interferometry, respectively. Interestingly, as pointed out in \cite{sjoeqvistPRL2000}, also the 
neutron-IFM experiment to investigate the $4\pi$ spinor-symmetry (see \cite{rauchPLA1975,wernerPRL1975} and Sec.\,\ref{sec:4pi}) can be interpreted in terms of mixed-state phases.

\begin{figure}
\begin{center} 
\scalebox{0.35}
{\includegraphics {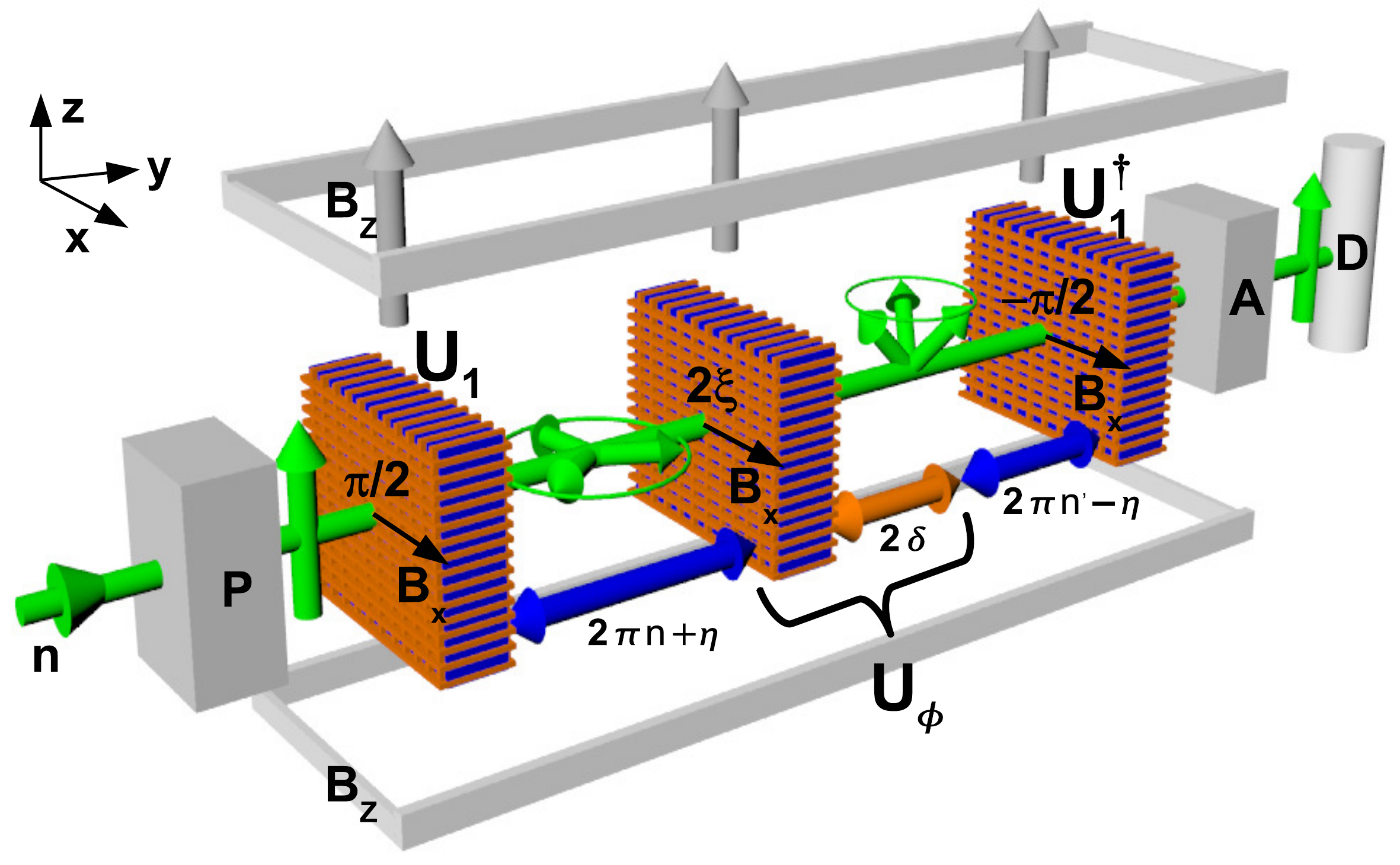}}
\caption{Sketch of neutron polarimeter setup for mixed-state phase measurement \cite{kleppPRL2008}.}
\label{fig:MixedStatePolarimeterSetup} 
\end{center}
\end{figure}

\begin{figure}
\begin{center}
    \scalebox{0.28}{\includegraphics{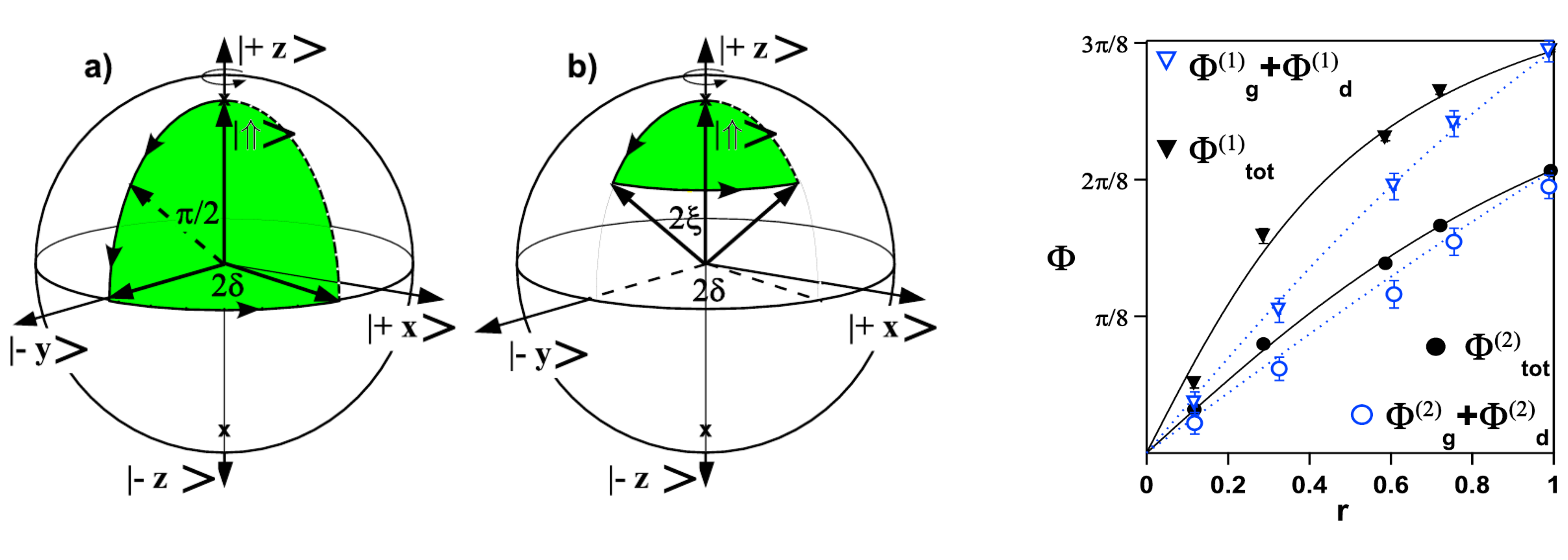}}
    \caption{Left: Transformations U$_\phi$ of $|\!\Uparrow\rangle$ 
				to achieve:
            a) Purely (non-cyclic) geometric phase ($2\xi=\pi/2$).
            b) A combination of dynamical and geometric phase ($0<2\xi<\pi/2$). 
Right: Filled markers are the measured total 
mixed state phase $\Phi_{\mbox{\scriptsize{tot}}}$ 
versus purity $r$.
Open markers show $\Phi_{\mbox{\scriptsize{g}}}^{(1,2)}
+\Phi_{\mbox{\scriptsize{d}}}^{(1,2)}$ 
as calculated 
from data measured in two separate experiments.	
The solid and dotted lines are theory
curves assuming either non-additivity and additivity, 
respectively (see text and Ref.\,\cite{kleppPRL2008}).}
    \label{fig:Klepp2008}
\end{center}
\end{figure}

The experimental arrangement for a related neutron-polarimeter experiment \cite{kleppPRL2008} is sketched in Fig.\,\ref{fig:MixedStatePolarimeterSetup}. 
A neutron beam passed the polarizer P 
preparing the beam in the up-state $|\!\Uparrow\rangle$ with
respect to a magnetic guide field in $z$-direction (B$_z$). 
The fields B$_x$ were chosen such that they carry out particular spin-evolutions. After U$_1$, the state of
the system was a coherent superposition of the two orthogonal spin
eigenstates, i.e. $|\psi_0\rangle=1/\sqrt{2}(|\!\Uparrow\rangle-i|\!\Downarrow\rangle)$. 
The subsequent DC coil and the
following propagation distance within B$_z$ 
defined a spin-evolution 
U$_\phi$ that can induce quantum phases of purely dynamic or geometric origin or combinations of both, depending on the values of $\xi$ and $\delta$ [see Fig.\,\ref{fig:Klepp2008}(left)].
Undergoing U$_\phi$, the two spin eigenstates acquired
opposite total phase
$\pm\delta$.
A third coil (corresponding to U$^{\dagger}_1$) was set to exactly reverse the action of the first one and $\phi$ could then be measured by applying an 
auxiliary dynamical phase shift proportional to $\eta$, projecting the spin state to $|\!\Uparrow\rangle$ and recording the intensity in detector D.
$\eta$ was varied by scanning the position of the second coil 
to yield intensity 
oscillations from which $\phi$ was calculated \cite{waghPLA1995b}.

The theoretical prediction for the mixed state phase is 
$\Phi(r)=\arctan\left(r\tan\delta\right)$ \cite{sjoeqvistPRL2000,larssonPLA2003}.
In neutron polarimetry, variation of $r$ was achieved in two manners: First, U$_\phi$ was applied in two separate experiments, namely using the incident state $|\!\Uparrow\rangle$ in the first and $|\!\Downarrow\rangle$ in the second run. From the resulting two datasets, various sums with different weighting were calculated \cite{kleppPLA2005}, thereby creating \lq{}mixtures\rq{}. 
Second, in addition to the DC current, random noise was applied to the first coil, thereby
changing B$_x$ in time. Although each separate transformation (at a particular time) was
unitary, due to the randomness of the signal we achieved a
non-unitary evolution to yield a mixed state
\cite{bertlmannPRA2006} of defined purity $r$ that corresponded to the noise-signal strength. 
The results confirmed the theoretical expectations ($r$-dependence) for both methods.

Since the mixed state phase is defined as a weighted average of phase
factors rather then a weighted average of phases,
separately measured mixed state phases cannot be
added up to a total phase (cf. also Sec.\,\ref{subsec:AvarietyOfTopAndGeoPh}). 
For example, suppose that two separate experiments are carried out: A pure state is subjected to a unitary transformation 
U$_{\mbox{\scriptsize{g}}}$ in the first and to
U$_{\mbox{\scriptsize{d}}}$ in the second experiment 
obtaining the pure state phases $\phi_{\mbox{\scriptsize{g}}}$ and $\phi_{\mbox{\scriptsize{d}}}$, respectively, with the (subsequently calculated) total phase $\phi_{\mbox{\scriptsize{g}}}+\phi_{\mbox{\scriptsize{d}}}$. 
Alternatively, we can also choose a combination 
of spin-rotation angles
$2\xi$ and $2\delta$ [see Figs.\,\ref{fig:MixedStatePolarimeterSetup} 
and \ref{fig:Klepp2008}\,(left)], leading to a transformation 
U$_{\mbox{\scriptsize{tot}}}$ 
such that we measure the same total pure state phase 
$\phi_{\mbox{\scriptsize{g}}}+\phi_{\mbox{\scriptsize{d}}}$ in one run (clearly, the three evolution paths induced by U$_{\mbox{\scriptsize{g}}}$, U$_{\mbox{\scriptsize{d}}}$ and U$_{\mbox{\scriptsize{tot}}}$ differ from each other).
However, the result of the latter experiment
for the system in a mixed input state would be
$\Phi_{\mbox{\scriptsize{tot}}}(r)
=\arctan{\left[r\tan(\phi_{\mbox{\scriptsize{g}}}
+\phi_{\mbox{\scriptsize{d}}})\right]}$. 
The total phase is then \emph{not} given by 
$\Phi_{\mbox{\scriptsize{g}}}(r)
+\Phi_{\mbox{\scriptsize{d}}}(r)$, with
$\Phi_{\mbox{\scriptsize{g}}}(r)
=\arctan{(r\tan\phi_{\mbox{\scriptsize{g}}})}$ and $\Phi_{\mbox{\scriptsize{d}}}(r)
=\arctan{(r\tan\phi_{\mbox{\scriptsize{d}}})}$.
In Fig.\,\ref{fig:Klepp2008}\,(right) the measured mixed state 
phases $\Phi^{(1,2)}_{\mbox{\scriptsize{tot}}}$ 
and the sum $\Phi_{\mbox{\scriptsize{g}}}^{(1,2)}
+\Phi_{\mbox{\scriptsize{d}}}^{(1,2)}$ for two settings of U$_{\mbox{\scriptsize{tot}}}$ are plotted. 

\subsection{Geometric phase in entangled system}

In this Section, the influence of the geometric phase on a Bell measurement is discussed \cite{bertlmannPRA2004}. The geometric phase in a single-particle system has been studied widely over almost three decades. Nevertheless, its effect on entangled quantum systems has been less investigated. Entanglement forms the basis for quantum communication and quantum information processing. Therefore, studies of entangled systems under evolutions leading to geometric phases are of great importance.

In accordance to the notation used in
\cite{bertlmannPRA2004}, the neutron wavefunction is defined via a tensor product of two Hilbert spaces: one Hilbert
space is spanned by the two possible paths in the interferometer given by $\ket{\textrm{I}}$ and $\ket{\textrm{II}}$; the other one by spin-eigenstates $\ket{\!\Uparrow}$ and $\ket{\!\Downarrow}$ that are defined with respect to the quantization axis along a static magnetic field. Interacting with a time-dependent magnetic field, the entangled Bell state acquires a geometric phase $\gamma$ tied to the evolution within the spin subspace: 
$
\ket{\psi (\gamma)}=1/\sqrt2(\ket{\textrm{I}}\ket{\!\Uparrow}+ e^{i\gamma}\ket{\textrm{II}}
\ket{\!\Downarrow}).
$

\begin{figure}
  \includegraphics[width=150mm]{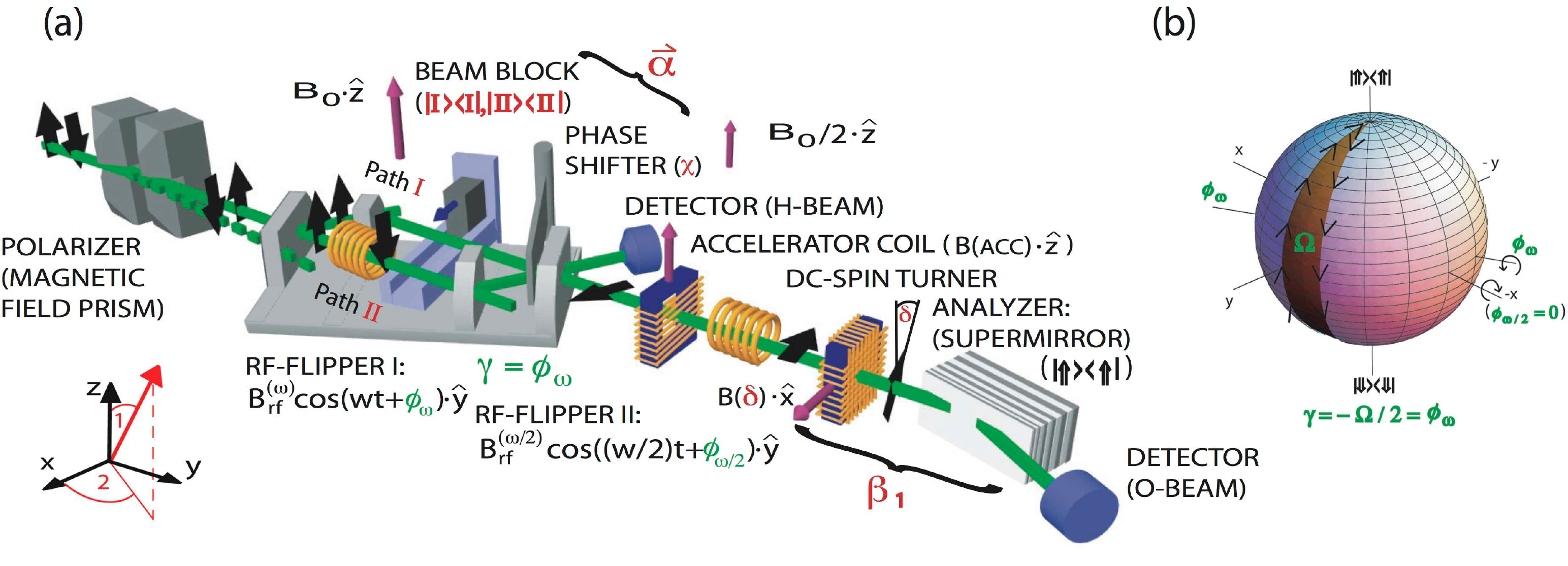}
  \caption{(a) The experimental apparatus for observation of the effect of geometric phase on Bell measurements \cite{sponarPRA2010}. The spin state acquires
a geometric phase $\gamma$ during the interaction with the two 
RF-fields. 
(b) Bloch-sphere description of the acquired geometric phase, dependent on the phase $\phi_{{\omega}}$ of the RF-field.\label{fig:GeoBellSetup}}
\end{figure}

An illustration of the setup \cite{sponarPRA2010} and a Bloch-sphere description of the induced geometric phase are given in Fig.\,\ref{fig:GeoBellSetup}. 
As in common Bell experiments, a joint measurement for spin and path was performed, applying the projection operators for the path 
$\hat P^{\rm{p}}_\pm(\vec{\alpha})=\ketbra{\pm\vec{\alpha}}{\pm\vec{\alpha}},$  
with
\mbox{$
\ket{\!+\!\vec\alpha}
=\cos(\alpha_1/2)\ket{\textrm{I}}
+e^{i\alpha_2}\sin(\alpha_1/2)\ket{\textrm{II}}$} 
and 
\mbox{$
\ket{\!-\!\vec\alpha}=
-\sin(\alpha_1/2)\ket{\textrm{I}}
+e^{i\alpha_2}\cos(\alpha_1/2)\ket{\textrm{II}},
$}
where $\alpha_1$ denotes the polar angle and $\alpha_2$ the azimuthal angle for the state on the path Bloch-sphere. In analogous manner,
${\beta_1}$ and ${\beta_2}$ were defined for the spin Bloch-sphere. Using these angles, one can define an expectation value for a joint measurement along directions $\vec\alpha$ and $\vec\beta$ and consequently a sum of four expectation values, dependent on the measurement directions and the geometric phase $\gamma$. That sum provides an $S$-value denoted as $S(\vec\alpha',\vec\beta,\vec\beta',\gamma)$ (without loss of generality, we set $\vec\alpha=0$). The upper boundary of $S(\vec\alpha',\vec\beta,\vec\beta',\gamma)$ is given by the value 2 for any NCHVT \cite{basuPLA2001}. 

\begin{figure}
  \includegraphics[width=150mm]{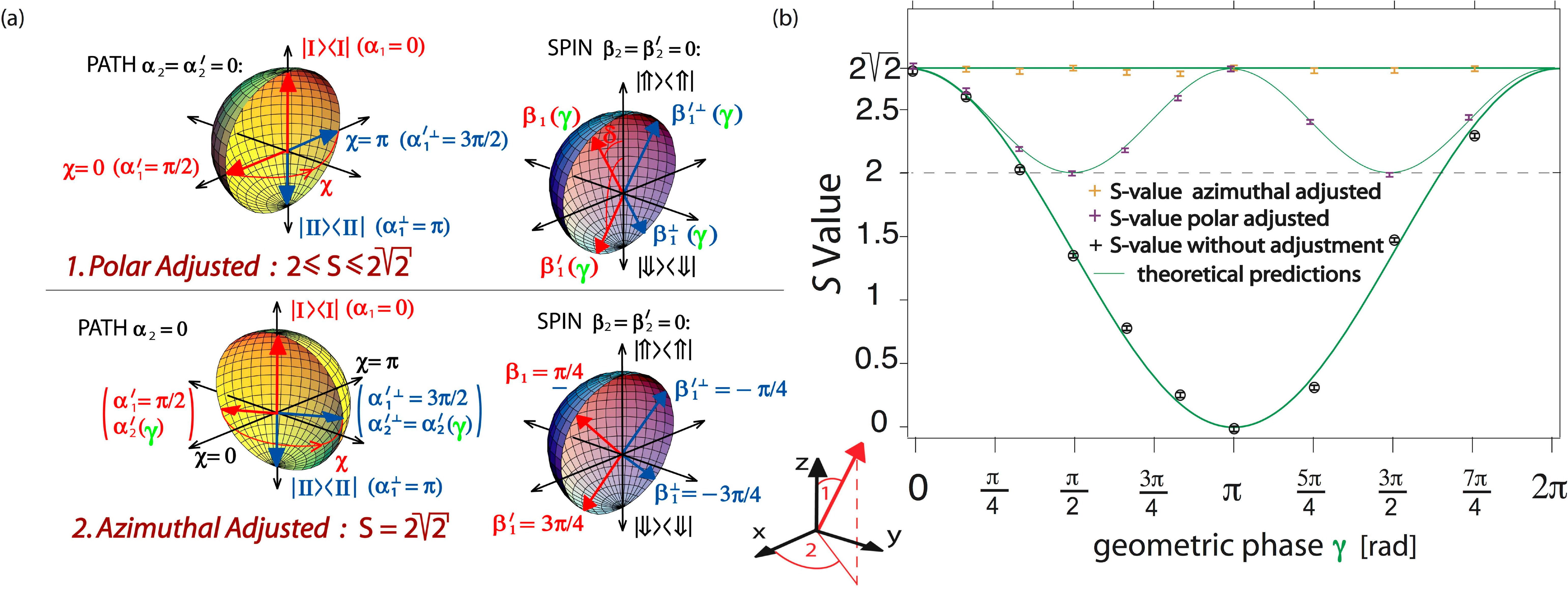}
  \caption{(a) Bloch-sphere description including measurement settings $\alpha$ and $\beta(\delta)$. (b) Experimental results for polar- and azimuthal-adjusted $S$-values vs. geometric phase $\gamma$ \cite{sponarPRA2010}.  \label{fig:PolAzz}}
\end{figure}

First, the case when the azimuthal angles were kept
constant is considered [see Fig.\,\ref{fig:PolAzz}\,(a)]. 
Here, one has 
\begin{eqnarray}
S(\alpha'_1,\beta_1,\beta'_1,\gamma)
&=&|-\sin\alpha'_1 (\cos\gamma\sin\beta_1+\cos\gamma\sin\beta'_1 )\nonumber\\
&&-\cos\alpha'_1(\cos\beta_1+\cos\beta'_1)  -\cos\beta_1+\cos\beta'_1|, 
\end{eqnarray}
which  yields a maximum $S$-value for parameters $\beta_1 = \arctan\big(\cos\gamma\big)\label{eq:Cond1},\quad \beta'_1
= \pi - \beta_1\label{eq:Cond2}\quad\rm{and}\quad \alpha'_1 =
\frac{\pi}{2}.
$
With these angles the maximal $S$ decreases
for increasing $\gamma$ to reach $S=2$ at $\gamma =
\frac{\pi}{2}$ [see Fig.\,\ref{fig:PolAzz}\,(b)].

Next, the situation in which the polar angles $\alpha'_1$, $\beta_1$
and $\beta'_1$ were kept constant at the Bell angles $\alpha'_1=\frac{\pi}{2}$, $\beta_1=\frac{\pi}{4}$, $\beta'_1=\frac{3\pi}{4}$, ($\alpha_1=0$), while
the azimuthal angles $\alpha'_2$, $\beta_2$ and $\beta'_2$ ($\alpha_2=0$) were varied, is discussed [see Fig.\,\ref{fig:PolAzz}\,(a)]. 
The corresponding S function is denoted as
\begin{eqnarray}
S(\alpha'_2,\beta_2,\beta'_2,\gamma)
=\left|-\sqrt{2}-\frac{\sqrt{2}}{2}\left[\cos(\alpha'_2-\beta_2-\gamma)
+\cos(\alpha'_2-\beta'_2-\gamma)\right]\right|.
\end{eqnarray}
The maximum value $S=2\sqrt{2}$ is reached for
$\beta_2=\beta'_2$ and $\alpha'_2-\beta'_2=\gamma$ (mod $\pi$).
For convenience, $\beta_2=0$ is chosen. 
For the spin measurement the directions were fixed to $\beta_1=\pi/4$, $\beta_2=0$ and $\beta'_1=3\pi/4$, $\beta'_2=0$. 
As predicted, the constant maximal $S$ value was found for $\alpha'_2=\gamma$, as shown in Fig.\,\ref{fig:PolAzz}\,(b). In the same plot, the case is
included for which no corrections were applied to the Bell angles.
Then the familiar maximum value of
$2\sqrt{2}$ was reached only for $\gamma=0$. At $\gamma=\pi$ the
value of $S=0$ was found.

In this experiment it was demonstrated that a geometric phase in one subspace does influence $S$, but does not lead to a loss of entanglement. Two schemes, namely polar- and azimuthal-adjustment of the Bell angles, were realized and showed to balance the influence of the geometric phase. 

\subsection{Robustness of the Berry phase}\label{subsec:RobustBerry}

A first attempt to treat non-unitary geometric phases theoretically was made in \cite{garrisonPLA1988} and \cite{mukundaAOP1993}. Using a quantum jump approach, it was found in \cite{carolloPRL2003} that the geometric phase could be independent of certain forms of noise. 
Following \cite{sjoeqvistPRL2000} and \cite{singhPRA2003}, the authors of \cite{tongPRL2004} found that only in the special case where the induced phase is of purely geometric origin, it is independent of fluctuations in direction parallel to the precession axis. 
The approach in \cite{peixotoDeFariaEPL2003} employs the concept of completely positive maps. The ideas developed in \cite{ericssonPRA2003} yield different values of geometric phases for different Kraus representations. The case in which the system undergoes an adiabatic evolution driven by a slowly varying magnetic field and is weakly coupled to a dissipative environment is studied in \cite{whitneyPRL2005}. Here, the acquired Berry phase suffers a purely geometric modification. The problem is tackled in more general frameworks in \cite{lombardoPRA2006}, \cite{yiPRA2005} and \cite{zhuPRA2005}, the latter putting much emphasis on the aspect of robustness of the non-adiabatic geometric phase from the point of view of quantum computation. There, maximum phase gate fidelities in the presence of stochastic control errors are found for vanishing dynamical phase. 

\begin{figure}
\begin{center}
    \scalebox{0.35}{\includegraphics{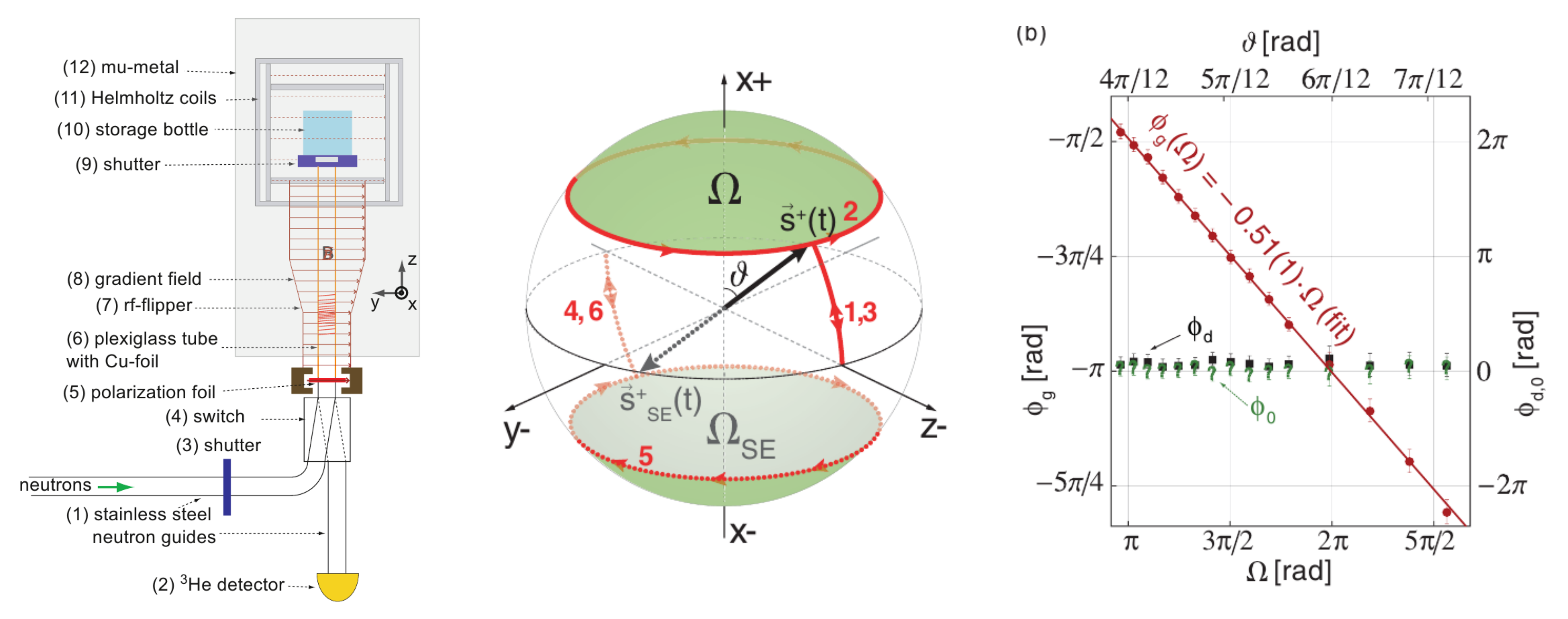}}
    \caption{Left: UCN-setup to measure the robustness of the Berry phase. Center: A spin-echo approach was used to compensate for dynamical phase, as is shown here for the up-spin state. Right: Measured geometric phase $\phi_g$, dynamical phase $\phi_d$ and $\phi_0$ without noise added \cite{filippNIMA2009,filippPRL2009}.}
    \label{fig:StabilityBerry1}
\end{center}
\end{figure}

In \cite{deChiaraPRL2003}, calculations suggest that the variance of the Berry phase is inverse proportional to the evolution time, an aspect which was also confirmed in a dedicated experiment with UCN \cite{filippPRL2009}. The experimental setup [see Fig.\,\ref{fig:StabilityBerry1}\,(left)] was conceptually similar to the ones used for 
measurements of the neutron electric-dipole-moment and the test of the Berry phase described in Sec.\,\ref{subsec:GeoPhase}. UCN were spin-filtered by a polarizing foil to store polarized UCN in a neutron bottle, mounted in the centre of three pairs of magnetic-field coils arranged in Helmholtz-geometry. The coils provided guide field, $\pi/2$- and $\pi$-pulses, the slowly varying magnetic field to carry out adiabatic spin evolutions and the adiabatic noise for the experiment. The evolution consisted of the paths on the Bloch-sphere as shown for the up-state (s$^+$) in Fig.\,\ref{fig:StabilityBerry1}\,(center). First, the magnetic field and the spin state were aligned with the negative $z$-direction. A $\pi/2$-pulse in $x$-direction was applied to generate an equal superposition of up- and down-spin state (aligned to $-z$- and $+z$-directions, respectively). An offset field in $x$-direction was turned on to set the opening angle $\vartheta$ of the cone and, thereby, the solid angle $\Omega$ (path 1). A circular evolution was carried out at constant $\vartheta$ (path 2) using the coil fields in $y$- and $z$-directions and s$^+$ was led back to the $-z$-direction (path 3) by turning off the field in $x$-direction. Now, a $\pi$-pulse was applied to exchange the directions of up- and down-spin states. The evolution was repeated with the up-component on the lower half of the Bloch-sphere such that the orientation of the area enclosed by the evolution path was the same as before (paths 4,5,6). Consequently, the geometric phase doubled, while the dynamical contributions canceled out. 
The resulting geometric phase $\phi_{g}$ could then be measured doing a spin-state analysis for various $\vartheta\propto\Omega$, i.e. measuring the polarization up and down in $x$-, $y$- and $z$-directions as a function of the enclosed solid angle. 

\begin{figure}
\begin{center}
    \scalebox{0.32}{\includegraphics{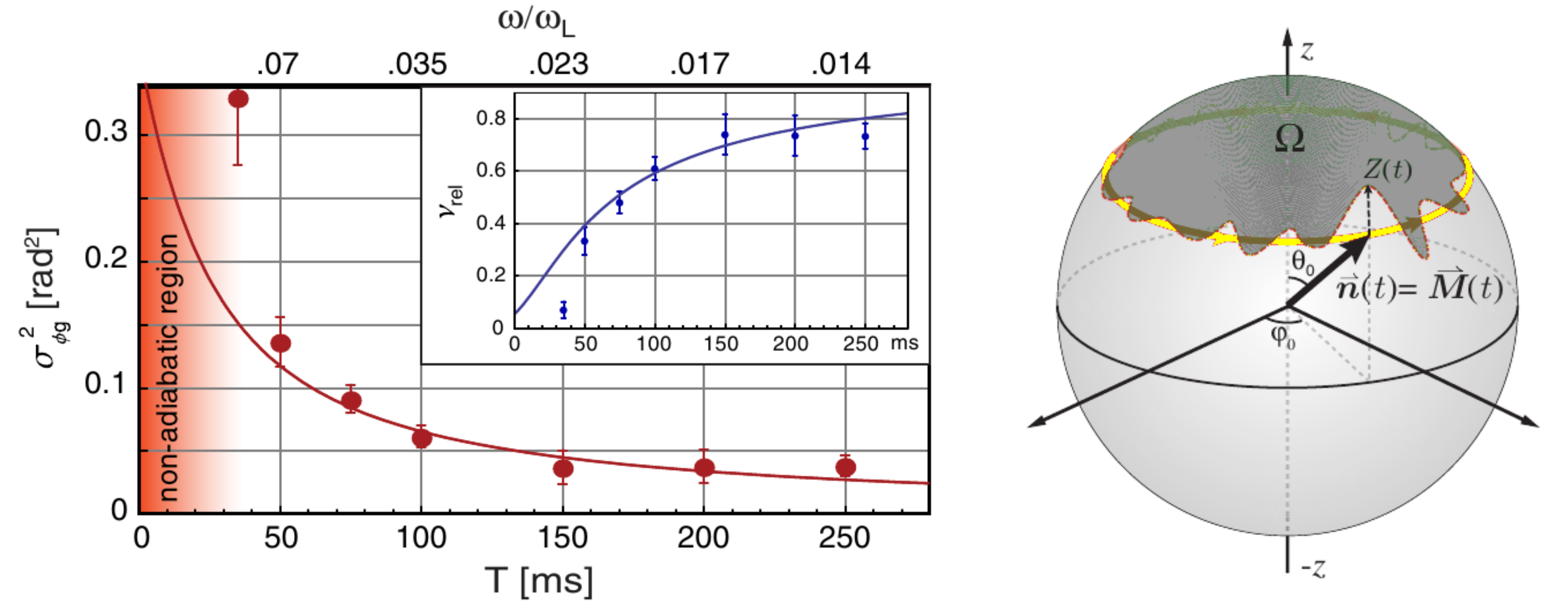}}
    \caption{Left: Measured decrease of the geometric-phase variance $\sigma^2_{\phi g}$ with increasing evolution time T \cite{filippPRL2009}. Right: Path of a circular evolution on the Bloch-sphere exhibiting fluctuations \cite{filippEPJST2008}.}
    \label{fig:StabilityBerry2}
\end{center}
\end{figure}

The results are seen in Fig.\,\ref{fig:StabilityBerry1}\,(right). The dynamical phase $\phi_d$ was measured by not reversing the sense of rotation after the $\pi$-pulse, so that the geometric phase cancels, while the dynamical doubles. 
As expected, the result was the same as for $\phi_0$, the phase measured without circular evolution with the same evolution time. 
The measurement of the geometric phase was repeated with fluctuations added to the magnetic field in $x$-direction. Measuring the geometric phase for many storage cycles at given evolution time T, one obtains the phase-average and variance $\sigma^2_{\phi g}$. The longer T, the smaller is the variance as can be seen from the data in Fig.\,\ref{fig:StabilityBerry2}\,(left), reflecting the resilience of the Berry phase against adiabatic noise. The effect can be understood by considering Fig.\,\ref{fig:StabilityBerry2}\,(right): for given noise spectrum, particular wiggles in the evolution path cannot cancel out if T is short, while for long T there is enough time for deviations from the noise-less path to occur also in the other direction, such that they are compensated and $\Omega$ remains determined by $\vartheta$ only. 

Adiabatic evolutions are comparably slow, so that the focus for applications in quantum computation was drawn to non-adiabatic evolutions. 
Newer developments focus on non-Abelian, non-adiabatic geometric phases to achieve fast quantum gates \cite{sjoeqvistNJP2012}. Relevant experimental results can be found in \cite{bergerPRA2013, abdumalikovNature2013}.

\section{Other quantum optical experiments with neutrons}\label{sec:others}

\subsection{Error-disturbance uncertainty relation in neutron spin-measurements}

The uncertainty principle is without any doubt one of the cornerstones of quantum physics. In his original paper from 1927 \cite{heisenbergZP1927} Heisenberg proposed a reciprocal relation for measurement \emph{noise} (nowadays referred to as  \emph{error}) and \emph{disturbance} in the famous $\gamma$-ray microscope thought experiment: ``\emph{At the instant when the position is determined - therefore, at the moment when the photon is scattered by the electron -  the electron undergoes a discontinuous change in momentum. This change is the greater the smaller the wavelength of the light employed - that is, the more exact the determination of the position\,\dots}'' \cite{heisenbergZP1927}. Heisenberg followed Einsteins realistic view, that is, to base a new physical theory only on observable quantities (\emph{elements of reality}), arguing that terms like velocity or position make no sense without defining an appropriate apparatus for a measurement. By solely considering the Compton-effect Heisenberg gave a rather heuristic estimate for the product of the error of a position measurement $p_1$ and the disturbance $q_1$ induced on the particle\rq{}s momentum, denoted as $p_1q_1\approx h$. Heisenberg's original formulation \cite{heisenbergZP1927,heisenbergBook1930} can be read in modern treatment as 
$
\epsilon(Q)\eta(P)\geq \hbar/2$, for error $\epsilon(Q)$ of a measurement of the position observable $Q$ and disturbance $\eta(P)$ of the momentum observable $P$ induced by the position measurement.
However, most modern textbooks introduce the uncertainty principle using the relation 
$
\sigma(Q)\sigma(P)\geq \hbar/2,
$
originally proved by Kennard in 1927 \cite{kennardZFP1927} for the standard deviations $\sigma (Q)$ and $\sigma(P)$ of the position observable $Q$ and the momentum observable $P$. Note that the physical situation is very different from the one discussed by Heisenberg: Here, statistical distributions of not a joint measurement of two observables but a measurement of either $Q$ or $P$ are considered. 
Kennard's relation addresses 
an intrinsic uncertainty every quantum system must possess -- no matter if it is measured or not. 
The disturbance caused by the measuring device is ignored here. Robertson generalized Kennard\rq{}s relation to standard deviations of arbitrary pairs of observables $A$ and $B$:
$\sigma(A)\sigma(B)\geq 1/2\vert\bra{\psi} [A,B] \ket{\psi}\vert$. Robertson's relation has been confirmed by many different experiments \cite{shullPR1969,yuenPRA1976,breitenbachNature1997} and is uncontroversial. 
The (formally) corresponding generalized form of Heisenberg's original error-disturbance uncertainty relation would read
\begin{equation}\label{eq:HeisGenaralized} 
\epsilon(A)\eta(B)\geq\frac{1}{2}\vert\bra{\psi} [A,B] \ket{\psi}\vert.
\end{equation}
However, certain measurements do not obey Eq.\,(\ref{eq:HeisGenaralized}) \cite{arthursPRL1988,ishikawaROMP1991,ozawaQAOOC1991}, proving it to be formally incorrect. In 2003, Ozawa introduced the correct form of a generalized error-disturbance uncertainty based on rigorous theoretical treatments of quantum measurements as
and proved the universal validity in the general theory of quantum measurements \cite{ozawaPRA2003}. 
Here, $\epsilon(A)$ denotes the root-mean-square (r.m.s.) error of an arbitrary measurement for an observable $A$, $\eta(B)$ is the r.m.s. disturbance on another observable $B$ induced by the measurement, and $\sigma(A)$ and $\sigma (B)$ are the standard deviations of $A$ and $B$ in the state $\ket{\psi}$. Error $\epsilon(A)$ and disturbance $\eta(B)$ are defined via an indirect measurement model for an apparatus $\boldsymbol{A(x)}$ measuring an observable $A$ of an object system $\boldsymbol{S}$ as
$
\epsilon(A)=\|(U^{\dagger}(\1\otimes M)U-A\otimes \1)\ket{\psi}\ket{\xi}\|$ and 
$
\eta(B)=\|(U^{\dagger}(B\otimes \1)U-B\otimes \1)\ket{\psi}\ket{\xi}\|
$. $\ket{\psi}$ is the state before the measurement of system $\boldsymbol{S}$, which is an element of the Hilbert space $\mathcal H^{\rm obj}$. $\ket{\xi}$ and $M$ are the initial state of the probe system $\boldsymbol{P}$ (in Hilbert space $\mathcal H^{\rm pro}$) and an observable $M$ -- referred to as \emph{meter observable} -- of $\boldsymbol{P}$. A unitary operator $U$ on  $\mathcal H^{\rm obj}\otimes\mathcal H^{\rm pro}$ describes the time evolution of the composite system $\boldsymbol{S}+\boldsymbol{P}$ during the measurement interaction. Here, the Euclidean norm is used, which is given by the square root of the inner product $\|X\ket{\psi}\|=\bra{\psi}X^\dagger X \ket{\psi}^{1/2}$ for a state vector $X\ket{\psi}$. In our experiment  the universally valid uncertainty relation error, as defined in Eq.\,(\ref{eq:Ozawa}), is tested via successive measurement of spin observables $A$ and $B$. Though claimed to be experimentally inaccessible \cite{wernerQIC2004,koshinoPRep2005}, in case of projective measurements the error $\epsilon(A)$ and the disturbance $\eta(B)$ can be expressed as a sum of expectation values for three different states \cite{ozawaAP2004}:

 \begin{subequations}
\begin{eqnarray}
 \epsilon(A)^{2}&= &2+ \langle \psi \vert O_{A}\vert\psi  \rangle +\langle \psi\vert A O_{A}  A\vert\psi \rangle 
 - \langle \psi\vert  (A+\1)O_{A}(A+\1)\vert\psi\rangle,\\
\eta(B)^{2}&=& 2+ \langle \psi \vert X_{B}\vert\psi  \rangle +\langle \psi\vert B O_{B}  B\vert\psi \rangle 
 - \langle \psi\vert  (B+\1)X_{B}(B+\1)\vert\psi\rangle.
 \end{eqnarray}
 \end{subequations}

\begin{figure}
\begin{center}
  \includegraphics[width=115mm]{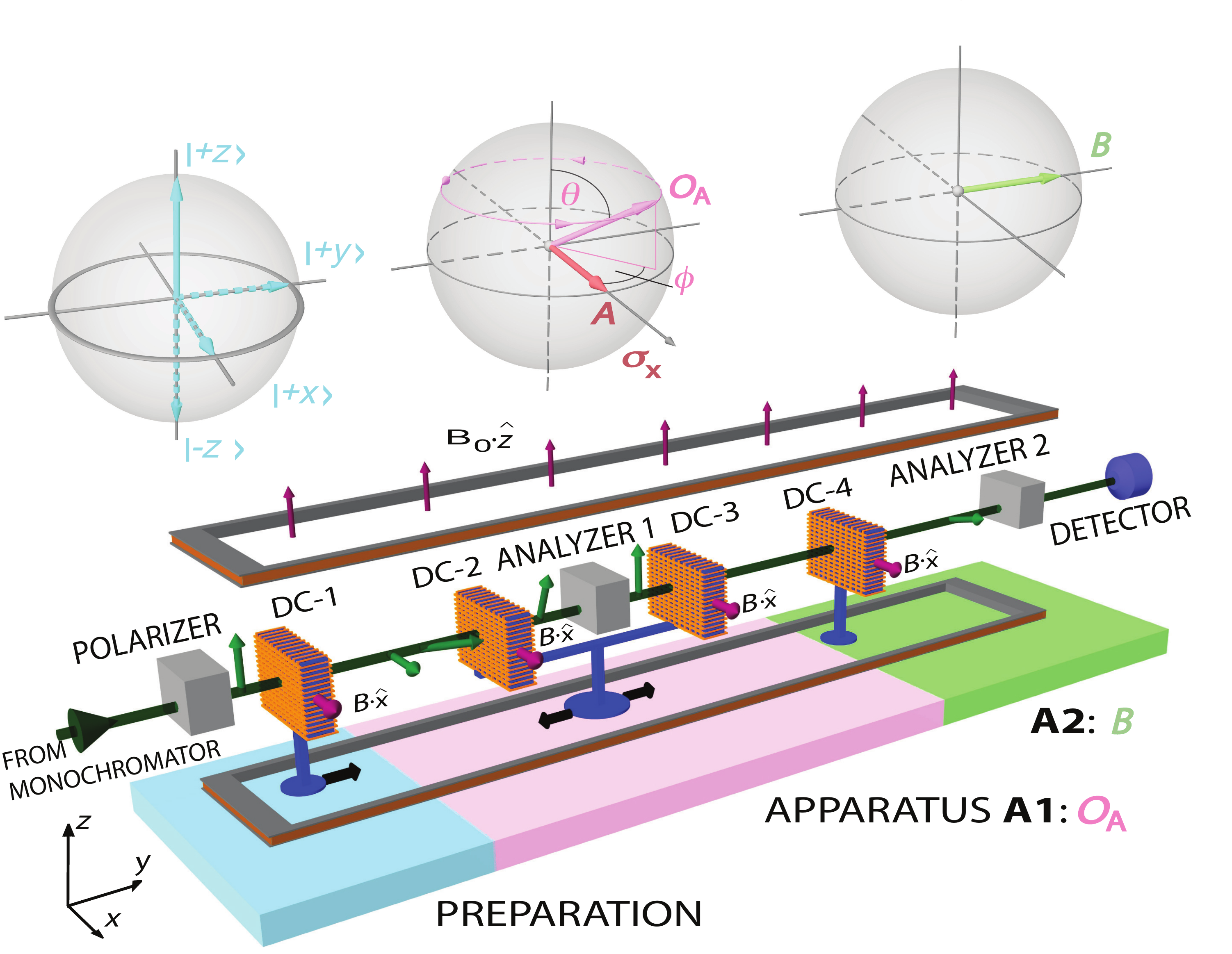}
  \caption{Neutron polarimetric setup for demonstration of the universally valid uncertainty relation for error
and disturbance in neutron-spin measurements used in \cite{erhartNatPhys2012} and  \cite{sulyokPRA2013}\label{fig:UncertaintySetup}. }
\end{center}
\end{figure}
\begin{equation}\label{eq:Ozawa} 
\epsilon(A)\eta(B)+ \epsilon(A)\sigma(B)+\sigma(A)\eta(B)\geq\frac{1}{2}\vert\bra{\psi} [A,B] \ket{\psi}\vert,
\end{equation}

The measuring apparatus $\bold{A1}$ is considered to carry out the projective measurement along an axis $\vec o_a(\theta,\phi)$ denoted as $O_A=\vec o_a(\theta,\phi)\vec\sigma$. Here, $\theta$ and $\phi$ denote polar and azimuthal angles of the measurement direction $\vec o_a$ and are experimentally controlled parameters. In order to detect the disturbance $\eta(B)$ on the observable $B$ induced by measuring $O_A$, apparatus $\bold{A2}$ carries out the projective measurement of $B$ on the state just after the first measurement. 
 
Consequently, all expectation values necessary to determine error $\epsilon(A)$ and disturbance $\eta(B)$ can be derived from the intensities in the input states $\ket{\psi}$, $A\ket{\psi}$, $(A+\1)\ket{\psi}$ and $\ket{\psi}$, $B\ket{\psi}$, $(B+\1)\ket{\psi}$. These states were generated by spin rotations within coil DC-1 at an appropriate position within the preparation section (blue) of the neutron optical setup depicted in Fig.\,\ref{fig:UncertaintySetup}. 
The projective measurement of $OA$ (apparatus $\bold{A1}$ - light red in Fig.\,\ref{fig:UncertaintySetup}) consisted of two sequential steps: First, the initially prepared state was projected onto the eigenstates of $OA$ by DC-2, which rotates the respective spin component of $\vec\sigma_a$ (associated to $O_A$) to the $+z$-direction. Second, in order to complete the projective measurement, the spin -- which is pointing to the $+z$-direction after the analyzer -- has to be prepared in an eigenstate of $O_A$. This was achieved by properly positioning DC-3. 
Finally, the $B$ measurement was performed (apparatus $\bold{A2}$ - green in Fig.\,\ref{fig:UncertaintySetup}) utilizing DC-4 and the second analyzer. Unlike for the $O_A$-measurement, subsequent preparation of the eigenstates of $B$ was not necessary since the detector was insensitive to the spin state. For the measurement of the standard deviations of the observables $A$ and $B$, the two measurement apparatuses were used individually. 

\begin{figure}
\begin{center}  
\includegraphics[width=120mm]{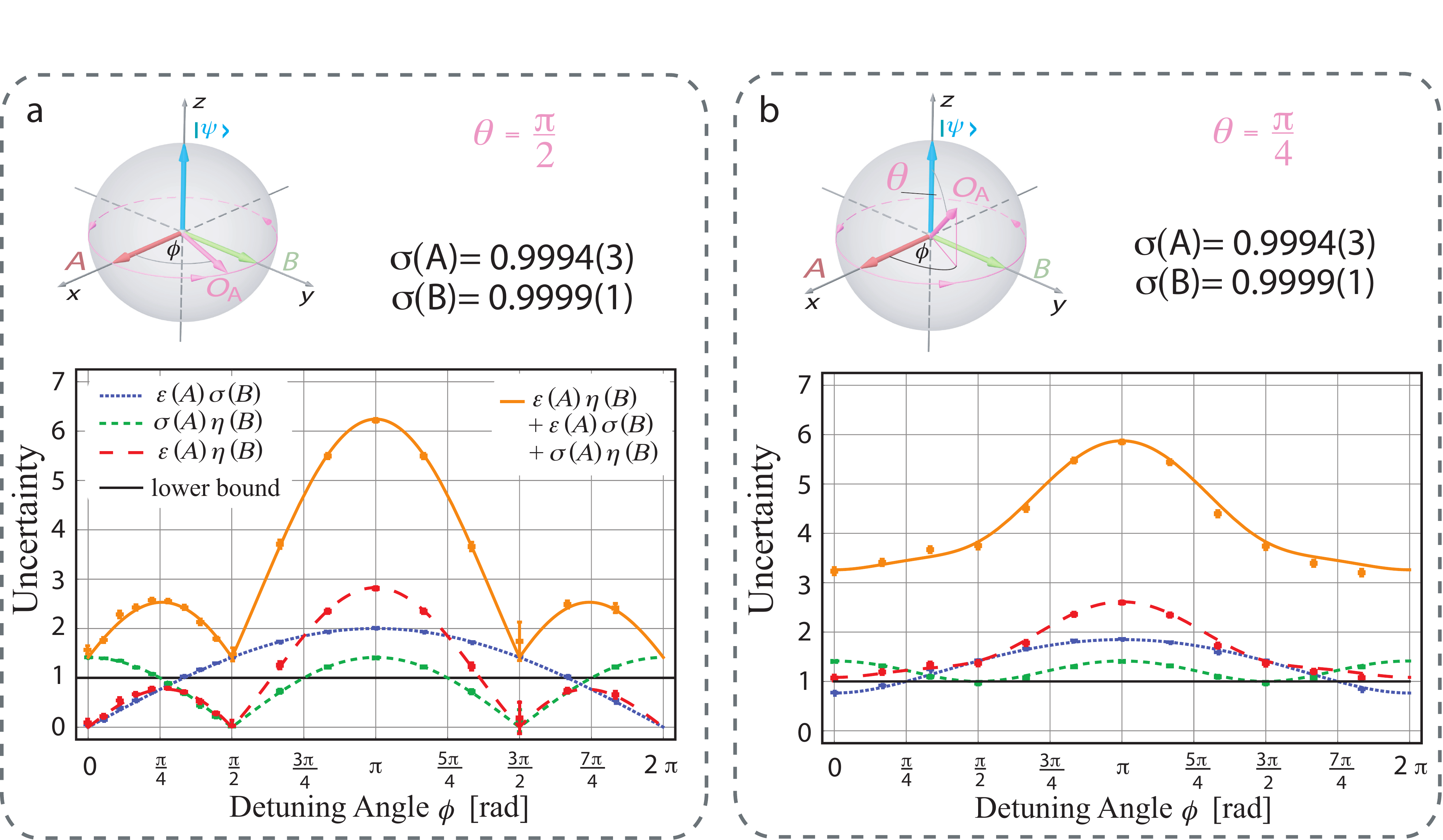}
  \caption{Experimentally determined values of  $\epsilon(A)\sigma(B)$, $\sigma(A)\eta(B)$ and $\epsilon(A)\eta(B)$ \cite{erhartNatPhys2012,sulyokPRA2013}.
 \label{fig:UncertaintyResultsStandardConfig}}
\end{center}
\end{figure}

The experimental settings for initial state $\ket{\psi}=\ket{\!\Uparrow}$ and observables $A=\sigma_x$ and $B=\sigma_y$ required the preparation of auxiliary input states $\ket{\!\Downarrow}$,  $\ket{\!\Uparrow}+\ket{\!\Downarrow}$ and $\ket{\!\Uparrow}+i\ket{\!\Downarrow}$. Standard deviations yielded $\sigma(A)=\sigma(B)=1$ and the right-hand sides of the uncertainty relations gave a lower bound of $1/2\bra{\psi}[A,B]\ket{\psi}=1$. In a first experimental run $O_A$ was varied along the equator ($\theta=\pi/2$), parameterized by its azimuthal angle $\phi$. For $\phi=0$, the error $\epsilon(A)$ vanished and the disturbance $\eta(B)$ was maximal. The disturbance $\eta(B)$ vanished for $O_A=B$ ($\phi=\pi/2)$ and reached a second maximum for $O_A=-A$. Note that at this point also the error $\epsilon(A)$ has its (only) maximum. The famous \emph{trade-off relation}, i.e. the reciprocal relation for error and disturbance, only holds for $-\pi/2\leq\phi\leq\pi/2$, which can be seen in Fig.\,\ref{fig:UncertaintyResultsStandardConfig}\,(a). The product of error and disturbance $\epsilon(A)\eta(B)$ -- the left hand side of Eq.\,(\ref{eq:HeisGenaralized})-- is below the limit given by $\frac{1}{2}\bra{\psi}[A,B]\ket{\psi}$ in a wide range of $\phi$-values. The latter reveals a violation of the generalized Heisenberg relation [Eq.\,(\ref{eq:HeisGenaralized})]. 
In particular, the left-hand side of Ozawa's relation $\epsilon(A)\eta(B)+ \epsilon(A)\sigma(B)+\sigma(A)\eta(B)$  [Eq.\,(\ref{eq:Ozawa})] is always larger than the lower bound defined by the expectation value of the commutator, demonstrating the validity of Ozawa's relation. 

In the following experiment, $O_A$ lay in the equatorial plane, the state evolved on circles of latitude of the Bloch-sphere (fixed polar angle $\theta$) as depicted in Fig.\,\ref{fig:UncertaintyResultsStandardConfig}\,(b). 
Neither the error $\epsilon(A)$ nor the disturbance $\eta(B)$ vanished, since they never coincided with $A$, $B$ or $-B$. 
Ozawa's inequality was again fulfilled over the entire range of $\phi$.

\subsection{Non-commutation of Pauli spin-operators in neutron polarimetry}

The uncertainty principle is closely related to one of the most
fundamental properties of QM, namely non-commutability of certain pairs of observables \cite{feynmanLectures1966}.
It is well known that pairs of different Pauli spin-operators do not commute, $[ \sigma_j , \sigma_k ] \neq 0$ for $j \neq k$, in particular:
\begin{equation}\label{63.1}
   [\sigma_z,\sigma_x]=\sigma_z \sigma_x - \sigma_x \sigma_z = 2i \sigma_y.
\end{equation}
This is one of the simplest examples of non-commutability for two-level quantum systems. 
As already mentioned before, the rotation of
a spin-1/2 system around the axis $\hat{\mathbf{\alpha}}$ by an angle $\alpha$ is written as
$U(\vec\alpha) = \exp (-i \vec{\sigma} \cdot \vec{\alpha}/2 )$ with the rotation vector $\vec{\alpha} = \alpha \hat{\mathbf{\alpha}}$.
For instance, by setting the rotation angle of the neutron spin to $\pi$, one can realize $U(\pi\hat l)=-i\sigma_l$, that is, operations represented by the Pauli operators $\sigma_l$ with $\hat l=\hat x,\hat y,\hat z$, in the lab.

In 1997, a neutron-IFM experiment \cite{waghPRL1997,allmanPRA1997} demonstrated non-commutability of the Pauli spin-operators for a special case: 
Two successive $\pi$-rotations transformed the initial spinor 
$|\!\Uparrow\rangle$ to the final one $|\!\Downarrow\rangle$ by applying both the spin rotations $-\sigma_A\sigma_B$ and $-\sigma_B\sigma_A$ in subsequent runs, where $A$ and $B$ are the \emph{mutually orthogonal} directions of two magnetic field rotation axes. 
A phase difference of $\pi$ between the results obtained with the above operations was measured using the IFM setup. The observed phase difference can be attributed to the geometric nature of spin rotations:
different evolution paths from the same initial state to the final state induce the observed phase difference by $\pi$. 

\begin{figure}
\centering\includegraphics[width=5.9in]{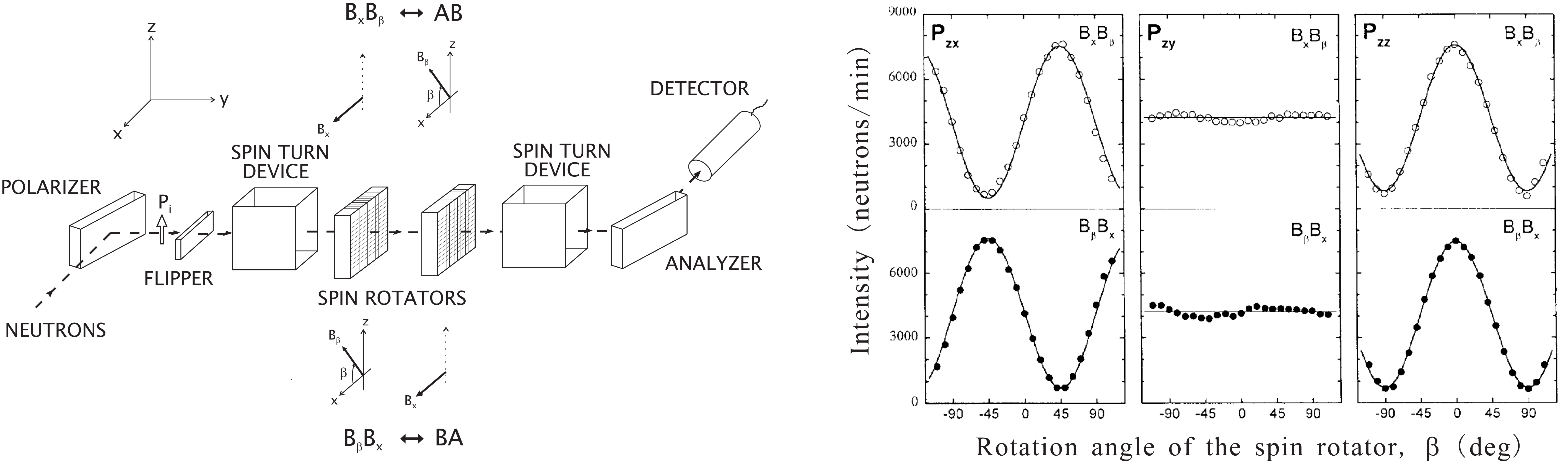}
\caption{
Sketch of the experimental setup (left) and the results (right) demonstrating the non-commutation relation of
Pauli spin-operators $\sigma_x$ and $\sigma_z$. The $x$-component of the final polarization vectors differed for configurations $AB$ and $BA$ \cite{hasegawaPLA1997}.}
\label{fig_nonc}
\end{figure}

Shortly thereafter, another approach using a neutron polarimeter revealed the non-commutability of Pauli spin-operators for non-orthogonal directions $A$ and $B$ \cite{hasegawaPLA1997}.
The experimental setup is depicted in Fig.\,\ref{fig_nonc}\,(left).  
A polarized neutron beam successively propagated through two $\pi$ spin-rotators. One was oriented in direction $A$ given by the $+x$-direction, in particular $\hat\alpha_A =(1,0,0)$. The other was oriented in direction $B$ in the 
$xz$-plane, specified by $\hat\alpha_B=(\cos\beta,0,\sin\beta)$.
The operators for the corresponding $\pi$-rotations are given by
$A=-i\sigma_x$ and $B=-i(\sigma_x\cos\beta+\sigma_z\sin\beta)$. 
Thus, for the spin operations $AB$ and $BA$, i.e., A followed by B and B followed by A, respectively, one obtains
\begin{equation}\label{63.3}
{AB = - \left( {\begin{array}{*{20}c}
   \cos\beta & -\sin\beta  \\
   \sin\beta & \cos\beta  \\
\end{array}} \right)\,{\rm{and}}~BA = -\left( {\begin{array}{*{20}c}
   \cos\beta & \sin\beta  \\
   -\sin\beta & \cos\beta  \\
\end{array}} \right).}
\end{equation}
The difference between $AB$ and $BA$ arises due to the fact that $\sigma_x$ and $\sigma_z$ do not commute. The above result suggests a possibility to observe differences 
in the final polarization due to non-commutability of the
Pauli spin-operators using a polarimeter setup. 
In particular, if the polarization vector of the incident beam is $P^i=(0,0,1)$, the final polarization vectors $P^f_{AB}$ and $P^f_{BA}$ -- after two $\pi$-rotations AB or BA -- become
\begin{equation}\label{63.4}
   {P^f_{AB}=(\sin 2\beta,0,\cos 2\beta)~{\rm{and}}~{P^f_{BA}= (-\sin 2\beta,0,\cos 2\beta)}}.
\end{equation}
Here, it is clearly seen that, while the $y$- and the $z$-components of the final polarization vectors are the same,
the non-commutability of the Pauli operators $\sigma_x$ and $\sigma_z$ leads to a difference in the $x$-components.

In the experiments \cite{hasegawaPLA1997}, specially designed spin-turn devices served to orient and analyze the polarization vector to the desired directions.
In order to show the non-commutation of the operators $A$ and $B$, the spin-rotators in the center of the setup were adjusted so that the rotator upstream represents $A$ and the downstream one $B$, and vice versa for the subsequent run. 
Neutron intensities were recorded as the function of the angle $\beta$ for polarization analysis in all three directions of space. 
Intensity modulations together with theory curves are shown in Fig.\,\ref{fig_nonc}\,(right). 
Here, one can see that interchange of $A$ and $B$ led to inverse modulation of the $x$-component of the final polarization vectors. 
Although the emerging polarization should have no $y$-component, a slight residual intensity modulation persisted upon variation of $\beta$. This was due to a small misalignment of the polarization-analysis direction.

It is well known that, while rotations about the same axis commute, rotations about different axes do not \cite{sakuraiBook1994}. 
The results of the measurement mentioned above, at first sight, seem to be a consequence of this fact. However, rotations in three-dimensional space are described by elements of the SO(3) group, while spin-1/2 operations are carried out by elements of the SU(2) group. 
The SU(2) group has richer resources
than SO(3): For instance, $4\pi$-symmetry of the spin-1/2 wave function only appears in SU(2). Thus, while the resulting difference of 
the final polarization vectors can be interpreted within 
SO(3)-terminology in some sense, non-commutability of Pauli-operators can have other aspects. 
For instance, the factor -1, or rather, the phase factor $e^{i\pi}$ in the relation 
$\sigma_z \sigma_x = - \sigma_x \sigma_z$ was examined carefully in \cite{hasegawaPRA1999}: influences of the dynamical and geometrical phases on the non-commutation relation of the Pauli operators were investigated there.

\subsection{Holographic-grating neutron interferometry}\label{subsec:HoloGratings}

As is well known, neutron optics is based upon the one-particle Schr\"{o}dinger equation \cite{searsBook1989}, which contains the optical nuclear potential $V_N$ for a (non-magnetic) material or, equivalently, the neutron refractive-index  
$
n_0=\sqrt{1-V_N/E_0}\approx 1-\lambda^2\, Nb_c/(2\pi).
$
Here, $E_0$ is the energy of the incident (free) neutron. 
If one can tune the value of the refractive index for neutrons and, moreover, imprint refractive-index structures to materials, tailor-made neutron-optical elements become feasible. 
For instance, a one-dimensional grating --  periodically modulated in the $x$-direction -- can be characterized by the refractive index $n(x)= n_0+\Delta n\cos(2\pi x/\Lambda)$, with the modulation amplitude
$\Delta n=\lambda^2\Delta Nb_c/(2\pi)$,
the average refractive index $n_0$ and the grating period $\Lambda$. The quantity $\Delta N$ is the number-density modulation amplitude. 
Note that $b_c$ is a real number, i.e. absorption is neglected in this description.

Artificial grating structures can also be produced by exploiting the light-induced change of the refractive index for light -- the photorefractive effect. 
Using an optical holography setup, signal and reference light-beams are superposed at the position of a recording material [photorefractive (poly)methylmethacrylate, for instance]. The superposition results in a periodic light-intensity pattern in the recording material -- modulating $N$ for a particular $b_c$ via an intensity-dependent photopolymerization process, say -- so that diffraction gratings for light and neutrons are recorded \cite{ruppPRL1990,fallyAPB2002}. 

\begin{figure}
\begin{center} 
\scalebox{0.42}
{\includegraphics {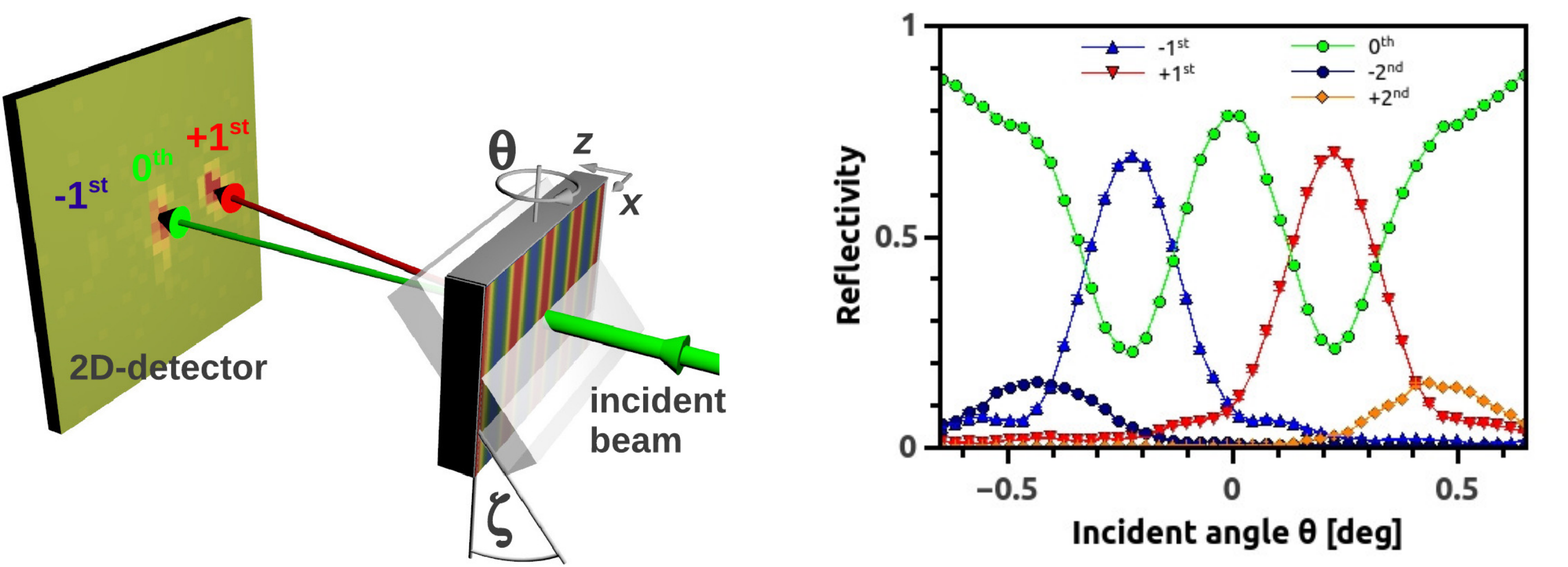}}
\caption{Left: Experimental setup to measure rocking curves of holographic gratings with neutrons. 
Right: Typical neutron rocking-curves measured with a holographic SiO$_2$ nanoparticle-polymer composite grating ($\Lambda=500$\,nm, $d_0\approx 100\,\mu$m) for $\zeta=40\,^\circ$ and $\lambda=39$\,\AA\,. The beam divergence is about 0.001\,rad. 
The sample size is typically a few cm$^2$.}
\label{fig:DiffrSetup} 
\end{center}
\end{figure}

After chemical synthesis, the recording material typically undergoes processes such as heating and drying for pre-polymerization and is deposited between two glass plates. 
Both coherent, s-polarized, plane light-waves enclose the (external) incidence angle $\theta^{(e)}$ with the sample surface normal with which they span the plane of incidence. Thus, the resulting light-intensity pattern exhibits a sinusoidal modulation: 
$I=|\psi_S+\psi_R|^2=A_0^2/2\left[1+\cos\left( Kx\right)\right]$, where $K=2\pi/\Lambda=2k_L\sin\theta^{(e)}$. For a given material, illumination time and intensity, the refractive-index change for light may be written as $\Delta n_{0,L} + \Delta n_L \cos (Kx)$. 
A neutron-diffraction experiment using VCN and holographic gratings is sketched in Fig.\,\ref{fig:DiffrSetup}. The gratings are mounted in Laue-geometry and tilted at the angle $\zeta$ about an axis parallel to the grating vector in order to adjust the effective thickness. The incident angle $\theta$ is varied to measure rocking curves in the vicinity of the Bragg-angle $\theta_B$ as defined by $\lambda=2\Lambda\sin\theta_B$. 

The most widely-used `grating-structures' in neutron-optics are probably crystals. 
For CN- and VCN-interferometry, also artificial structures have been implemented \cite{ioffePL1985,gruberPLA1989,vanDerZouwNIMA2000,
funahashiPRA1996,sekiJOPCS2012,pfeifferPRL2006,
stroblPRL2008,gruenzweigRScIntr2008,
mankeNatCom2010}.

\begin{figure}
\begin{center} 
\scalebox{0.40}
{\includegraphics {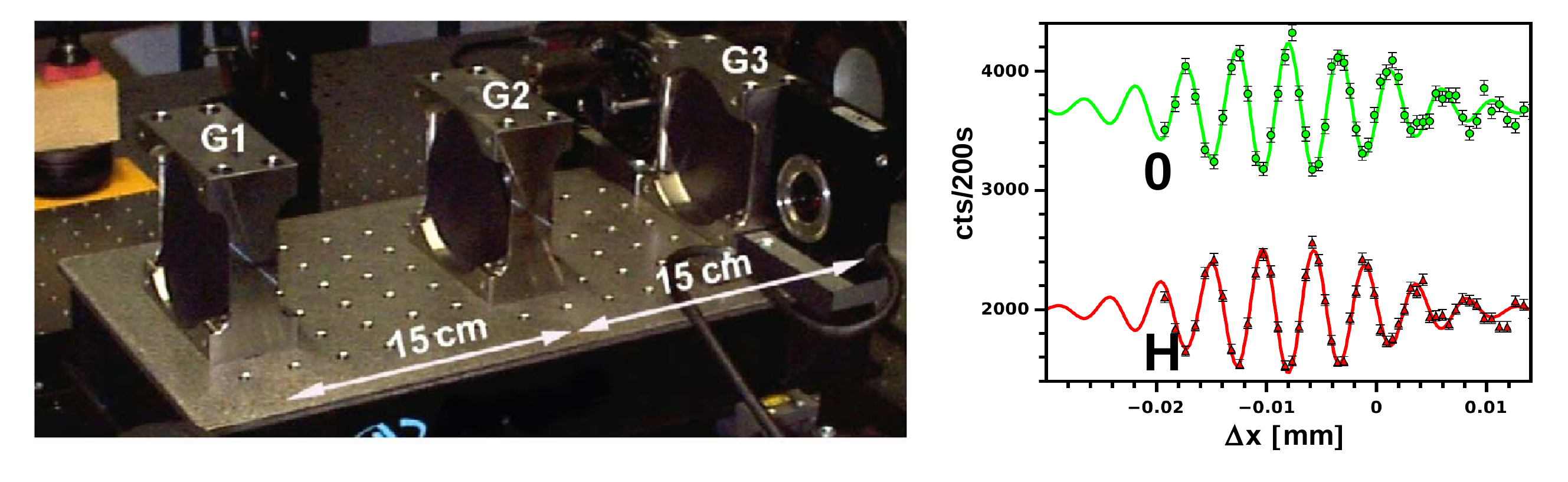}}
\caption{Left: Holographic grating IFM. The LLL-geometry is built up by three (poly)methylmethacrylate gratings G1, G2 and G3 with $\Lambda=381$\,nm. The beam separation at G2 is about 1\,mm for $\lambda\approx 26$\,\AA\,. Right: Intensity oscillation measured at the SANS-instrument D22 of Institut Laue Langevin (ILL) \cite{prunerNIMA2006}.}
\label{fig:PMMAIFM} 
\end{center}
\end{figure}

Several material classes have been investigated as candidates for holographic production of diffractive elements for neutron optics. 
The investigations of deuterated (poly)methylmethacrylate gratings led to successful tests of LLL neutron IFMs for CN by Schellhorn \emph{et al.} \cite{schellhornPhysB1997} and Pruner \emph{et al.} \cite{prunerNIMA2006} [see Fig.\,\ref{fig:PMMAIFM}\,(left)]. Those IFMs were operated in just the same manner as perfect-crystal IFMs (see Sec.\,\ref{sec:PerfectCrystalNIFM}). 
Typical intensity oscillations are shown in Fig.\,\ref{fig:PMMAIFM}\,(right). Also materials containing liquid crystals have led to promising results for neutron diffraction \cite{fallyPRL2006}.

Inorganic nanoparticles (NPs) embedded in a photopolymer matrix had already been investigated intensively
(see, for instance, \cite{suzukiAPL2002,sakhnoNT2007}) for light-optics applications. 
Including NPs in the polymer matrix increases the mechanical stability, i.e. shrinkage -- typical for polymerisation processes -- is reduced \cite{hataOL2010}. 
Long term mechanical stability has been confirmed by regularly checking the properties of certain samples with light.
The NPs so far used for neutron diffraction studies (mostly SiO$_2$) have an average core diameter of about 13\,nm with size distribution of approximately $\pm 5\,\%$. In the preparation process, a photoinitiator is added to enable the monomer to polymerize on illumination with light. Before recording, the photoinitiator, the monomer and the NPs are homogeneously distributed in the sample material. Via the photoinitiator the light-intensity pattern induces polymerization in the bright sample regions, a process that consumes monomers that diffuse from dark to bright regions \cite{tomitaOL2005}. As a consequence of the growing monomer-concentration gradient, NPs move from bright to dark regions, resulting in an approximately sinusoidal NP-concentration pattern, which can, thus, be used as neutron diffraction grating. 
An advantage of NP-polymer composites is that the refractive-index modulation could be tuned by choosing the species of NPs and the contained isotopes. 
In general, one will choose an isotope with high $b_c$. Furthermore, the optical potential could contain an absorptive or magnetic term, for instance. A grating with superparamagnetic NPs could be produced such that it reflects one spin state while being essentially transparent for the orthogonal \cite{kleppJOPConfSer2012}. 

The feasibility of a holographic-grating beamsplitter for CN has been demonstrated \cite{fallyPRL2010,kleppPRA2011}. 
Neutron diffraction experiments with free-standing NP-polymer film-gratings -- without glass plates, to decrease absorption and incoherent scattering -- have been carried out and demonstrated that 90\,\% reflectivity, i.e., mirror-like behavior, is achievable for a neutron wavelength of about 41\,\AA\,\cite{kleppAPL2012}. 
Also, a Zernike three-path IFM with high phase-sensitivity could be built \cite{kleppAPL2012b}. 
A review of the field is given in \cite{kleppMaterials2012}.

\subsection{Suppressed decoherence in neutron interferometry}

\begin{figure}
\centering\includegraphics[width=5.in]{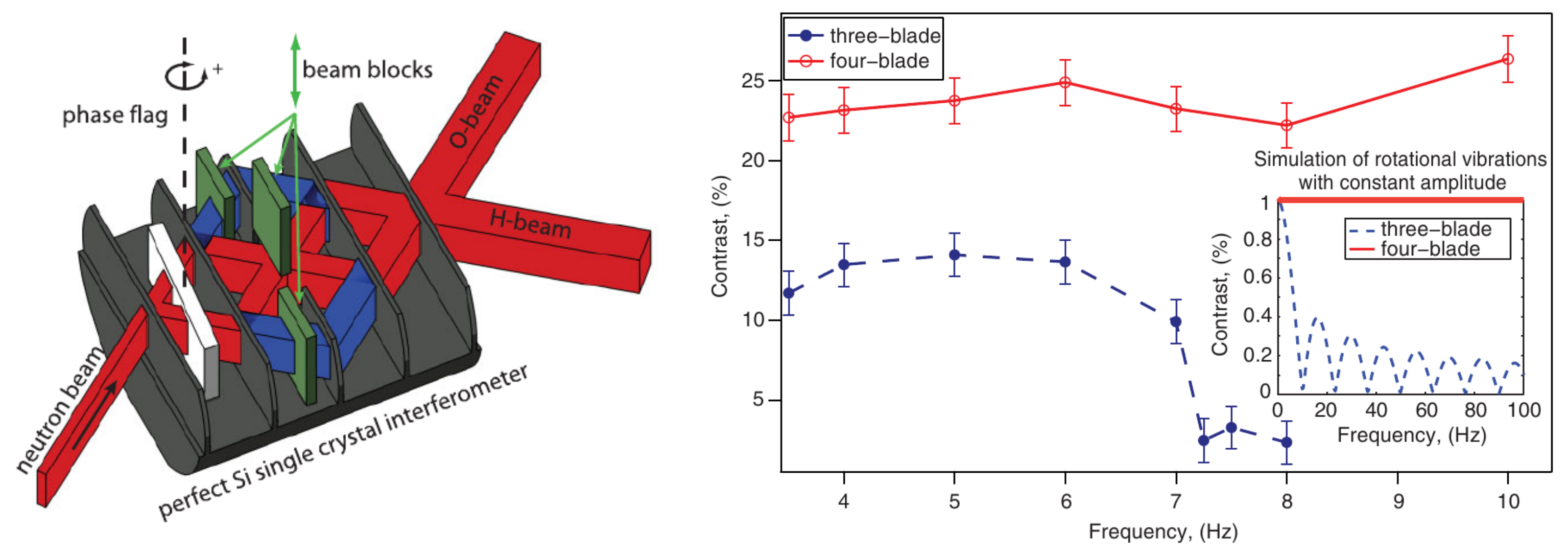}
\caption{Left: Perfect-crystal IFM that can be switched between the usual three-blade mode and four-blade mode. Right: Contrast of measured intensity oscillations versus vibrational frequency for three-blade (full symbols) and four-blade (empty symbols) modes. The inset shows a theoretical calculation for rotational vibrations. Reprinted with permission from \cite{pushinPRA2009,pushinPRL2011}. Copyright (2009,2011) by the American Physical Society.}
\label{fig:Pushin2011}
\end{figure}

More than two decades ago, a neutron-IFM layout that is insensitive to the Sagnac phase shift, an effect due to the neutron IFM being at rest in a non-inertial (rotating) frame of reference \cite{wernerPRL1979}, was conceived \cite{mashhoonPRL1988,ioffeNIMA1988}. The Sagnac effect is proportional to the area enclosed by the IFM paths. The resulting phase shift can be compensated by designing an IFM with crossed paths in the center such that the two paths enclose two equal areas with opposite orientations before and after crossing [see Fig.\,\ref{fig:Pushin2011}\,(left)].

Recently, it was noticed \cite{pushinPRA2009} that such an IFM design could also provide remedy for low-frequency, low-amplitude vibrations that usually make elaborate anti-vibration systems necessary for doing perfect-crystal interferometry. A perfect-crystal IFM was made that can be operated in standard Mach-Zehnder mode (three-blade mode) and in crossed-paths mode (four-blade mode) by removing/inserting certain Cd beam-blocks, as shown in Fig.\,\ref{fig:Pushin2011}\,(left). 
Vibrations of well-defined frequency were induced and the visibility of intensity oscillations was measured for both modes of operation \cite{pushinPRL2011}. The data shown in Fig.\,\ref{fig:Pushin2011}\,(right) demonstrates resilience of the crossed-paths-mode against such vibrations. In \cite{pushinPRL2011}, the neutron beam is modelled as a beam of particles bouncing off the Bragg-planes of the crystal slabs at a particular instant, acquiring a random phase shift. Since one measures expectation values, the obtained detector signal comprises an average over a distribution of such random phase shifts. As long as the vibrational spectrum contains only low frequencies, i.e. if the IFM is approximately in the same state of motion while a neutron passes it, all random momentum shifts accumulated from the first slab to the crossing are compensated after the crossing. 

The method can also be interpreted in terms of decoherence-free subspaces (see, for instance, \cite{lidarPRL1998}): The four possible paths in the IFM (without any beam blocks inserted) span a 4-dimensional state-space. In crossed-paths mode, only the subspace spanned by those two states that are insensitive to vibrations are used \cite{pushinPRA2009}.

\subsection{The Goos-H\"{a}nchen shift}\label{subsec:GoosHanchen}

A light beam -- considered as a superposition of plane waves -- experiences a displacement $\zeta$ in the plane of incidence along the surface of a totally reflecting medium. This phenomenon is referred to as Goos-H\"{a}nchen shift \cite{goosAP1947}. It is essentially an interference effect resulting from the slightly different phase shifts the incident plane-wave components undergo upon total reflection by a surface. These phase shifts depend on the complex reflection coefficient, the angle of incidence and the wavelength. Also, interpretations in terms of energy flux and evanescent waves are discussed \cite{renardJOSA1964,laiPRE2000,wangPRL2013}. 

\begin{figure}
\centering\includegraphics[width=5.in]{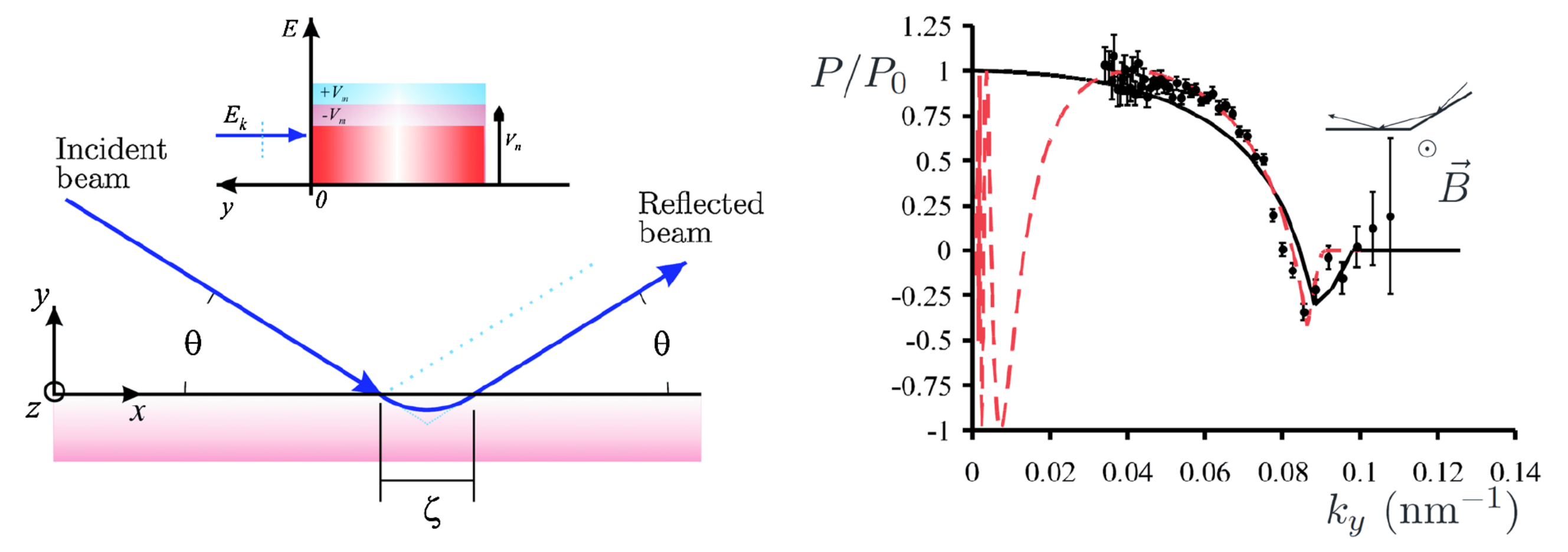}
\caption{Left: Sketch explaining the Goos-H\"{a}nchen shift $\zeta$ upon total refection of a neutron beam by a material surface. Inset: Energy of neutrons in orthogonal spin-states upon total reflection on magnetic mirror.  
Right: Normalized degree of polarization versus $k_y$. Reprinted with permission from \cite{deHaanPRL2010}. Copyright (2010) by the American Physical Society.}
\label{fig:DeHaan2010}
\end{figure}

For neutrons, there have been a couple of proposals on how to pin down the sub-micron effect (see, for instance, \cite{maazaOC1997,ignatovichPLA2004}). Experimental data was limited to measurements of the reflectivity of magnetized surfaces for up- and down-spin incident beams due to the difference in the optical potential for orthogonal spin states [see Fig.\,\ref{fig:DeHaan2010}\,(left,\,inset)] without explicit reference to the Goos-H\"{a}nchen shift \cite{topervergPB2005}.  

In \cite{deHaanPRL2010}, an experiment is described that measured the polarization rotation upon total reflection occurring for an incident beam prepared in a spin superposition. This rotation is a result of relative phase that the up- and down-spin components accumulate on reflection from a magnetic material. The relative phase can -- in turn -- be attributed to the different distances the up- and down-spin components cover within the magnetic material due to the difference in Goos-H\"{a}nchen shifts $\zeta_+$ and $\zeta_-$ for up- and down-spin states. 
The results for two subsequent reflections are shown in Fig.\,\ref{fig:DeHaan2010}\,(right). The solid line is the theoretical prediction, while data points follow a simulated curve (dashed line) that includes a small magnetic anisotropy of the surface as well as a small dispersive phase. 

The conclusions drawn were criticized in a comment by Ignatovich, who points out that the experiment does not constitute a direct measurement of the Goos-H\"{a}nchen shift because there is no clear-cut relation between Larmor precession and $\zeta_{\pm}$ \cite{ignatovichPRL2010,deHaanPRL2010b}.
The issue is still to be clarified by an experiment that does not rely on the neutron spin and can directly measure the Goos-H\"{a}nchen shift $\zeta$ of an unpolarized neutron beam upon total reflection by a non-magnetic surface.  

\subsection{Neutron whispering-gallery modes}\label{subsec:Whispering}

The idea that neutrons occupy discrete energy levels in particular forms of potentials like a series of narrow slits in a silicon slab \cite{rauchNature2002} or the gravitational potential \cite{nesvizhevskyNature2002,jenkeNatureP2011} has inspired a neutron-optical experiment on a phenomenon that is well-known from acoustics: words spoken in a low voice at one position in elliptically shaped galleries can -- under certain conditions -- be clearly heard at the far end. 
The effect is also observed in optics and was explained for the first time as long as a century ago \cite{rayleighPM1914}. Looking back at all the analogies between light- and matter-wave-optics it is only natural to expect whispering-gallery modes also for neutron waves travelling in a curved surface of a mirror. Indeed, such an experiment was carried out recently at the ILL \cite{nesvizhevskyNatureP2010,rauchNatureP2010}. 

\begin{figure}
\centering\includegraphics[width=5.0in]{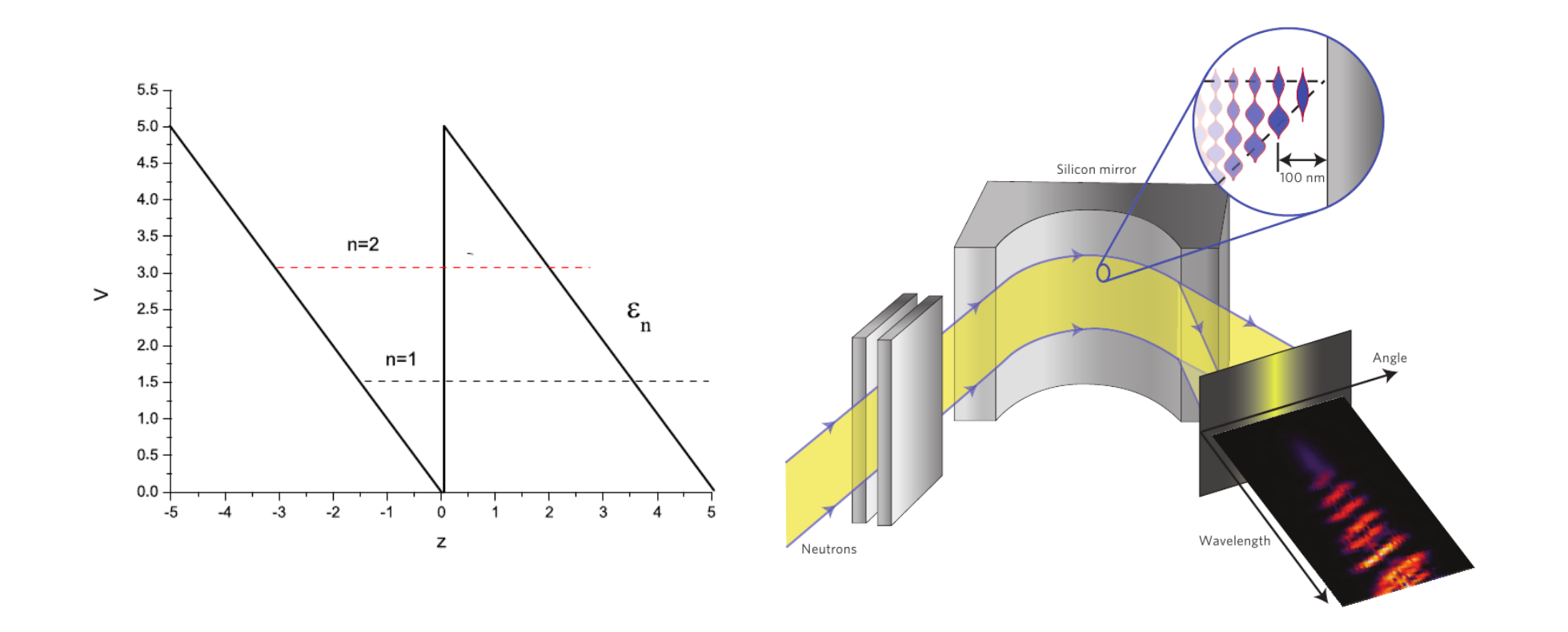}
\caption{Left: Effective potential V as formed by the combination of the curved-mirror Fermi potential and the centrifugal force close to the mirror surface. Right: Scheme of the experiment \cite{nesvizhevskyNatureP2010}. Reprinted with permission from \cite{rauchNatureP2010}. Copyright (2010) by Alain Filhol (ILL).}
\label{fig:Nesvizhevsky2010}
\end{figure}

The principle of the experiment is the following:
A collimated neutron beam is guided to a curved silicon perfect-crystal mirror [Fig.\,\ref{fig:Nesvizhevsky2010}\,(right)]. 
Its Fermi-potential together with the centrifugal force on the neutrons forms a triangular effective potential-well that results in bound neutron states near the mirror surface [see Fig.\,\ref{fig:Nesvizhevsky2010}\,(left); $n=1,2,\dots$ are the populated neutron energy levels]. The slopes of the effective potential are determined by the centrifugal acceleration $a=v/R^2$, where $v$ and $R$ are the neutron velocity and the radius of curvature of the silicon mirror, respectively \cite{nesvizhevskyPRA2008}.
At the exit of the mirror, interference fringes could be detected as a function of reflection angle and wavelength using a time-of-flight technique and a position-sensitive detector. A broad incident spectrum from thermal neutrons to VCN (2\,\AA\,to 30\,\AA ) was used for the experiment, so that interference could be observed starting from the wavelength cut-off (too short wavelengths) to very long-wavelengths bound-states. The effect holds some potential for novel 
surface-analysis methods with neutrons. 

\subsection{Concluding Remarks and Outlook}\label{sec:concl}
In this paper, we have presented a survey of neutron-optical experiments investigating quantum-mechanical phenomena of fundamental nature. Peculiarities of quantum theory, such as quantum contextuality, multi-partite entanglement of single-neutrons, topological phases or decoherence effects are covered. Starting from historical experiments such as the verification of the 4$\pi$-symmetry of the spin-1/2 wave function or the famous COW experiment, subjects of current scientific debates, as, for instance, violation of Heisenberg's error-disturbance uncertainty relation, are discussed. The methods utilized in the presented experiments range from neutron interferometry and polarimetry with thermal neutrons or holographic-grating interferometry with VCN to storage experiments using UCN. The experiments have been performed on different neutron sources worldwide, just to mention the ILL (Grenoble, France), Atominstitut (Vienna, Austria), ISIS (Oxfordshire, UK), NIST (Gaithersburg, USA) or KURRI (Kyoto, Japan), reflecting an unquenchable thirst for insight into fundamental issues of quantum mechanics across borders and continents. 

Upcoming neutron-optical experiments concern so-called weak measurements, a new measurement concept introduced by Aharonov, Albert and Vaidman \cite{aharonovPRL1988}. Such experiments may help to further broaden our understanding of quantum phenomena. The procedure involves three steps: \emph{(i)} quantum state preparation (preselection), \emph{(ii)} a weak perturbation, i.e., measurement of an observable that disturbs the system only weakly, and \emph{(iii)} postselection of the final quantum state. Weak measurements are under discussion to be used for amplification of minute effects for precision measurements or to illuminate quantum paradoxes such as the quantum Ceshire-Cat (disembodiment of a particle and its properties \cite{aharonovNJP2013}) and the three-box paradox \cite{reschPLA2004}.
Also, based on the accomplishments of neutron interferometry and polarimetry described in the present review, new aspects of wave-particle duality could come to light.

\section*{Acknowledgment}

The authors would like to thank their colleagues who helped to realize many of the experiments described here. Furthermore, the authors are grateful for being -- in many different ways -- supported by Gerald Badurek, Martin Fally, Peter Geltenbort, Hartmut Lemmel, Masanao Ozawa, Christian Pruner, Helmut Rauch, Georg Sulyok and Yasuo Tomita.
Y.H. would like to thank Prof. A. Hosoya for long-lasting encouragement.
This work was partly supported by the Austrian Science Fund (FWF), Project No. P25795-N20 and P24973-N20.


%

\vfill\pagebreak


\end{document}